\documentclass[a4paper, 12pt]{elsarticle}

\pdfoutput=1

\usepackage{graphicx}
\usepackage{amssymb}
\usepackage{amsmath}
\usepackage{bm}

\usepackage{appendix}

\usepackage{color}

\usepackage{hyperref}

\makeatletter
\def\ps@pprintTitle{%
     \let\@oddhead\@empty
     \let\@evenhead\@empty
     \def\@oddfoot{\footnotesize\itshape
     {\copyright 2019. This manuscript version is made available under the
      \href{https://creativecommons.org/licenses/by-nc-nd/4.0/}{CC-BY-NC-ND 4.0 license}.}
      \hfill}
     \let\@evenfoot\@oddfoot}

\newcommand\Appx[1]{Appendix~\ref{#1}}

\newcommand\chE[1]{{#1}}
\newcommand\chV[2]{{{#1}}}

\def\norm#1{\left\lVert#1\right\rVert}
\def\shp{\varsigma}

\newcommand\dlt{\delta}          

\newcommand\dd{\mathrm{d}}
\newcommand\nrm[2]{\left\| {#1}\right\|_{#2}}  

\newcommand\hh{\bar{h}}

\newcommand\cwto{\rightharpoonup}
\newcommand\csto{\rightarrow}

\newcommand\MeanY[1]{\Mcal_{Y}{\left (#1\right )}}
\newcommand\MeanYe[1]{\Mcal_{Y}^\veps{\left (#1\right )}}
\newcommand\plper{{\ul{\#}}} 

\newcommand\ipGm[2]{\left \langle{#1},\,{#2}\right\rangle_{\Gamma_0}}
\newcommand\trace[2]{{\rm{trace}}_{#1}({#2})}

\newcommand\Lpar{[\kern-0.17em{[}}
\newcommand\Rpar{]\kern-0.18em{]}}

\newcommand\Jump[2]{\Lpar {#1} \Rpar_{{#2}}^\pm}

\def\nablad{\hat{\nabla}}
\def\gradpl{\ol{\nabla}}
\def\gradplx{\ol{\nabla}_x}
\def\gradplxS{\ol{\nabla}_x^S}
\def\gradply{\ol{\nabla}_y}
\def\gradplS{\ol{\nabla}^S}
\def\gradplyS{\ol{\nabla}_y^S}
\def\hom{{\rm{H}}}
\def\ext{{\rm{ext}}}

\def\Hpdb{{\bf{H}}_\#^1}

\def\wrt{{w.r.t.{~}}}
\def\rhs{{r.h.s.{~}}}
\def\lhs{{l.h.s.{~}}}
\def\ie{{\it i.e.~}}
\def\eg{{\it e.g.~}}

\newcommand{\eq}[1]{(\ref{#1})}

\def\bmi#1{\textbf{\textit{#1}}}

\newcommand{\ul}[1]{\underline{#1}}
\newcommand{\ol}[1]{\overline{#1}}

\def\pdiff#1#2{\frac{\partial {#1}}{\partial {#2}}}

\def\TL{{\textit{TL}}}

\def\psibf{{\mbox{\boldmath$\psi$\unboldmath}}}
\def\chibf{{\mbox{\boldmath$\chi$\unboldmath}}}
\def\sigmabf{{\mbox{\boldmath$\sigma$\unboldmath}}}
\def\vthetabf{{\mbox{\boldmath$\vartheta$\unboldmath}}}
\def\xibf{{\mbox{\boldmath$\xi$\unboldmath}}}
\def\thetabf{{\mbox{\boldmath$\theta$\unboldmath}}}
\def\Pibf{{\mbox{\boldmath$\Pi$\unboldmath}}}

\def\R{\hbox{\rm I\kern-0.2em R}}
\def\Z{\hbox{\rm Z\kern-0.3em Z}}

\def\vtheta{\vartheta}
\def\vphi{\varphi}
\def\veps{\varepsilon}
\def\vepsdel{{\varepsilon\delta}}
\def\vkappa{\varkappa}

\def\imu{{\rm{i}}} 

\def\Om{\Omega}
\def\om{\omega}
\def\pd{\partial}

\def\RR{{\mathbb{R}}}
\def\ZZ{{\mathbb{Z}}}

\def\ab{{\bmi{a}}}
\def\bb{{\bmi{b}}}

\def\eb{{\bmi{e}}}
\def\fb{{\bmi{f}}}

\def\kb{{\bmi{k}}}
\def\mb{{\bmi{m}}}
\def\nb{{\bmi{n}}}

\def\ub{{\bmi{u}}}
\def\vb{{\bmi{v}}}
\def\wb{{\bmi{w}}}

\def\Ab{{\bmi{A}}}
\def\Bb{{\bmi{B}}}
\def\Cb{{\bmi{C}}}
\def\Db{{\bmi{D}}}

\def\Hb{{\bmi{H}}}
\def\Ib{{\bmi{I}}}

\def\Pb{{\bmi{P}}}

\def\Sb{{\bmi{S}}}
\def\Tb{{\bmi{T}}}

\def\Wb{{\bmi{W}}}
\def\Zb{{\bmi{Z}}}

\def\Acal{\mathcal{A}}
\def\Bcal{\mathcal{B}}
\def\Ccal{\mathcal{C}}
\def\Dcal{\mathcal{D}}
\def\Ecal{\mathcal{E}}
\def\Fcal{\mathcal{F}}

\def\Hcal{\mathcal{H}}

\def\Jcal{\mathcal{J}}
\def\Kcal{\mathcal{K}}
\def\Lcal{\mathcal{L}}
\def\Mcal{\mathcal{M}}
\def\Ncal{\mathcal{N}}

\def\Scal{\mathcal{S}}
\def\Tcal{\mathcal{T}}

\def\pop#1{\widetilde{#1}}  
\def\Tuf#1{{\mathcal{T}}_\veps{\left ({#1}\right )}} 
\newcommand\Tuftxt{\Tcal_\veps\,}

\def\ipYs#1#2{\left ({#1},\,{#2}\right )_{Y^*}}
\def\intY{\sim \kern-1.2em \int}

\def\Eop{{{\rm I} \kern-0.2em{\rm E}}}%

\newcounter{theorem}
\newenvironment{mytheorem}[1]%
{\vspace{7pt} \noindent{\bf Theorem
  \refstepcounter{theorem}\thetheorem. \protect\label{#1}
\hspace{-2mm}}\it}%
{\vspace*{7pt}}

\newcounter{remark}
\newenvironment{myremark}[1]%
{\vspace{7pt} \noindent{\bf Remark
  \refstepcounter{remark}\theremark. \protect\label{#1} \hspace{-2mm}}\rm }%
{\vspace*{-7pt} \flushright $\triangle$\\ } 

\begin{document}

\begin{frontmatter}

\title{Homogenization of the vibro--acoustic transmission on perforated plates}

\author[NTIS]{E.~Rohan\corref{cor1}}
\ead{rohan@kme.zcu.cz}
\cortext[cor1]{Corresponding author}
\author[NTIS]{V.~Luke\v{s}}
\ead{vlukes@kme.zcu.cz}
%
%
\address[NTIS]{European Centre of Excellence,
NTIS New Technologies for Information Society,
Faculty of Applied Sciences,\\ University of West Bohemia, \\
Univerzitn\'\i~22, 30614 Plze\v{n}, Czech Republic}

\begin{abstract}
The paper deals with modelling of acoustic waves which propagate in inviscid
fluids interacting with perforated elastic plates. The plate can be replaced by
an interface on which transmission conditions are derived by homogenization of
a problem describing vibroacoustic fluid-structure interactions in a
transmission layer in which the plate is embedded. The Reissner-Mindlin theory
of plates is adopted for periodic perforations designed by arbitrary
cylindrical holes with axes orthogonal to the plate midplane. The homogenized
model of the vibroacoustic transmission is obtained using the two-scale
asymptotic analysis with respect to the layer thickness which is proportional
to the plate thickness and to the perforation period. The nonlocal, implicit
transmission conditions involve a jump in the acoustic potential and its normal
one-side derivatives across the interface which represents the plate with a
given thickness. The homogenized model was implemented using the finite element
method and validated using direct numerical simulations of the non-homogenized
problem. Numerical illustrations of the vibroacoustic transmission are
presented.
\end{abstract}


\end{frontmatter}


\section{Introduction}\label{sec-intro}

The noise and vibration reduction belongs to important issues in design of
structures used in the automotive industry, or civil engineering. The engine
silencer used to reduce the noise emitted by the exhaust gas presents an
important and well known example. However, there are many similar solid structures
which can influence the acoustic wave propagation in fluid. Usually they involve porous,
or perforated plates, or panels, such that they are permeable for the gas flow.
The straightforward approach to modelling the acoustic wave propagation through
vibrating perforated plates consists in solving directly the vibroacoustic
problem with a 3D elastic structure describing the plate. However, its
numerical treatment using the finite element method can lead to an intractable
problem because of the prohibitive number of DOFs corresponding to the
geometric complexity of the perforated structure. Therefore, it is reasonable
to replace the elastic plate by an interface on which coupling transmission
conditions are prescribed.

In this paper, we consider the acoustic wave propagation in an inviscid fluid interacting with elastic structures designed as periodically perforated plates.
The aim is to derive non-local vibro-acoustic transmission
conditions using the periodic homogenization method. Although similar problems
have been treated in the literature, cf. \cite{DELOURME201228}, in this
context, the plate elasticity has not been considered yet. As for the rigid
structures, semi-empirical formulae for the acoustic impedance exist which were
tuned by experiments, or developed using the electro-acoustic equivalent
circuit theory \cite{Jung-etal-2007-JKPS,Sakagami2010,Stremtan2012}, or the
Helmholtz-Kirchhoff integral theory \cite{Zhou2013}. During the last decade, a
number of works appeared which are based on a homogenization strategy. For a
thin rigid perforated plate represented by interface $\Gamma_0$ and
characterized by the thickness $\approx \dlt$ it has been shown in
\cite{bb2005,DELOURME201228} that this interface is totally transparent for the
acoustic field at the zero order $\dlt^0$ terms of the model which describes
the limit behaviour for $\dlt \rightarrow 0$, cf. \cite{Dorlemann2017}. For a
higher order approximation, an approach based on the so-called inner and outer
asymptotic expansions has been developed, such that two associated acoustic
fields are coupled, one being relevant in the proximity of the perforations,
the other at a distance from the limit interface, see \eg
\cite{Clayes-Delourm-AA2013,Marigo-Maurel-JASA2016,Marigo-Maurel-PRSA2016}. In
contrast with \cite{bb2005} dealing with thin perforated interfaces only, in
\cite{rohan-lukes-waves07} we were concerned with homogenization of a
fictitious layer in which rigid periodically distributed obstacles were placed.
In particular, a rigid plate perforated by arbitrary shaped pores could be
considered. Therein nonlocal transmission conditions were obtained as the
two-scale homogenization limit of a standard acoustic problem imposed in the
layer.

Here we follow the approach reported in
\cite{rohan-lukes-waves07} to develop vibroacoustic transmission conditions
which substitute the vibroacoustic interaction on an
elastic perforated plate immersed in the acoustic fluid. Up to our knowledge, despite some numerical studies, see \eg.
\cite{Takahashi-Tanaka2002}, a rigorous treatment of such a problem has not
been treated using the homogenization method so far. As the result we obtain
vibroacoustic transmission conditions in a form of an implicit
Dirichlet-to-Neumann operator. Due to this operator, the elastic perforated
plate can be replaced by an interface on which a jump of the global acoustic
pressure is linked to the acoustic momenta associated with two faces of the
homogenized plate. It allows us to obtain an efficient numerical model which
takes into account geometrical details of the periodic perforation without need
of discretizing the vibroacoustic problem at the global level. In other words,
the homogenized interface provides a reduced model in which a complex 3D
elastic structure is replaced by a 2D perforated plate model whose coefficients
retain information about the perforation geometry. To do so, we rely on the
homogenized Reissner-Mindlin plate tailor-made for the ``simple'' perforation
represented by general cylindrical holes with axes orthogonal to the mid-plane
of the plate. Elastic strongly heterogeneous plates were treated in
\cite{rohan-miara-CRAS2011,rohan-miara-ZAMM2015} where the framework of the Reissner-Mindlin theory
was used to derive a model of phononic plates, cf. \cite{Rohan2015bg-plates}, but without the interaction with
an exterior acoustic field.
  
The proposed modelling conception based on the problem decomposition and using
the homogenization provides an alternative framework for modelling of
microporous panels which are known for their capabilities of acoustic
attenuation \cite{Toyoda-JSV2005,Zhou2013,LIU2016149}. In \cite{Maxit-JASA2012}
the so-called patch transfer functions were developed for numerical modelling
of compliant micro-perforated panels.


The plan of the paper is as follows. In Section~\ref{sec-problem} the vibroacoustic problem of the wave propagation in a waveguide containing the perforated plate is decomposed into the problem in a fictitious transmission layer (the ``in-layer'' problem) and the ``outer'' problem governing the acoustic field out of the layer. The ``in-layer'' vibroacoustic problem is treated using the homogenization method in Section~\ref{sec-homog},
where the local problems imposed in the representative periodic cell are introduced and
formulae for the homogenized coefficients are given.  In
Section~\ref{sec-global}, as the main result of this paper, the global acoustic problem is established using the limit ``in-layer'' and the ``outer'' problems which are coupled using additional conditions derived by an additional integration and averaging procedure.
The limit two-scale model of the homogenized layer is validated in Section~\ref{sec-valid} using direct numerical simulations of the original problem. Finally, in Section~\ref{sec-simul}, the proposed model is employed to simulate wave propagation in a waveguide equipped with the perforated plate. Some technical auxiliary derivations are presented in the Appendix.

\paragraph{Notation.}
In the paper, the mathematical models are
formulated in a Cartesian coordinate system $\mathcal{R}(\text{O};\eb_1,\eb_2,\eb_3)$ where $O$ is the origin of the space and
$(\eb_1,\eb_2,\eb_3)$ is a orthonormal basis for this space. The spatial position $x$ in the medium is specified through the coordinates $(x_1,x_2,x_3)$ with respect to a Cartesian reference frame $\mathcal{R}$. The boldface
notation for vectors, $\ab = (a_i)$, and for tensors,  $\bb = (b_{ij})$,
is used. 
 The gradient and divergence operators applied to a vector $\ab$ are
denoted by $\nabla\ab$ and $\nabla \cdot \ab$, respectively. By $\nabla^S\ub$ we denote the symmetrized gradient $\nabla\ub$, \ie the strain tensor. When these operators have a subscript
which is space variable, it is for indicating that the operator acts relatively
at this space variable, for instance $\nabla_x = (\pd_i^x)$. 
The symbol dot `$\cdot$' denotes the scalar product
between two vectors and the symbol colon `$:$' stands for scalar (inner) product
of two second-order tensors.
Throughout the paper, $x$ denotes the global (``macroscopic'') coordinates, while the ``local''
coordinates $y$ describe positions within the representative unit cell
$Y\subset\RR^3$ where $\RR$ is the set of real numbers. By latin subscripts $i,j,k,l \in\{1,2,3\}$  we refere to vectorial/tensorial components in $\RR^3$, whereas subscripts $\alpha,\beta \in \{1,2\}$ are reserved for the tangential components with respect to the plate midsurface, \ie coordinates $x_\alpha$ of vector represented by $x' = (x_1,x_2) = (x_\alpha)$ are associated with directions $(\eb_1,\eb_2)$. Moreover, $\gradplx = (\pd_\alpha)$ is the ``in-plane'' gradient.
The gradient in the so-called dilated configuration with coordinates $(x',z)$ is denoted by $\hat \nabla = (\gradpl, \frac{1}{\veps} \pd_z)$.
{We also use the jump \wrt the transversal coordinate, $\Jump{q(\cdot,x_3)}{r} = q(\cdot,r/2) - q(\cdot,-r/2)$.}

\begin{figure}[t]
\centerline{\includegraphics[width=0.95\linewidth]{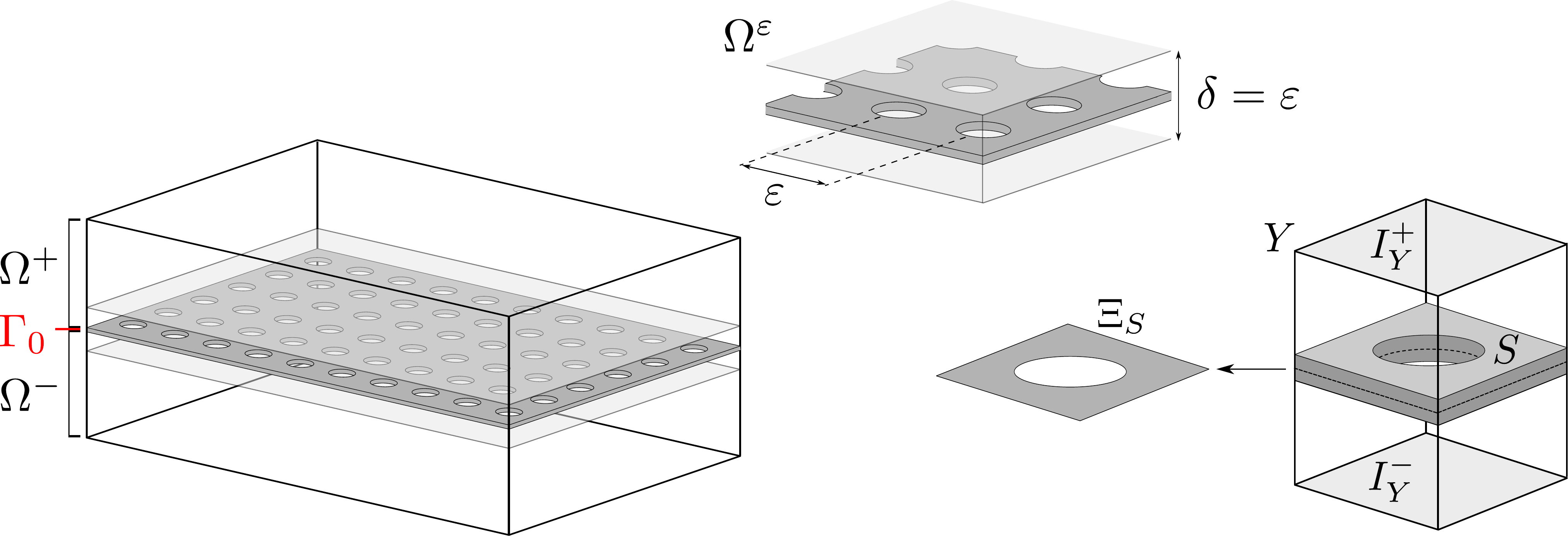}}
\caption{Left: The global domain decomposition into $\Om^+$ and
  $\Om^-$ separated by the homogenized perforated plate represented by
  the interface $\Gamma_0$. Center: detail of the layer structure; the
  layer thickness is proportional to the plate thickness and to the
  perforation period. Right: perforated interface and the representative periodic
  cell $Y = Y^*\cup\ol{S}$.}\label{fig-1}
\end{figure}

\section{Formulation and decomposition of the vibroacoustic transmission problem}
\label{sec-problem}
The aim of the paper is to find a representation of the vibro-acoustic
interaction on a perforated plate. For this, homogenized vibroacoustic
transmission conditions are derived using the asymptotisc analysis \wrt a scale
parameter $\veps$ which has a double role: on one hand it deals with the
thickness of an elastic plate when considered as a 3D object, on the other hand
it describes the size and spacing of holes periodically drilled in the plate
structure.

The flowchart of deriving the transmission conditions for a limit global
problem consists of the following steps:

\begin{itemize}

\item The vibro-acoustic problem (later called the ``global problem'') is
formulated in a domain $\Om^G\subset \RR^3$ in which the perforated elastic
plate is embedded, being represented by a planar surface -- the plate
midsurface.

\item A transmission layer $\Om_\dlt$ of the thickness $\dlt$ is introduced in
terms of $\Gamma_0$ which constitutes its midsurface. This will allow to
decompose the global problem into two subproblems: the vibroacoustic
interaction in the layer $\Om_\dlt$ and the outer acoustic problems in
$\Om^G\setminus \Om_\dlt$. The two subproblems are coupled by natural
transmission conditions on the ``fictitious'' interfaces $\Gamma_\dlt^\pm$.

\item We consider the layer thickness being proportional to the scale
parameter, thus, $\delta = \vkappa \veps$, where $\vkappa>0$ is fixed. The
asymptotic analysis $\dlt\approx \veps\rightarrow 0$ considered for the problem
in $\Om_\dlt$ with the Neumann type boundary conditions on $\Gamma_\dlt^\pm$
leads to the homogenized vibroacoustic transmission problem defined on
$\Gamma_0$. In this analysis, $\veps$ has the double role announced above and
the plate is described using its 2D representation in the framework of the
Reissner-Mindlin plate theory. In Remark~\ref{rem-h} we explain the dual
interpretation of the plate thickness used in the asymptotic analysis of the
vibro-acoustic problem.

\item The final step is to derive the limit global problem for the acoustic
waves in the fluid interacting with the homogenized perforated plate
represented by $\Gamma_0$. For this, with a few modifications we follow the
approach used in \cite{rohan-lukes-waves07}, where the rigid plate was considered;
a given plate thickness $h$ corresponds to given finite thickness $\dlt_0$ of
the transmission layer. Then the continuity of the acoustic fields on
interfaces $\Gamma_{\dlt_0}^\pm$ yields the homogenized vibroacoustic
transmission conditions which hold on $\Gamma_0$.

\end{itemize}

\begin{figure}
  \chV{\centering
  \includegraphics[width=0.98\linewidth]{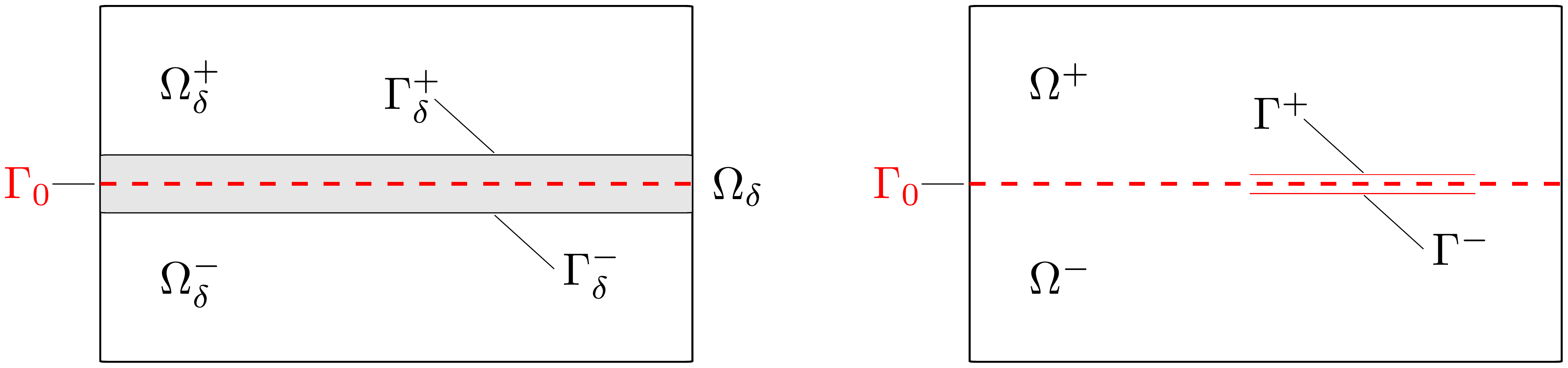}}{}
\caption{Left: Transmission layer $\Om_\dlt$ of thickness $\dlt$ embedded in the global domain $\Om^G$. Right: Interface $\Gamma_0$ representing the homogenized transmission layer.}\label{fig-problem-decomposition}
\end{figure}

\subsection{Global problem with transmission layer}

\begin{figure}
\centerline{\includegraphics[width=0.8\linewidth]{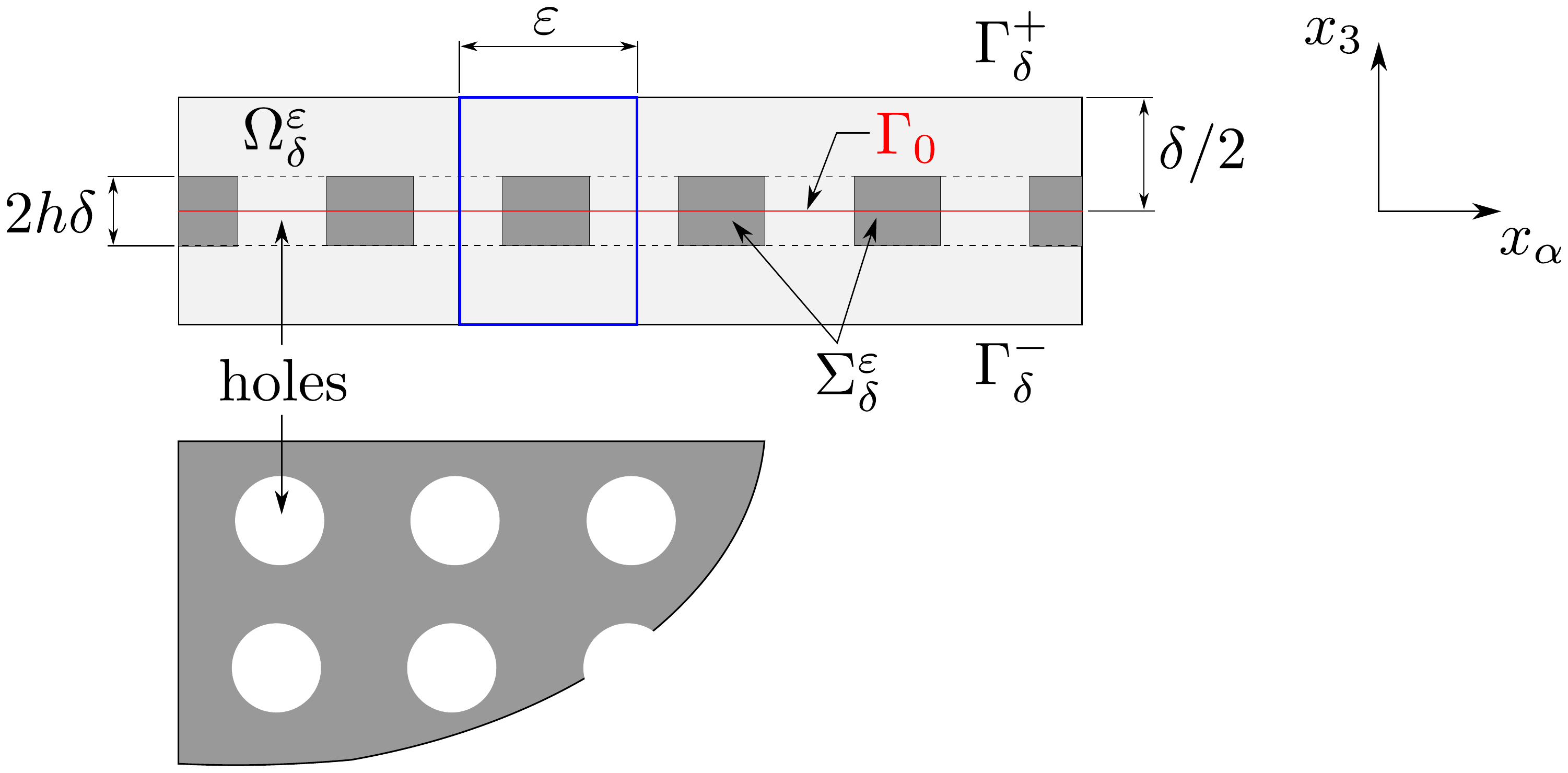}} 
\caption{Scheme of the transmission layer $\Om_\dlt$ in which the perforated plate $\Sigma_\dlt^\veps$ (dark gray) is embedded. The complementary domain $\Om_\dlt^\veps$ is occupied by the acoustic fluid (light gray).}\label{fig-scheme-layer}
\end{figure}

In this section, we introduce the problem of acoustic waves in a domain $\Om^G$
with embedded perforated elastic plate $\Sigma^\veps$, see Fig.~\ref{fig-problem-decomposition} and  Fig.~\ref{fig-scheme-layer}. The acoustic fluid
occupies the domain $\Om^{*\veps}$. We consider a fictitious transimssion layer
$\Om_\dlt$ with a thickness $\dlt > 0$, such that $\Sigma^\veps \subset
\Om_\dlt$. The plate thickness is $h^\veps = \veps\bar{h}$, while the layer
thickness $\dlt = \vkappa \veps$ for a given fixed $\vkappa > 0$.

For a fixed parameter $\veps > 0$, correspondingly to $\Om^{*\veps}$ and
$\Sigma^\veps$, we use a simplified notation $\Om^*$ and $\Sigma$. The acoustic
harmonic wave with the frequency $\omega$ is described by the acoustic
potential $p:\Om^{*}\ni x \mapsto\RR^3$ in the fluid, the corresponding wave in
the elastic body is described by the displacement field $\ub:\Sigma^{}\ni x
\mapsto\RR^3$. Assuming the body is fixed to a rigid frame on the boundary
$\pd_\ub\Sigma$ and interacting with the fluid on $\pd_*\Sigma =
\pd\Sigma\setminus\pd_\ub\Sigma$, these fields satisfy the following
equalities:
\begin{equation}\label{eq-G1}
\begin{split}
c^2 \nabla^2 p + \om^2 p & = 0 \quad \mbox{ in } \Om^{*}\;,\\
\nabla\cdot \sigmabf(\ub) + \om^2 \rho \ub & = 0 \quad \mbox{ in } \Sigma^{}\;,\\
\mbox{ acoustic transmission: }& \\
\left.
\begin{array}{rcl}
\imu \om \nb\cdot\ub & = & \nb\cdot\nabla p\\
\nb \cdot \sigmabf(\ub) & = & \bb(p)  = \imu \om \rho_0 p \nb
\end{array}
\right\}& \quad \mbox{ on } \pd_* \Sigma\;,\\
 \mbox{ incident, or reflected acoustic waves in the fluid: }& \\
r \imu\om c p + c^2 \pdiff{p}{n} & = s 2\imu\om c\bar p
\quad  \mbox{ on } \pd_\ext \Om^G\setminus \pd_\ub\Sigma\;, \\
 \mbox{ clamped elastic structure: }& \\
\ub & = 0 \quad  \mbox{ on } \pd_\ub\Sigma\;.
\end{split}
\end{equation}
Above, $c$ is the sound speed in the acoustic fluid, $\sigmabf(\ub)$ is the
stress in the lienar elastic solid, $\rho_0$ is reference fluid density, and by
$\nb = (n_i)$ we denote the normal vector. The constants $r,s \in \{0,1\}$ and
$\bar p$ are defined to describe incident, reflected, or absorbed acoustic
waves in the fluid, according to a selected part of the boundary.

\subsection{Geometry of the perforated layer}

Given a bounded 2D manifold $\Gamma_0 \subset \{x \in \Om^G| x_3 = 0\}$
representing the plate mid-plane, we introduce $\Om_\delta = \Gamma_0 \times
]-\delta/2,\delta/2[ \subset \Om^G$, an open domain representing the
transmission layer. This enables to decompose $\Om^G$ into three nonoverlapping
parts, as follows: $\Om^{G} =\Om_\dlt \cup \Om_\dlt^+ \cup \Om_\dlt^-$. Thus,
the transmission layer is bounded by $\pd \Om_\delta$ which splits into three
parts:
\begin{equation}\label{eq-1}
\begin{split}
\pd \Om_\delta = \Gamma_\delta^+ \cup  \Gamma_\delta^- \cup \pd_\ext \Om_\delta\;,
\quad \Gamma_\delta^\pm = \Gamma_0 \pm {\delta\over{2}} \vec{e_3}\;,
\quad \pd_\ext \Om_\delta = \pd \Gamma_0 \times  ]-\delta/2,\delta/2[\;,
\end{split}
\end{equation}
where $\delta > 0$ is the layer thickness and $\vec{e_3} = (0,0,1)$, see
Fig.~\ref{fig-1}. In the context of the transmission layer definition, we
consider the plate as a 3D domain $\Sigma^{\veps}$ defined in terms of the
perforated midsurface $\Gamma^{\veps}$; the following definitions are employed:
\begin{equation}\label{eq-hh}
\begin{split}
\Sigma^{\veps} & = \Gamma^{\veps}\times \veps \hh ]-1/2, +1/2[\;,\\
\pd\Sigma^{\veps} & = \pd_\circ\Sigma^{\veps} \cup \pd_+ \Sigma^{\veps} \cup  \pd_- \Sigma^{\veps} \cup \pd_\ub\Sigma^{\veps}\;,\\
\mbox{ where } & \\
\pd_\circ\Sigma^{\veps} & = \pd_\circ\Gamma^\veps\times \veps \hh ]-1/2, +1/2[ \;,\\
\pd_\pm \Sigma^{\veps} & = \Gamma^{\veps} \pm \veps \hh/2\;,
\end{split}
\end{equation}
where $\pd_\ub\Sigma^{\veps}$ is the surface where the plate is clamped.

The midsurface $\Gamma^{\veps}$ representing the perforated plate is generated
using a representative cell $\Xi_S \subset \RR^2$, as a periodic lattice. Let
$\Xi = {]0,\ell_1[}\times{]0,\ell_2[}$, where $\ell_1,\ell_2 > 0$ are given
(usually $\ell_1=\ell_2=1$) and consider the hole $\Xi^* \subset \Xi$, whereas
its complement $\Xi_S =\Xi\setminus \ol{\Xi^*}$ defines the solid plate
segment. Then
\begin{equation}\label{100}
\begin{split}
\Gamma^\veps & = \bigcup_{k \in \ZZ^2}  \veps\left(\Xi_K + \sum_{i=1,2}k_i\ell_i\vec{e_i} \right)\cap \Gamma_0 \;,
\end{split}
\end{equation}
Further we introduce the representative periodic cell $Y$ and define its solid part
$S \subset Y$, 
\begin{equation}\label{101}
\begin{split}
Y & = \Xi \times {]-\vkappa/2,+\vkappa/2]}\;,\\
S & = \Xi_S \times{\hh }  ]-1/2,+1/2[\;,
\end{split}
\end{equation}
so that $Y^* = Y\setminus \ol{S}$ is the fluid part. Obviously, in the
transmission layer $\Om_\dlt$, the fluid occupies the part
\begin{equation}\label{102}
\begin{split}
\Om^{*\veps} & = \bigcup_{k \in \ZZ^2}\veps(Y^* + \sum_{i=1,2}k_i\ell_i\vec{e_i})\cap \Om_\dlt\;,
\end{split}
\end{equation}
where $\vec{e_1} = (1,0,0)$ and $\vec{e_2} = (0,1,0)$.

For completeness, by virtue of \eq{eq-hh} we can introduce the decomposition of
boundary $\pd S = \pd_\circ S \cup \pd_\pm S \cup \pd_\# S$. For this we need
the boundary $\pd \Xi_S = \pd_\circ \Xi_S \cup \pd_\#\Xi_S$, where $\pd_\#\Xi_S
\equiv \pd \Xi$, so that the closed curve $\pd_\circ \Xi_S = \pd \Xi^*$
generates the cylindrical boundary $\pd_\circ S$:
\begin{equation}\label{103}
\begin{split}
\pd_\circ S & = \pd_\circ\Xi_S \times{\hh }  ]-1/2,+1/2[\;,\\
\pd_\pm S & = \Xi_S \pm \vec{e_3}\hh/2\;,\\
\pd_\# S & = \pd\Xi \times{\hh }  ]-1/2,+1/2[\;.
\end{split}
\end{equation}
For the sake of simplicity, by $\pd \Xi_S$ we shall refere to $\pd_\circ  \Xi_S$.

\subsection{Problem decomposition}

The domain split allows us to decompose problem \eq{eq-G1} into three parts. By
$P^\dlt$ we denote the acoustic potential in $\Om_\dlt^+$ and $\Om_\dlt^-$,
whereas $p^\veps$ is the acoustic potential in the transmission layer
$\Om_\delta$, see Fig.~\ref{fig-scheme-layer}. Further, by $\imu\om g^{\veps\pm}$ we denote the acoustic fluid
velocity projected into the normal of the interfaces $\Gamma_\dlt^\pm$. The
following subproblems are considered:
\begin{list}{}{}
  \item 1,2: Given $\hat p^\veps$ on $\Gamma_\dlt^\pm$, find $P^\dlt$ defined
  in $\Om_\dlt^{G} =\Om_\dlt^+ \cup \Om_\dlt^-$, such that
\begin{equation}\label{eq-G2-01}
\begin{split}
c^2 \nabla^2 P^\dlt + \om^2 P^\dlt & = 0 \quad \mbox{ in } \Om_\dlt^+\cup\Om_\dlt^-\;,\\
\mbox{ \textbf{ interface condition:} } & \\
 P^\dlt  & = \hat p^\veps \quad \mbox{ on } \Gamma_\dlt^\pm\;,\\
 \mbox{ other boundary conditions }& \\
r \imu\om c P^\dlt + c^2 \pdiff{P^\dlt}{n} & = s 2\imu\om c\bar p
\quad  \mbox{ on } \pd_\ext \Om_\dlt^G\;, \\
\end{split}
\end{equation}
where $\pd_\ext \Om_\dlt^G = \pd\Om^G \cap (\pd\Om_\dlt^+\cup\pd\Om_\dlt^-)$ is
the ``external'' boundary. As in problem \eq{eq-G1}, $r$ and $s$ are constants
attaining values 0, or 1, whereas $\bar p$ is the amplitude of an incident
wave. In the context of a waveguide, we consider $\pd_\ext \Om_\dlt^G$ to be
decomposed into three parts, $\pd_{\rm in} \Om_\dlt^G$, $\pd_{\rm out}\Om_\dlt^G$, and $\pd_w \Om_\dlt^G$, denoting the input, the output and walls,
respectively. By the constants $r,s$ in \eq{eq-G2-01}$_3$ different conditions
on $\pd \Om_\dlt^G$ are respected: $r=s=0$ on the walls $\pd_w \Om_\dlt^G$,
whereas $r=s = 1$ on $\pd_{\rm in} \Om_\dlt^G$ and $r=1$, $s = 0$ on $\pd_{\rm out} \Om_\dlt^G$, which accounts for the non-reflection condition.

\item 3: Given $g^{\veps\pm}$ on $\Gamma_\dlt^\pm$, find $p^\veps$ in
$\Om_\dlt^{*\veps}$ and $\ub^\veps$ in $\Sigma^\veps$, such that

\begin{equation}\label{eq-G2-02}
\begin{split}
c^2 \nabla^2 p^\veps + \om^2 p^\veps & = 0 \quad \mbox{ in } \Om_\dlt^{*\veps} \;, \\
\mbox{  \textbf{interface conditions}  } \quad \pdiff{p^\veps}{n} & = - \imu \om g^{\veps\pm} \quad \mbox{ on } \Gamma_\dlt^{\pm}\;,\\
\nabla\cdot \sigmabf(\ub^\veps) + \om^2 \rho \ub^\veps & = 0 \quad \mbox{ in } \Sigma_\dlt^\veps\;,\\
\mbox{ acoustic transmission: }& \\
\left.
\begin{array}{rcl}
\imu \om \nb\cdot\ub^\veps & = & \nb\cdot\nabla p^\veps\\
\nb \cdot \sigmabf(\ub^\veps) & = & \bb(p^\veps)  = \imu \om \rho_0 p^\veps \nb
\end{array}
\right\}& \quad \mbox{ on } \pd_* \Sigma_\dlt^\veps\;,\\
\mbox{ other boundary conditions }
& \\
r \imu\om c p^\veps + c^2 \pdiff{p^\veps}{n} & = s 2\imu\om c\bar p
\quad  \mbox{ on } \pd_\ext \Om^G\setminus \pd_\ub\Sigma_\dlt^\veps\;, \\
 \mbox{ clamped elastic structure: }& \\
\ub^\veps & = 0 \quad  \mbox{ on } \pd_\ub\Sigma_\dlt^\veps\;.
\end{split}
\end{equation}
where $\pd_* \Sigma_\dlt^\veps = \pd\Sigma_\dlt^\veps \cap
\pd\Om_\dlt^{*\veps}$ is the surface of the elastic structure in contact with
the fluid, thus, $\pd\Sigma_\dlt^\veps = \pd_* \Sigma_\dlt^\veps \cup
\pd_\ub\Sigma_\dlt^\veps$.

\item 4. For a fixed $\veps$ and $\dlt$, solutions to problems \eq{eq-G2-01}
and \eq{eq-G2-02} are equivalent to the solution of \eq{eq-G1}, if the coupling
conditions hold:
\begin{equation}\label{eq-G2-03}
\begin{split}
\imu\om g^{\veps\pm} & = \pdiff{P^\dlt}{n^\pm}  \quad \mbox{ on } \Gamma_\dlt^\pm\;,\\
p^\veps & = \hat p^\veps = P^\dlt \quad \mbox{ on } \Gamma_\dlt^\pm\;,
\end{split}
\end{equation}
where $n^\pm$ referes to normals $\nb^\pm$ outer to domains $\Om_\dlt^\pm$.
\end{list}

\subsection{Plate model}

The 3D model of an elastic plate involved in problem \eq{eq-G1} can be replaced
by a plate model which describes a thin structure. We assume a small
$\veps_0>0$ for which the limit model of acoustic transmission can be
interpreted. In this paper we shall approximate behaviour of the thin elastic
structure by the Reissner-Mindlin (R-M) plate model, which allows us to
consider the effects related to shear stresses induced by rotations of the
plate crossections \wrt the mid-plane.

The R-M plate model can be obtained by the asymptotic analysis of the
corresponding 3D elastic structure while its thickness $ \tilde h\rightarrow
0$. However, the obtained limit model is then interpreted in terms of a given
thickness $h > 0$. We shall discus this point in Remark~\ref{rem-h}.

The plate is represented by its perforated mean surface $\Gamma^\veps$,
therefore all involved variables depend on $\veps$. However, for a while we
drop the superscript $^\veps$ related to these variables. The plate deflections
are described by amplitude of the membrane elastic wave $\ub= (u_1,u_2)$, of
the transverse wave $u_3$ and of the rotation wave $\thetabf=
(\theta_1,\theta_2)$. Two linear constitutive laws are involved, which depend
upon the {second order} tensor $\Sb=(S_{ij}) = \shp \delta_{ij}$, where
$\shp>0$ is the shear coefficient, and the {fourth order} elasticity tensor
$\Eop=(E_{ijkl})$ which is given by the Hooke law adapted for the plane stress
constraint; we define (all indices $i,j,k,l = 1,2$)
\begin{equation}\label{eq-va3}
\begin{split}
  {W}_{ij}(\vb) &:= E_{ijkl}\pd_l v_k =[\Eop \gradplS\vb]_{ij},\\
 Z_i(u_3,\thetabf) &:=S_{ij}(\pd_j u_3-\theta_j)=[\Sb (\gradpl
      u_3 - \thetabf)]_i.
\end{split}
\end{equation}

The Reissner-Mindlin plate model is derived  using the following kinematic
ansatz confining the displacement $\wb = (\ol{\wb},w_3)$ in a plate with the
actual thickness $h$,
\begin{equation}\label{eq-va4}
\begin{split}
\ol{\wb}(x',z) & = \ol{\ub}(x') + h z \thetabf(x')\;,\quad w_3(x',z) = u_3(x')\;,
z \in {[-1,1]}\;, x' \in \Gamma^\veps\;,
\end{split}
\end{equation}
where $\ol{\ub} = (u_1,u_2)$ is ``membrane-mode'' displacements, i.e. vector
$\ub = (\ol{\ub},u_3)$ involves also the ``transversal mode'' (the deflection).
The vector fields $(\ub,\thetabf)$ satisfy the following equations in
$\Gamma^\veps$,
\begin{equation}\label{eq-va5}
\begin{split}\om^2h\rho \;\ol{\ub} +h\gradpl \cdot\Wb(\ol{\ub})& = - \ol{\fb}(p)\;,
\quad \mbox{ in } \Gamma^\veps\;,\\
\om^2 h\rho \;u_3+h\gradpl\cdot \Zb (u_3 ,\thetabf )& =-f_3(p)\;,\quad \mbox{ in } \Gamma^\veps\;,\\
\om^2\frac{h^3}{12}\rho\; \thetabf+\frac{h^3}{12}\gradpl \cdot\Wb(\thetabf ) & = -\ol{\mb}\;\;,\quad \mbox{ in } \Gamma^\veps\;,\\
h \ol{\nb}\cdot\Wb(\ol{\ub}) & = \ol{\fb}^\pd\;\quad \mbox{ on } \pd_\circ\Gamma^\veps\;,\\
h \ol{\nb}\cdot\Zb(u_3 ,\thetabf ) & = f_3^\pd\;\quad \mbox{ on } \pd_\circ\Gamma^\veps\;,\\
\frac{h^3}{12}\ol{\nb}\cdot \Wb(\ol{\ub})& = \ol{\mb}^\pd\;\quad \mbox{ on } \pd_\circ\Gamma^\veps\;,\\
\ub & = 0\;\quad \mbox{ on } \pd_\ext\Gamma^\veps\;,\\
\thetabf & = 0 \;\quad \mbox{ on } \pd_\ext\Gamma^\veps\;,
\end{split} 
\end{equation}
where $\pd_\circ\Gamma^\veps = \pd\Gamma^\veps\setminus\pd_\ext\Gamma^\veps$
describes the perforations. Above the applied forces $\fb = (\ol{\fb},f_3)$,
$\ol{\fb}^\pd$ and moments $\ol{\mb}$, $\ol{\mb}^\pd$ depend on the acoustic
potential $p$. The crucial step in deriving the model of vibroacoustic
transmission consists in describing these forces in terms of $p$ imposed on
surface $\pd\Sigma^\veps$ in the 3D plate representation.



\begin{myremark}{rem-h}
  In our asymptotic analysis of the acoustic transmission layer, we shall use the
  plate thickness in two contexts:
  \begin{itemize}
    \item The periodically perforated plate model defined in terms of the 2D domain
    $\Gamma^\veps \subset \Gamma_0$ representing the mid-plane and the thickness $h
    = \veps_0 \hh$ with $\veps_0>0$ being fixed. In fact, for a given thickness $h$
    and the perforation design (a given size of the holes yielding $\veps_0$) we
    can obtain $\hh$.

    \item To describe the interaction between the 3D elastic structure and the
    acoustic fluid, the thickness must be proportional to $\veps$ which is also
    related to the transmission layer thickness $\dlt = \vkappa\veps$, thus, we
    consider $h^\veps = \veps\hh$ and the elastic body occupying domain
    $\Sigma^{\veps}$, see \eq{eq-hh}.
  \end{itemize}

  Thus, the homogenization of the periodically perforated plate is done by
  pursuing the asymptotic analysis $\veps\rightarrow 0$ applied to the 2D plate
  model \eq{eq-va5} divided by $h$. Whereas $h$ is fixed in the plate equation
  operator, beeing independent of $\veps$, at the \rhs terms we get
  $1/(\veps\hh)$ which is coherent with the dilation operation applied when
  dealing with fluid equation, see Section~\ref{sec-dil}.
\end{myremark}

\subsection{Variational formulation of the vibroacoustic problem in the layer}

In order to derive the homogenized model of the transmission layer, we shall
need the variational formulation of problem \eq{eq-G2-02} with the plate model
\eq{eq-va5}.

Find $p^\veps \in H^1(\Om^{*\veps})$ and $(\ub^\veps,\thetabf^\veps) \in
(H^1_0(\Om))^5$ such that
\begin{equation}\label{eq-va6}
\begin{split}
c^2 \int_{\Om^{*\veps}} \nabla p^\veps \cdot \nabla q^\veps
- \om^2 \int_{\Om^{*\veps}} p^\veps q & =
-\imu \om c^2\left ( \int_{\Gamma^{\pm\veps}} g^{\veps\pm} q^\veps\,d\Gamma
+ \int_{\pd \Sigma^{\veps}} \nb\cdot\wb^\veps q^\veps\,d\Gamma \right ),
\end{split}
\end{equation}
for all $q \in H^1(\Om^{*\veps})$, where $\nb$ is outward normal to
domain $\Sigma^\veps$, and 
%
\begin{equation}\label{eq-va7}
\begin{split}
& \om^2 h \int_{\Gamma^\veps} \rho \ub^\veps \cdot \vb^\veps 
  \om^2 \frac{h^3}{12} \int_{\Gamma^\veps} \rho\thetabf^\veps\cdot \psibf^\veps
  -h \int_{\Gamma^\veps} [\Eop^\veps \gradplS \ol{\ub}^\veps ]: \gradplS \ol{\vb}^\veps\\
&-h \int_{\Gamma^\veps} [\Sb^\veps (\gradpl u_3^\veps - \thetabf^\veps)]\cdot
(\gradpl v_3^\veps - \psibf^\veps) - \frac{h^3}{12}
\int_{\Gamma^\veps} [\Eop^\veps \gradplS  \thetabf^\veps]: \gradplS \psibf^\veps\\
 = &
 \int_{\Gamma^\veps}   \fb^\veps(p^\veps)\cdot\vb^\veps  +
\int_{\Gamma^\veps} \ol{\mb}^\veps(p^\veps)\cdot \psibf^\veps
+\int_{\pd_\circ\Gamma^\veps}   \ol{\fb}^{\pd,\veps}(p^\veps)\cdot\ol{\vb}^\veps 
+\int_{\pd_\circ\Gamma^\veps}   \ol{\mb}^{\pd,\veps}(p^\veps)\cdot\psibf^\veps
\;,
\end{split}
\end{equation}
for all test functions $(\vb^\veps,\psibf^\veps)\in (H^1_0(\Om))^5$. In
\eq{eq-va6}, the displacements $\wb^\veps$ defined on the surface
$\pd\Sigma^{\veps} $ are expressed using the mid-plane kinematic fields. Due to
\eq{eq-va4}, it holds that
\begin{equation}\label{eq-wf4}
\begin{split}
\wb^\veps(x',x_3) &  = ( w_k^\veps(x',x_3))\;,\\
w_\alpha^\veps(x',x_3) & = u_\alpha^\veps(x') - x_3 \theta_\alpha^\veps(x')\;,\quad\alpha = 1,2 \;,\\
w_3^\veps(x',x_3) & = u_3^\veps(x')\;,
\end{split}
\end{equation}
where $x' \in \Gamma^\veps$, $x_3 \in \veps \hh ]-1/2,1/2[$. In analogy, the
test displacements $\tilde w_k^\veps(x',x_3)$ $k=1,2,3$ can \chV{be} introduced in terms
of the test functions $(\vb^\veps,\psibf^\veps)$ involved in \eq{eq-va7}; in
this equation, the \rhs integrals express the virtual power
\begin{equation}\label{eq-fba}
\begin{split}
\int_{\pd_*\Sigma^\veps} \bb^\veps\cdot \tilde \wb^\veps\;,
\end{split}
\end{equation}
where the traction stress $\bb^\veps = \imu\om\rho_0 \nb p^\veps$ is induced by
the acoustic pressure in the fluid.

\subsection{Fluid structure interaction on the plate surface $\pd\Sigma$}

The forces and moments involved in the \rhs of \eq{eq-va7} can be identified
using the 3D representation of the plate surface $\pd_*\Sigma^\veps$ decomposed
according to \eq{eq-hh}. The actual surface traction $b_i^\veps = \imu\om\rho_0
n_i p^\veps$ is given by the acoustic potential and by the surface normal $\nb
= (n_i)$; note that $n_\alpha = 0$, $\alpha = 1,2$ on $\pd_\pm\Sigma^\veps$,
whereas $n_3 = 0$ on $\pd_\circ\Sigma^\veps$. Hence, it can be shown that the
following expressions hold:
\begin{equation}\label{eq-fb}
\begin{split}
\ol{f}_\alpha^\veps & = 0\;,\quad \ol{m}_\alpha^\veps  = 0\;,\quad f_3^\pd  = 0\;,\\
\ol{f}_3^\veps & = \sum_{s = +,-}b_3(x',s \veps\hh/2) = \imu\om\rho_0 (p^\veps(x',\veps\hh/2)- p^\veps(x',-\veps\hh/2))\;,\\
{f}_\alpha^{\pd,\veps} & =  \int_{-h^\veps/2}^{h^\veps/2}b_\alpha(x',x_3)\dd x_3
 = \imu\om\rho_0 \int_{-h^\veps/2}^{h^\veps/2} n_\alpha(x') p^\veps(x',x_3)\dd x_3
\;,\\
{m}_\alpha^{\pd,\veps} & = -\int_{-h^\veps/2}^{h^\veps/2}x_3 b_\alpha(x',x_3)\dd x_3= -\imu\om\rho_0\int_{-h^\veps/2}^{h^\veps/2}x_3 n_\alpha(x') p^\veps(x',x_3)\dd x_3 \;.
\end{split}
\end{equation}

Then we consider the fluid equation. In \eq{eq-va6}, in the integral on
$\pd\Sigma^\veps$, the displacement field $\wb$ must be expressed in terms of
the mid-plane displacements and rotations $\ub^\veps$ and $\thetabf^\veps$, as
introduced in \eq{eq-wf4}. This yields
\begin{equation}\label{eq-wf5}
\begin{split}
\int_{\pd\Sigma^\veps} \nb\cdot\wb^\veps q^\veps = 
\int_{\Gamma^\veps} u_3^\veps \Jump{q^\veps(\cdot,x_3)}{\veps \hh} + \int_{\pd\Gamma^\veps}
\veps\hh \int_{-1/2}^{1/2} \bar \nb \cdot (\ub^\veps - \veps\hh \zeta \thetabf^\veps) q^\veps(\cdot,\veps\hh \zeta) \dd \zeta\;.
\end{split}
\end{equation}



\subsection{Dilated formulation}\label{sec-dil}

We can now state the vibro-acoustic problem in the dilated layer $\hat\Om =
\Gamma_0 \times ]-\vkappa/2, +\vkappa/2[$, where the fluid occupies domain
$\hat\Om^* = \{(x',\veps^{-1}x_3) \in \RR^3|x\in\Om^{*\veps}\}$, see \eq{102}.
Using $z = \veps^{-1}x_3$, while $x' = (x_\alpha)$, with new coordinates
$(x',z)$, the gradients are $\nablad = (\pd_\alpha,\veps^{-1}\pd_z)$, thus
$\nabla p(x) = \nablad p(x_\alpha,z)$; to simplify the notation, we shall use
the same notation for functions depending on $x_3$, but expressed in terms of
$z$.

By virtue of the dilation and the periodic unfolding, the vibroacoustic problem
can be transformed in the domain which does not change with $\veps$.
Consequently the standard means of convergence can be used to obtain the limit
model.

Equation \eq{eq-va6} with the substitution \eq{eq-wf5} can now be transformed
by the dilatation (the same notation for all variables is adhered, but should
be interpreted in this new context of this dilated formulation):
\begin{equation}\label{eq-wf6}
\begin{split}
\int_{\hat\Om^\veps} \hat\nabla p^\veps \cdot \hat\nabla q^\veps - \frac{\om^2}{c^2} \int_{\hat\Om^\veps} p^\veps q^\veps 
 = -\frac{\imu \om}{ \veps} 
\int_{\Gamma^\pm} \hat g^{\veps\pm} q^\veps \\
 -\frac{\imu \om}{ \veps} \left[\int_{\Gamma^\veps} u_3^{\veps} \Jump{q^\veps(\cdot,x_3)}{\veps \hh}
+ \int_{\pd_\circ\Gamma^\veps}
\veps\hh \int_{-1/2}^{1/2} \bar \nb \cdot (\ol{\ub}^{\veps} - \veps\hh \zeta \thetabf^{\veps}) q^\veps(\cdot, \veps\hh \zeta) \dd \zeta
\right]\;.
\end{split}
\end{equation}
Further we employ \eq{eq-fb} to rewrite \eq{eq-va7} which is divided by
$h^\veps$; by virtue of Remark~\ref{rem-h}, the plate thickness is given, \ie
$h = \veps_0\hh$, however, when dealing with the \rhs interaction terms, $h :=
h^\veps = \veps\hh$ in accordance with the dilation transformation. Thus we get
%
\begin{equation}\label{eq-wp7} 
\begin{split}
& \om^2 \int_{\Gamma^\veps} {\rho} \ub^\veps \cdot \vb^\veps +
\om^2 \frac{h^2}{12} \int_{\Gamma^\veps} {\rho}\thetabf^\veps\cdot \psibf^\veps\\
& - \int_{\Gamma^\veps} [\Eop^\veps \gradplS \ol{\ub}^\veps ]: \gradplS \ol{\vb}^\veps
-\int_{\Gamma^\veps} [\Sb^\veps (\gradpl u_3^\veps - \thetabf^\veps)]\cdot
(\gradpl v_3^\veps - \psibf^\veps) - \frac{h^2}{12}
\int_{\Gamma^\veps} [\Eop^\veps \gradplS  \thetabf^\veps]: \gradplS \psibf^\veps\\
 = &
\frac{\imu \om\rho_0}{\veps\hh} \left[
\int_{\Gamma^\veps} v_3^\veps  \Jump{p^\veps(\cdot,x_3)}{\veps \hh}
+ \veps\hh \int_{\pd_\circ\Gamma^\veps}\int_{-1/2}^{1/2}p^\veps(\cdot, \veps\hh \zeta) \bar\nb \cdot (\ol{\vb}^\veps - \veps\hh\zeta \psibf^\veps)\dd \zeta
\right]
\;.
\end{split}
\end{equation}
It is worth noting that, in \eq{eq-wf6} and \eq{eq-wp7}, the \rhs integrals
provide a symmetry of the following formulation.
\paragraph{The problem formulation} The vibroacoustic interaction in the
dilated layer $\hat\Om_\dlt$ is described by
$(p^\veps,\ub^\veps,\thetabf^\veps) \in H^1(\hat\Om^{*\veps})\times
(H_0^1(\Gamma^\veps))^5$ which satisfy equations \eq{eq-wf6}-\eq{eq-wp7} for
any test fields $(q^\veps,\vb^\veps,\psibf^\veps) \in
H^1(\hat\Om^{*\veps})\times (H_0^1(\Gamma^\veps))^5$.

Let $g^{0} \in L^2(\Gamma_0)$ and $g^{1\pm}(x',y') \in L^2(\Gamma_0 \times
\RR^2)$, whereby $g^{1\pm}(x',\cdot)$ being $\Xi$-periodic in the second
variable; we define
\begin{equation}\label{eq-aes2}
\begin{split}
\hat g^{\veps+}(x') & = g^{0}(x')  + \veps g^{1+}(x',\frac{x'}{\veps}) \;,\\
\hat g^{\veps-}(x')  & = -g^{0}(x')  - \veps g^{1-}(x',\frac{x'}{\veps})\;.
\end{split}
\end{equation}
%
For any $\veps > 0$ and $\hat g^{\veps\pm}$ defined according to \eq{eq-aes2},
the vibroacoustic interaction problem constituted by equations
\eq{eq-wf6}-\eq{eq-wp7} possesses a unique solution
$(p^\veps,\ub^\veps,\thetabf^\veps)$. As an essential step of the proof, the
\emph{a~priori} estimates are derived in the \Appx{sec-appendixA}.

\section{Homogenization of the transmission layer}
\label{sec-homog}
In this section, we introduce the convergence result which yields the limit
acoustic pressure and the plate displacements and rotations. These are involved
in the limit two-scale equations of the vibroacoustic problem imposed in the
transmission layer. The asymptotic analysis is based on the unfolding method which was inaugurated in the seminal paper \cite{Cioranescu-etal-2008} and elaborated further for thin structures in \cite{Cioranescu2008-Neumann-sieve}.
In our setting, the unfolding operator $\Tuftxt{}:  L^2(\Om_\delta;\RR) \rightarrow L^2(\Gamma_0 \times Y;\RR)$ transforms a function $f(x)$ defined in $\Om_\delta$ into a function of two variables, $x' \in \Gamma_0$ and $y \in Y$. 
For any $f \in L^1(Y)$, the cell average involved in all unfolding intergartion formulae will be abreviated by
\begin{equation}\label{eq-uf4}
\begin{split}
\frac{1}{|\Xi|}\int_\Xi f= \intY_\Xi f\;\quad \frac{1}{|\Xi|}\int_{D} f =:  \intY_D f\;,
\end{split}
\end{equation}
whatever the domain $D \subset \ol{Y}$ of the the integral is (\ie volume, or surface).
\subsection{The convergence results}

Based on the a~priori estimates derived in the Apendix~A, the following theorem
holds.

\begin{mytheorem}{thm-aes}
Let us assume
\begin{equation}\label{eq-aes3}
\begin{split}
\nrm{p^\veps}{L^2(\hat\Om^{*\veps})} \leq C\;,\quad \nrm{\ol{\ub}^{\veps}}{[L^2(\Gamma^\veps)]^2} \leq C\;,\\
\nrm{u_3^{\veps}}{L^2(\Gamma^\veps)}\leq C\;,\quad
\nrm{\theta^{\veps}}{[L^2(\Gamma^\veps)]^2}\leq C\;,
\end{split}
\end{equation}
then the folloving estimates can be obtained:
\begin{equation}\label{eq-aes4}
\begin{split}
\nrm{\hat\nabla p^\veps}{[L^2(\hat\Om^{*\veps})]^3} \leq C\;,\quad \nrm{\gradpl \ol{\ub}^{\veps}}{[L^2(\Gamma^\veps)]^4} \leq C\;,\\
\nrm{\gradpl u_3^{\veps}}{[L^2(\Gamma^\veps)]^2}\leq C\;,\quad
\nrm{\gradpl \theta^{\veps}}{[L^2(\Gamma^\veps)]^4}\leq C\;.
\end{split}
\end{equation}
Since $\hat\nabla p^\veps = (\gradpl p^\veps,  \veps^{-1} \pd_z p^\veps)$,
we have 
\begin{equation}\label{eq-aes4a}
\begin{split}
\nrm{\gradpl p^\veps}{[L^2(\hat\Om^{*\veps})]^2} \leq C\;,\quad 
\nrm{\pd_z p^\veps}{[L^2(\hat\Om^{*\veps})]^2}\leq \veps C\;.
\end{split}
\end{equation}
\end{mytheorem}

Due to Theorem~\ref{thm-aes} providing the estimates \eq{eq-aes3}-\eq{eq-aes4a}
we obtain the convergence of the unfolded functions(For the definition of the
unfolding operator we refere \eg to \cite{Cioranescu-etal-2008}). First we
observe (note \eq{eq-aes4a}$_2$):
\begin{equation}\label{eq-aes5a}
\begin{split}
p^\veps & \cwto p^0 \quad \mbox{ w. in } L^2(\hat\Om)\;,\\
\pd_z p^\veps & \cwto 0\quad \mbox{ w. in } L^2(\hat\Om)\;,
\end{split}
\end{equation}
thus, $\pd_z p^0 = 0$. The classical results of the unfolding method of
homogenization yield
\begin{equation}\label{eq-aes5}
\begin{split}
\Tuf{p^\veps} & \cwto p^0 \quad \mbox{ w. in } L^2(\Gamma_0 \times Y^*)\;,\\
\Tuf{\gradpl p^\veps} & \cwto \gradpl_{x'} p^0 + \gradpl_{y'}p^1 \quad \mbox{ w. in } L^2(\Gamma_0 \times Y^*)\;,\\
\frac{1}{\veps}\Tuf{\pd_z p^\veps} & \cwto  \pd_z p^1 \quad \mbox{ w. in } L^2(\Gamma_0 \times Y^*)\;.
\end{split}
\end{equation}
Above $p^0 \in H^1(\Gamma_0)$ and $p^1 \in L^2(\Gamma_0; H_\plper^1(Y^*))$,
where $H_\plper^1(Y^*)$ is the subspace of $H^1(Y)$ generated by $\Xi$-periodic
functions (thus, the periodicity in $y_\alpha$ holds, but not in $y_3 = z$), with vanishing average in $Y$.

For the plate responses we get
\begin{equation}\label{eq-aes7}
\begin{split}
\ub^{\veps} & \csto \ub^{0} \quad \mbox{ s. in } [L^2(\Gamma_K)]^2\;,\\
\Tuf{\ub^{\veps}} & \cwto \ub^{0} \quad \mbox{ w. in } [L^2(\Gamma_K \times \Xi_K)]^3\;,\\
\Tuf{\gradpl \ub^{\veps}} & \cwto \gradpl_{x'} \ub^{0} + \gradpl_{y'}\ub^{1} \quad \mbox{ w. in } [L^2(\Gamma_0 \times \Xi_S)]^6\;,
\end{split}
\end{equation}
where $\ub^{0} \in [H^1(\Gamma_K)]^3$ and $\ub^{1} \in L^2(\Gamma_K; [H_\#^1(\Xi_K)]^3)$. Here 
$H_\#^1(\Xi_K)$ is subspace of$H^1(\Xi)$ involving only  $\Xi$-periodic functions with vanishing average in $\Xi_K$. For the rotations we obtain
 \begin{equation}\label{eq-aes8}
\begin{split}
\Tuf{\thetabf^{\veps}} & \cwto \thetabf^0 \quad \mbox{ w. in } [L^2(\Gamma_0 \times \Xi_S)]^2\;,\\
\Tuf{\gradpl \thetabf^{\veps}} & \cwto \gradpl_{x'}\thetabf^0 + \gradpl_{y'} \thetabf^1 \quad \mbox{ w. in } [L^2(\Gamma_0 \times \Xi)]^4\;,
\end{split}
\end{equation}
where $\thetabf^0 \in [H^1(\Gamma_K)]^2$ and $\thetabf^1 \in L^2(\Gamma_0;
[H_\#^1(\Xi_S)]^2)$.

The limit vibro-acoustic problem can be derived by a formal approach which
relies on the recovery sequences (w.r.t. $\veps$) constructed in accordance
with the convergence result. Neglecting the higher order terms in $\veps$, the
following approximate expansions for unfolded vibroacoustic fields
$(p^\veps,\ub^\veps,\thetabf^\veps)$ are considered:
\begin{equation}\label{eq-va13a}
\begin{split}
\Tuf{p^\veps} & = p^0(x') + \veps p^1(x',y)\;, \\
\Tuf{\ol{\ub}^\veps} & =
\ol{\ub}^0(x') + \veps\ol{\ub}^1(x',y')\;,\\
\Tuf{u_3^\veps} & =
u_3^0(x') + \veps u_3^1(x',y')\;,\\
\Tuf{\thetabf^\veps} & = \thetabf^0(x') + \veps \thetabf^1(x',y')\;,
\end{split}
\end{equation}
where $x' \in \Gamma_0$, $ y' \in \Xi$ and $y = (y',z) \in Y$; in~\eq{eq-va13a},
all the two-scale functions are $\Xi$-periodic in the second variable.
%
Analogous expansions involving two-scale functions periodic in $y'$ will be
employed as the test functions involved in
\eq{eq-wf6}-\eq{eq-wp7},
\begin{equation}\label{eq-va13b}
\begin{split}
\Tuf{q^\veps} & = q^0(x') + \veps q^1(x',y)\;, \\
\Tuf{\ol{\vb}^\veps} & =
\ol{\vb}^0(x') + \veps\ol{\vb}^1(x',y')\;,\\
\Tuf{v_3^\veps} & =
v_3^0(x') + \veps v_3^1(x',y')\;,\\
\Tuf{\psibf^\veps} & = \psibf^0(x') + \veps \psibf^1(x',y')\;.
\end{split}
\end{equation}

It is worth to note that the use of the recovery sequences simplifies the
derivation of limit equations of the vibro-acoustic model which, however, can
be obtained more rigorously using the asymptotic analysis applied directley to
equations \eq{eq-wf6}-\eq{eq-wp7}. We shall substitute the ansatz
\eq{eq-va13a}-\eq{eq-va13b} in unfolded equations \eq{eq-wf6}-\eq{eq-wp7} and
explore the limit form for $\veps\rightarrow 0$.

\subsection{Limit fluid equation}

The unfolded left hand side of \eq{eq-wf6} yields the following limit form:
\begin{equation}\label{eq-f1}
\begin{split}
c^2 \int_{\Gamma_0} \intY_{Y^*} \Tuf{\gradpl p^\veps} \cdot \Tuf{\gradpl q^\veps}
+ \frac{c^2}{\veps^2} \int_{\Gamma_0} \intY_{Y^*}\pd_z \Tuf{p^\veps}
\pd_z \Tuf{q^\veps} - \om^2 \int_{\Gamma_0} \intY_{Y^*} \Tuf{p^\veps} \Tuf{q^\veps}\\
\rightarrow
c^2 \int_{\Gamma_0} \intY_{Y^*} (\gradplx p^0 + \gradply p^1)\cdot
(\gradplx q^0 + \gradply q^1) + c^2 \int_{\Gamma_0} \intY_{Y^*}\pd_z p^1 \pd_z q^1 - \om^2 \int_{\Gamma_0} \intY_{Y^*} p^0 q^0\;.
\end{split}
\end{equation}
The unfolded right hand side integrals can be written, as follows:
\begin{equation}\label{eq-f2}
\begin{split}
-\frac{\imu\om c^2}{\veps} \int_{\Gamma_0} & \left[ 
\intY_{I_y^+}(g^{0} + \veps g^{1+})(q^0 +\veps q^1)
+ \intY_{I_y^+}(-g^{0} - \veps g^{1-})(q^0 +\veps q^1)\right]\\
-\frac{\imu \om c^2}{\veps} \int_{\Gamma_0} & \left[  \intY_{\Xi_S}\Tuf{u_3^{\veps}} \Jump{\veps q^1(\cdot,z)}{\hh} \right. \\
 & \left. \quad
+ \frac{\veps\hh}{\veps}\intY_{\pd \Xi_S}
 \int_{-1/2}^{1/2} \bar \nb \cdot (\Tuf{\ol{\ub}^{\veps}} - \veps\hh \zeta \Tuf{\thetabf^{\veps}}) \big(q^0(x') + \veps q^1(x',y',\hh \zeta)\big) \dd \zeta
\right]\;.
\end{split}
\end{equation}
In the limit, the first integral related to the dilated fictitious interfaces
$\Gamma^\pm$ yields
\begin{equation}\label{eq-f3}
\begin{split}
-\imu\om c^2 \int_{\Gamma_0} \left(q^0 \intY_\Xi (g^{1+}-g^{1-}) + g^0\left(\intY_{I_y^+}q^1 -  \intY_{I_y^-}q^1\right)\right)\;.
\end{split}
\end{equation}
The second integral in \eq{eq-f2} can be rewritten, as follows (omitting the
factor $-\imu\om c^2$)
\begin{equation}\label{eq-f4}
\begin{split}
& \frac{1}{\veps}\int_{\Gamma_0}\intY_{\Xi_S}(u_3^0+\veps u_3^1) \veps \Jump{q^1}{\hh}\\
& +\frac{\hh}{\veps}\int_{\Gamma_0} \intY_{\pd\Xi_S} \int_{-1/2}^{1/2} \ol{\nb}\cdot \left(\ol{\ub}^0 + \veps \ol{\ub}^1 -\veps\hh \zeta(\thetabf^0 + \veps\thetabf^1)\right)(q^0 +\veps q^1)\;,
\end{split}
\end{equation}
where only $q^1(x',y',\hh \zeta)$ depends on $\zeta \in ]-1/2,+1/2[$. Hence,
since \mbox{$\int_{-1/2}^{1/2} \zeta = 0$}, in the limit, the second integral
in \eq{eq-f2} yields
\begin{equation}\label{eq-f5}
\begin{split}
-\imu\om c^2 \int_{\Gamma_0}\left(u_3^0\intY_{\Xi_S} \Jump{q^1}{\hh}
+ \ol{\ub}^0\cdot\intY_{\pd\Xi_S}\ol{\nb}\hh\int_{-1/2}^{1/2}q^1 \dd\zeta + 
q^0 \hh \intY_{\pd\Xi_S}\ol{\nb}\cdot\ol{\ub}^1\right)\;.
\end{split}
\end{equation}
Now the limit fluid equation constituted using \eq{eq-f1}\eq{eq-f3} and
\eq{eq-f5} attains the following form:
\begin{equation}\label{eq-f6}
\begin{split}
& c^2 \int_{\Gamma_0} \intY_{Y^*} (\gradplx p^0 + \gradply p^1)\cdot
(\gradplx q^0 + \gradply q^1) + c^2 \int_{\Gamma_0} \intY_{Y^*}\pd_z p^1 \pd_z q^1 - \om^2 \int_{\Gamma_0} \intY_{Y^*} p^0 q^0\\
& = 
-\imu\om c^2 \int_{\Gamma_0} \left[
q^0 \intY_\Xi \Delta g^{1} + g^0\left(\intY_{I_y^+}q^1 -  \intY_{I_y^-}q^1\right)\right.\\
& \left. + u_3^0\intY_{\Xi_S} \Jump{q^1}{\hh}
+ \ol{\ub}^0\cdot\intY_{\pd\Xi_S}\ol{\nb}\hh\int_{-1/2}^{1/2}q^1 \dd\zeta + 
q^0 \hh \intY_{\pd\Xi_S}\ol{\nb}\cdot\ol{\ub}^1\right]
\end{split}
\end{equation}
where $\Delta g^{1}=g^{1+}-g^{1-}$.

\subsection{Limit plate equation}

The unfolded left hand side of \eq{eq-wp7} yields the following limit form:
\begin{equation}\label{eq-p1}
\begin{split}
-\om^2 \int_{\Gamma_0} \rho_S \left( \ub^0\cdot\vb^0 + 
\frac{h^2}{12} \thetabf^0\cdot\psibf^0\right)
+  \int_{\Gamma_0}\intY_{\Xi_S}[\Eop(\gradplxS\ol{\ub}^0 + \gradplyS\ol{\ub}^1)]:(\gradplxS\ol{\vb}^0 + \gradplyS\ol{\vb}^1) \\
+ \int_{\Gamma_0} \intY_{\Xi_S}\Sb(\gradpl_x u_3^0 + \gradpl_y u_3^1 - \thetabf^0)\cdot
(\gradpl_x v_3^0 + \gradpl_y v_3^1 - \psibf^0)\\
+ \frac{h^2}{12}\int_{\Gamma_0} \intY_{\Xi_S}[\Eop(\gradplxS \thetabf^0 + \gradplyS \thetabf^1)]:(\gradplxS \psibf^0 + \gradplyS \psibf^1)
\end{split}
\end{equation}
The unfolded right hand side integrals can be written in analogy with the ones
involved in the fluid equation, see \eq{eq-f4}. Since the role of the solution
and the test function switches, the unfolded form of \eq{eq-wp7} yields
\begin{equation}\label{eq-p2}
\begin{split}
& \frac{\imu\om\rho_0}{\veps\hh}\int_{\Gamma_0}\intY_{\Xi_S}(v_3^0+\veps v_3^1) \veps \Jump{p^1}{\hh}\\
& +\frac{\imu\om\rho_0}{\veps}\int_{\Gamma_0} \intY_{\pd\Xi_S} \int_{-1/2}^{1/2} \ol{\nb}\cdot \left(\ol{\vb}^0 + \veps \ol{\vb}^1 -\veps\hh \zeta(\psibf^0 + \veps\psibf^1)\right)(p^0 +\veps p^1)\;,
\end{split}
\end{equation}
which, in the limit, yields an analogous expression as the one of \eq{eq-f5}.
Thus, the limit of the plate equation \eq{eq-wp7} is constituted by \eq{eq-p1}
which equals to
\begin{equation}\label{eq-p3}
\begin{split}
\eq{eq-p1} = \imu\om \rho_0 \int_{\Gamma_0}\left(v_3^0\frac{1}{\hh}\intY_{\Xi_S}  \Jump{p^1}{\hh}
+ \ol{\vb}^0\cdot\intY_{\pd\Xi_S}\ol{\nb}\int_{-1/2}^{1/2}p^1 \dd\zeta + 
p^0 \intY_{\pd\Xi_S}\ol{\nb}\cdot\ol{\vb}^1\right)\;.
\end{split}
\end{equation}

\begin{myremark}{rem-S}
Integrals over the plate surface involving $q^1$ and $p^1$ in \eq{eq-f5} and
\eq{eq-p3}, respectively, can be written in a more compact form; for any
two-scale function $\vphi(x',y',\zeta)$ it holds that
\begin{equation}\label{eq-dSint}
\begin{split}
v_3^0\frac{1}{\hh}\intY_{\Xi_S} \Jump{\vphi}{\hh}\dd y'
+ \ol{\vb}^0\cdot\intY_{\pd\Xi_S}\ol{\nb}\int_{-1/2}^{1/2}\vphi \dd\zeta 
= \frac{1}{\hh} \vb^0\cdot \intY_{\pd S}\nb \vphi\;.
\end{split}
\end{equation}

\end{myremark}

\subsection{Local problems in $Y^*$}

When testing the limit equation \eq{eq-f6} with $q^1\not=0$ while $q^0 = 0$,
the local problem in the fluid part is obtained which reveals linear dependence
of $p^1$ on the ``macroscopic'' functions $\ub^0,p^0$ and $g^0$. Therefore, we
can introduce the following split:
\begin{equation}\label{eq-f7}
\begin{split}
p^1(x',y) = \pi^\beta(y)\pd_\beta^x p^0(x') + \imu\om \xi(y) g^0(x') +  \imu\om \eta^k(y) u_k^0(x')\;,
\end{split}
\end{equation}
where $y = (y',z) \in Y^*$,
and introduce the following 3 autonomous problems for
$\pi^\beta,\xi,\eta^k \in H_\plper^1(Y^*)$:
\begin{equation}\label{eq-va20}
\begin{split}
\ipYs{\nabla_y \pi^\beta}{\nabla_y \psi} & = - \int_{Y^*} \pd_\beta^y \psi\;,
\quad \forall \psi \in  H_\plper^1(Y^*)\;,\quad \beta = 1,2\;,\\
\ipYs{\nabla_y \xi}{\nabla_y \psi} & = -\left(\int_{I_y^+} \psi
- \int_{I_y^-} \psi\right) \;,
\quad \forall \psi \in  H_\plper^1(Y^*)\;,\\
\ipYs{\nabla_y \eta^k}{\nabla_y \psi} & = -\int_{\pd S}n_k \psi
\quad \forall \psi \in  H_\plper^1(Y^*)\;, \quad k = 1,2,3\;.
\end{split}
\end{equation}

\subsection{Local problems on $\Xi_S$}

We consider the limit equation governing the plate response; its left and right
hand sides are constituted by \eq{eq-p1} and \eq{eq-p3}, respectively. Upon
testing there subsequently by $\psibf^1$, $\ol{\vb}^1$ and $v_3^1$, whereas all
$\psibf^0$, $\ol{\vb}^0$ and $v_3^0$ vanish, the following local
``microscopic'' equations are obtained,
\begin{equation}\label{eq-va30}
\begin{split}
\frac{h^2}{12}\int_{\Gamma_0} \intY_{\Xi_S}[\Eop(\gradplxS \thetabf^0 + \gradplyS \thetabf^1)]: \gradplyS \psibf^1 = 0\;,\quad \forall\psibf^1  \in L^2(\Gamma_0;(H_\#^1(\Xi_S))^2)\;,\\
\int_{\Gamma_0} \intY_{\Xi_S}\Eop (\gradplS_x  \ol{\ub}^0 + \gradplS_y  \ol{\ub}^1)
: \gradplS_y  \ol{\vb}^1 = \imu \om \rho_0\int_{\Gamma_0}p^0\intY_{\pd\Xi_S} \ol{\nb}\cdot\ol{\vb}^1
\;,\quad \forall \ol{\vb}^1 \in L^2(\Gamma_0;(H_\#^1(\Xi_S))^2)\;,\\
\int_{\Gamma_0} \intY_{\Xi_S}\Sb (\gradpl_x u_3^0 + \gradpl_y u_3^1
- \thetabf^0)\cdot\gradpl_y v_3^1 = 0\;,\quad \forall v_3^1 \in  L^2(\Gamma_0;H_\#^1(\Xi_S))\;.
\end{split}
\end{equation}
Due to the linearity of \eq{eq-va30}, the following split of the two-scale
functions can be introduced
\begin{eqnarray}
\ol{\ub}^1 & =& \ol{\chibf}^{\alpha\beta} (\gradplS \ol{\ub}^0)_{\alpha\beta} + \ol{\chibf}^* \imu\om\rho_0 p^0\;,\label{eq-ms1}\\
u_3^1 & =& \chi^k \left( (\gradpl u_3)_k - \theta_k\right) 
\;,\label{eq-ms2}\\
\thetabf^1 & =& \ol{\chibf}^{\alpha\beta} (\gradplS \thetabf^0)_{\alpha\beta} 
\;,\label{eq-ms3}
\end{eqnarray}
where $\ol{\chibf}^{\alpha\beta},\ol{\chibf}^*\in\Hpdb(\Xi_S)$, and $ \chi^k
\in H_\#^1(\Xi_S)$ are the corrector functions. They express the local
characteristic responses of the plate which can be computed independently of
the macroscopic responses $\gradplS \ol{\ub}^0$, $\gradplS \thetabf^0$,
$\gradpl u_3$, and $p^0$. It is worth noting that the same functions
$\ol{\chibf}^{\alpha\beta}$ are involved in both $\ol{\ub}^1$ and $\thetabf^1$
due to the similar structure of \eq{eq-va30}$_1$ and \eq{eq-va30}$_2$. The
following three local autonomous problems have to be solved,
\begin{itemize}
\item Find  $\ol{\chibf}^{\alpha\beta}\in \Hpdb(\Xi_S)/\RR^2$ such that
\begin{equation}\label{eq-mic3}
\begin{split}
\intY_{\Xi_S} (\Eop \gradplS_y(\ol{\chibf}^{\alpha\beta} + \Pibf^{\alpha\beta}):
\gradplS_y\ol{\vb} & = 0 \quad \forall \ol{\vb} \in \Hpdb(\Xi_S)\;,
\end{split}
\end{equation}
where $\Pibf^{\alpha\beta} = (\Pi_\nu^{\alpha\beta})$, $\Pi_\nu^{\alpha\beta} = y_\beta\delta_{\alpha i}$ with $\nu,\alpha,\beta = 1,2$.

\item Find $\chi^\alpha \in H_\#^1(\Xi_S)/\RR$ such that
\begin{equation}\label{eq-mic4}
\begin{split}
\intY_{\Xi_S} (\Sb \nabla_y (\chi^\alpha + y_\alpha))\cdot \nabla_y \tilde z & = 0
\quad \forall \tilde z \in H_\#^1(\Xi_S)\;, \quad \alpha - 1,2\;.
\end{split}
\end{equation}
\item Find $\ol{\chibf}^* \in \Hpdb(\Xi_S)/\RR^2$ such that
\begin{equation}\label{eq-mic5}
\begin{split}
 \intY_{\Xi_S}\Eop \gradplS_y \ol{\chibf}^*
: \gradplS_y \ol{\vb} = \intY_{\pd \Xi_S} \ol{\nb}\cdot\ol{\vb}
\;,\quad \forall \ol{\vb} \in \Hpdb(\Xi_S)\;.
\end{split}
\end{equation}
\end{itemize}

\subsection{Homogenized equations associated with the fluid}

The macroscopic equation governing the acoustic potential $p^0$ distributed on
the homogenized interface $\Gamma_0$ is obtained upon testing \eq{eq-wf6} with
$q^0\not=0$ while $q^1 = 0$, which yields
\begin{equation}\label{eq-f10}
\begin{split}
& c^2 \int_{\Gamma_0} \intY_{Y^*} (\gradplx p^0 + \gradply p^1)\cdot
\gradplx q^0   - \om^2 \int_{\Gamma_0} \intY_{Y^*} p^0 q^0\\
& = 
-\imu\om c^2 \int_{\Gamma_0} q_0 \left(
\intY_\Xi \Delta g^{1} + \hh \intY_{\pd\Xi_S}\ol{\nb}\cdot\ol{\ub}^1\right)\;,
\end{split}
\end{equation}
where $\Delta g^{1}$ was introduced in \eq{eq-f6}.
By virtue of the multiplicative splits \eq{eq-f7} and \eq{eq-ms1}, the integrals over $Y^*$, $\Xi_S$ and $\pd \Xi_S$ involving the two-scale functions can be expressed in terms of the macroscopic variables and using the homogenized coefficients
$\Ab = (A_{\alpha\beta})$, $\Bb = (B_\alpha)$, $\Db = (D_k)$, $\Hb = (H_{\alpha\beta})$ and $K$,
\begin{equation}\label{eq-f11}
\begin{split}
A_{\alpha\beta} & = \intY_{Y^*}{\nabla_y (\pi^\beta + y_\beta)}\cdot{\nabla_y (\pi^\alpha + y_\alpha)}\;,\\
B_\alpha & = \intY_{Y^*} \pd_\alpha^y \xi\;,\\
D_k^\alpha & =  \intY_{Y^*} \pd_\alpha^y \eta^k =  \int_{\pd S} n_k \pi^\alpha\;,\\
H_{\alpha\beta}&  = \intY_{\pd \Xi_S}\ol{\nb}\cdot\ol{\chibf}^{\alpha\beta}\;,\\
K & = \intY_{\pd \Xi_S}\ol{\nb}\cdot\ol{\chibf}^{*}\;.
\end{split}
\end{equation}
The alternative expression of $P_k^\alpha$ by $\pi^\alpha$ is obtained due to
the local problems \eq{eq-va20}.

Finally we can obtain the following \emph{extended equations of the  acoustic
transmission} satisfied by $(p^0,g^0)\in H^1(\Gamma_0)\times L^2(\Gamma_0)$
\begin{equation}\label{eq-f12}
\begin{split}
& c^2\int_{\Gamma_0} ( \Ab \gradpl_x p^0 )\cdot \gradpl_x q^0
- \zeta^* \om^2 \int_{\Gamma_0} p^0 q^0 + \imu \om c^2\int_{\Gamma_0} g^0 \Bb\cdot\gradpl_x q^0 + \imu \om c^2\int_{\Gamma_0}\ \gradpl_x q^0 \cdot \Db\ub^0\\
= & -\imu \om c^2\hh \int_{\Gamma_0} q^0\left (\frac{1}{\hh}\intY_\Xi \Delta g^{1} + \imu \om\rho_0
 K p^0  +  \Hb:\gradplS_x\ol{\ub}^0
\right )\;,
\end{split}
\end{equation}
for all $q^0 \in H^1(\Gamma_0)$, where $\zeta^* =
|Y^*|/|\Xi|$.  

\subsection{Homogenized equations associated with the plate}

In the limit equation governing the plate response, see \eq{eq-p1} and
\eq{eq-p3}, we apply the macroscopic test functions $\vb^0,\psibf^0$, whereas
we put ${\vb}^1=0$ and $\vthetabf^1=0$. Thus we obtain
\begin{equation}\label{eq-p5}
\begin{split}
-\om^2\left(
\int_{\Gamma_0}\intY_{\Xi_S}\rho_S \ub^0\cdot \vb^0 +
\frac{h^2}{12} \int_{\Gamma_0}\intY_{\Xi_S}\rho_S \thetabf^0\cdot \vthetabf^0
\right)\\
+
 \int_{\Gamma_0}\intY_{\Xi_S}
(\Sb (\gradpl_y u_3^1 + \gradpl_x u_3^0 - \thetabf^0))\cdot
(\gradpl_x  v_3^0 - \psibf^0)  \\
+ \frac{h^2}{12} \int_{\Gamma_0}\intY_{\Xi_S}(\Eop (\gradplS_y\thetabf^1 + \gradplS_x\thetabf^0)):\gradplS_x\psibf^0
+  \int_{\Gamma_0}\intY_{\Xi_S}(\Eop (\gradplS_y\ol{\ub}^1 + \gradplS_x\ol{\ub}^0)): \gradplS_x{\ol{\vb}}^0 \\
= \imu\om \rho_0 \int_{\Gamma_0}\left(\vb^3\frac{1}{\hh}\cdot\intY_{\pd S}{\nb}p^1 \right)
\;,
\end{split}
\end{equation}
Upon substituting $(\ub^1,\theta^1)$ and $p^1$ by the splits
\eq{eq-ms1}-\eq{eq-ms3} and \eq{eq-f7}, the effective model parameters can be
introduced. Symmetric expressions for the two elasticity tensors are derived
using the local problems \eq{eq-mic3}-\eq{eq-mic4},
\begin{equation}\label{eq-p6}
\begin{split}
E_{\alpha\beta\mu\nu}^\hom & =  \intY_{\Xi_S} \Eop \gradplS_y({\ol{\chibf}^{\mu\nu} + \Pibf^{\mu\nu}}):\gradplS_y({\ol{\chibf}^{\alpha\beta} + \Pibf^{\alpha\beta}})\;,\\
S_{\alpha\beta}^\hom & = \intY_{\Xi_S} \left[\Sb\nabla_y (\chi^\alpha + y_\alpha)\right]\cdot
\nabla_y (\chi^\beta + y_\beta)\;.
\end{split}
\end{equation}
Due to the presence of $p^0$ in the expression of $\ol{\ub}^1$, see \eq{eq-ms1}, a
pressure-coupling term appears which involves coefficient $\Hb$ introduced in \eq{eq-f11},
\begin{equation}\label{eq-p7}
\begin{split}
 \imu\om\rho_0 p^0\intY_{\Xi_S} (\Eop \gradplS_y \ol{\chibf}^*):\gradplS_x\ol{\vb}^0 =
 -\imu\om\rho_0 p^0 \Hb:\gradplS_x\ol{\vb}^0\;,
\end{split}
\end{equation}
where the identity follows upon substituting the test functions in \eq{eq-mic3}
and \eq{eq-mic5} by $\ol{\chibf}^*$ and $\ol{\chibf}^{\alpha\beta}$,
respectively.

In the \rhs integrals of \eq{eq-p5}, the following coefficients $\Cb = (C_k)$
and $\Tb = (T_j^k)$ are introduced,
\begin{equation}\label{eq-p8}
\begin{split}
C_k  & = \intY_{I_y^+} \eta^k - \intY_{I_y^-}\eta^k \;,\\
T_j^k & = - \intY_{\pd S}\eta^k n_j = \ipYs{\nabla_y \eta^k}{\nabla_y \eta^j} = T_k^j\;,\quad k,j = 1,2,3\;,
\end{split}
\end{equation}
so that $\Tb$ is symmetric.

Using the homogenized coefficients, the macroscopic (homogenized) plate
equation \eq{eq-p5} can be rewritten:
\begin{equation}\label{eq-p9}
\begin{split}
& -\om^2 \int_{\Gamma_0}\ol{\rho}_S\left( \ub^0\cdot \vb + \frac{h^2}{12}
\thetabf^0\cdot\vthetabf \right) +
 \int_{\Gamma_0} (\Sb^\hom(\gradpl_x u_3^0 - \thetabf^0))\cdot
(\gradpl_x  v_3 - \vthetabf)\\
& + \frac{h^2}{12} \int_{\Gamma_0}(\Eop^\hom \gradplS_x\thetabf^0):\gradplS_x\vthetabf + \int_{\Gamma_0}(\Eop^\hom \gradplS_x\ol{\ub}^0):\gradplS_x{\ol{\vb}}
- \imu\om\rho_0 \int_{\Gamma_0} p^0\Hb:\gradplS_x{\ol{\vb}}\\
&  =
  \frac{\rho_0}{\hh} \int_{\Gamma_0} \vb \cdot\left(
\imu \om\Db^T\gradpl p^0 - \om^2 \Cb g^0 + \om^2 \Tb\ub^0
\right)\;,
\end{split}
\end{equation}
for all $(\vb,\vthetabf) \in (H_0^1(\Gamma_0))^5$. Above $\ol{\rho}_S$ is the
effective plate density, $\ol{\rho}_S = |\Xi|^{-1}\int_{\Xi_S}\rho$.

\section{Global problem with vibroacoustic transmission conditions}
\label{sec-global}
We recall the problem decomposition according to \eq{eq-G2-01}, \eq{eq-G2-02}
and conditions \eq{eq-G2-03}. The problem \eq{eq-G2-02} describing the
vibro-acoustic response in the layer has been homogenized, yielding equations
\eq{eq-mac-inpl1} and \eq{eq-mac-trans1}. Our further effort will focus on the
coupling the acoustic field in the layer with the surrounding environment. We
consider dilated domains $\hat\Om^\pm$, such that $\Gamma_0$ is their common
boundary, \ie $\pd\hat\Om^+\cap\pd\hat\Om^+ = \Gamma_0$. By $\hat P$ we denote
the dilated solutions in domains $\hat\Om^\pm$, the traces of $\hat P$ on
$\Gamma_0$ are denoted by $\hat P^\pm$; obviously $\hat P^+ =
\trace{\Gamma_0}{\hat P|_{\hat\Om^+}}$ and $\hat P^- = \trace{\Gamma_0}{\hat
P|_{\hat\Om^-}}$.

\subsection{Coupling of the layer with external acoustic fields}

In this section, we use the convergence result concerning the acoustic potential
$p^\veps$, namely \eq{eq-aes5}, and consider coupling of the acoustic fields
``inside'' the layer with the ones ``outside'' the layer. 
In particular, below we
introduce a coupling equation \eq{eq-cc1} which is associated with the limit
equations in the homogenized layer and provide the transmission conditions for
the global problem.

We recall the condition $P^\dlt = \hat p^\veps$ on $\Gamma_\dlt^\pm$ defined in
\eq{eq-G2-01}which is now treated in a weak sense. The jump of the exterior
field across the layer with finite $\delta>0$ is expressed, as follows,
\begin{equation}\label{eq-cc1}
\begin{split}
 \int_{\Gamma_\delta^+} \psi P^\dlt  -  \int_{\Gamma_\delta^-}\psi P^\dlt   =  \int_{\Gamma_0} \psi \int_{-\delta/2}^{\delta/2} \pd_{x_3} \tilde p^\vepsdel \quad \forall \psi \in L^2(\Gamma_0)\;,
\end{split}
\end{equation}
where we assume $\psi = \psi(x')$, $x' \in \Gamma_0$, where by $\tilde{}$ we
denote an extension of $p^\vepsdel$ to the whole $\Om_\delta$ We may apply the
dilation transformation; let $\hat P^{\dlt+}$ is defined on $\Gamma_0$, such
that $\hat P^{\dlt+}(x',0) = P^\dlt(x',\dlt/2)$ and, in analogy, we introduce
$\hat P^{\dlt-}$, consequently \eq{eq-cc1} can be written,
\begin{equation}\label{eq-cc2}
\begin{split}
  \int_{\Gamma_0} \psi (\hat P^{\dlt+} - \hat P^{\dlt-}) 
=  \veps \int_{\Gamma_0} \psi \int_{-\vkappa/2}^{\vkappa/2} \frac{1}{ \veps}\pd_{z} \tilde p^\veps \quad \forall \psi \in L^2(\Gamma_0)\;.
\end{split}
\end{equation}
With reference to Remark~\ref{rem-h}, we now consider a finite layer thickness
$\delta_0 = \vkappa\veps_0 >0$ in the \lhs expression of \eq{eq-cc2}, whereas
we pass to the limit on the \rhs; this yields the following approximation
\begin{equation}\label{eq-cc3}
\begin{split}
\frac{1}{\veps_0} \int_{\Gamma_0} \psi (\hat P^{\dlt+} - \hat P^{\dlt-} ) 
& \approx  \lim_{\veps\rightarrow 0}\int_{\Gamma_0} \psi \int_{-\vkappa/2}^{\vkappa/2} \frac{1}{\veps}
\intY_{\Xi}\pd_{z} \Tuf{\tilde p^\veps} \\
& = \int_{\Gamma_0} \psi \int_{-\vkappa/2}^{\vkappa/2} \intY_{\Xi}\pd_z \tilde p^1 = \int_{\Gamma_0} \psi \Jump{\tilde p^1}{\vkappa}\\
& = \int_{\Gamma_0} \psi \left(\intY_{I_y^+} p^1 - \intY_{I_y^-} p^1\right)
 \quad \forall \psi \in L^2(\Gamma_0)\;.
\end{split}
\end{equation}

We substitute the split form of $p^1$ in \eq{eq-cc3} which yields new positive
coefficient $F > 0$ and two other coefficients $\Bb',\Cb'$:
\begin{equation}\label{eq-cc4}
\begin{split}
F & = - \intY_{I_y^+} \xi + \intY_{I_y^-} \xi = \intY_{Y^*}\nabla_y \xi \cdot \nabla_y \xi\;.
,\\
{B'}_\alpha & = \intY_{I_y^+} \pi^\alpha - \intY_{I_y^-}  \pi^\alpha = \intY_{Y_\vkappa^*} \pd_\alpha^y \xi = B_\alpha\;,\quad \alpha = 1,2\;,\\
{C'}_k  &  = \intY_{I_y^+} \eta^k - \intY_{I_y^-}\eta^k = \intY_{\pd S} n_k \xi = C_k\;,\quad k = 1,2,3\;,
\end{split}
\end{equation}
where the equalities $\Bb = \Bb'$ and $\Cb = \Cb'$ are obtained due to the
local microscopic problems \eq{eq-va20}.
Now the limit coupling condition \eq{eq-cc3} can be written in terms of the
homogenized coefficients $F,\Cb$ and $\Bb$
\begin{equation}\label{eq-cc5}
\begin{split}
\int_{\Gamma_0} \psi \left(
\Bb'\cdot\gradpl_x p^0 - \imu \om F g^0 + \imu \om \Cb'\cdot \ub^0
\right) & = \frac{1}{\veps_0} \int_{\Gamma_0} \psi (\hat P^{\dlt_0+} - \hat P^{\dlt_0-})\quad \forall \psi \in L^2(\Gamma_0)\;.
\end{split}
\end{equation}
Due to the above mentioned symmetry, in \eq{eq-cc5}, coefficients $\Bb'$ and
$\Cb'$ can be replaced symply by $\Bb$ and $\Cb$, which reveals the symmetry of
the system of equations \eq{eq-f12}, \eq{eq-p9} and \eq{eq-cc5}.

A question which arizes naturally is how the limit field $p^0$ defined on
$\Gamma_0$ is related to traces $\hat P^{\dlt_0\pm}$ of the global solution in
$\Om_{\dlt_0}^\pm$. Recalling again the 2nd condition in \eq{eq-G2-03}, we can
establish a blending function: $\tilde P_\dlt(x',z):= (\vkappa/2 + z)\hat
P_\dlt^+ + (z - \vkappa/2) \hat P_\dlt^-$, where $(x',z) \in \Gamma_0
\times]-\vkappa/2,+\vkappa/2[$ and $\hat P_\dlt^\pm$ has been defined above.
Further we consider the following condition (recall that $\hat\Om$ is the
dilated layer):
\begin{equation}\label{eq-cc6}
\begin{split}
\int_{\hat\Om} (\tilde p^\veps - \tilde P_\dlt) \ol{\vphi} = 0 \quad \forall \vphi \in L^2(\Gamma_0),
\end{split}
\end{equation}
such that $\ol{\vphi}(x',z) = \vphi(x')$ for $x' \in \Gamma_0$. Due to the convergence result \eq{eq-aes5a}$_1$ and due to the construction of $\tilde P_\dlt$, in the limit $\veps\rightarrow 0$ we get
\begin{equation}\label{eq-cc7}
\begin{split}
\int_{\Gamma_0} \left(p^0- \frac{1}{2}\left[\hat P^++\hat P^-\right]\right) {\vphi} = 0 \quad \forall \vphi \in L^2(\Gamma_0)\;.
\end{split}
\end{equation}
Recalling the finite thickness $\dlt_0$, this equation can be interpretted for
$\hat P^\pm\approx \hat P_{\dlt_0}^\pm$, as in the case of the coupling
condition \eq{eq-cc3}. By virtue of the \rhs integral in \eq{eq-cc5}, for a
later use we introduce $\Delta P$, so that due to \eq{eq-cc7} $p^0$ is
expressed, as follows,
\begin{equation}\label{eq-G3-03}
\begin{split}
\Delta P & = \frac{1}{\veps_0}\left(\hat P^+ - \hat P^-\right)\;,\\
p^0 & = \frac{1}{2}\left(\hat P^+ + \hat P^-\right)\;.
\end{split}
\end{equation}

\subsection{Vibro-acoustic problem in the homogenized layer}

We shall summarize the limit equations of the homogenized transmission layer
problem arising from the problem constituted by equations
\eq{eq-wf6}-\eq{eq-wp7} and with the imposed acoustic momentum fluxes $\hat
g^{\veps\pm}$ given in the form \eq{eq-aes2}.

As announced in the introduction, the homogenized Reissner-Mindlin plate model
is valid for \textit{``simple'' perforations} generated by cylindrical holes
$\pd_\circ \Sigma$ defined according to \eq{eq-hh}, being generated by surface
$\pd_\circ S$, see \eq{103}. \chE{This leads to various simplifications which
have already been respected when deriving the homogenized model.} Moreover,
from the microscopic problems \eq{eq-va20}, due to Remark~\ref{rem-S}, the
following cancellations can be deduced due to the special geometry, in
particular
\begin{equation}\label{eq-vaZ1}
\begin{split}
C_1 = C_2 & = 0\;, \quad C_3 = \intY_{\pd S} n_3 \xi\;, \\
D_3^\alpha & = 0\;, \quad D_k^\alpha  =
 \intY_{\pd S} n_k \pi^\alpha\;, \quad k=1,2,\;\alpha = 1,2\;,\\
B_\alpha & = 0\;, \quad \alpha = 1,2\;, 
\end{split}
\end{equation}
where the last identity has allready been observed in
\cite{rohan-lukes-waves07}, dealing with rigid plates.

\begin{myremark}{rem-B}
For rigid plates, the plate deflections and rotations disappear. Thus, equation
\eq{eq-p9} is irrelevant and, hence, coefficients $\Pb$ and $\Cb$ and $\Hb$ are
not involved in \eq{eq-cc5} and \eq{eq-f12}. However, quite general types of
perforations can be considered for which coefficients $B_\alpha$ do not vanish.
\end{myremark}

To rewrite the resulting equations in a more convenient form,  we
introduce the inner product notation:
$$
\ipGm{\phi}{\psi} = \int_{\Gamma_0}\phi\psi\;.
$$

By virtue of the \rhs integral in \eq{eq-f12}, we shall use 
\begin{equation}\label{eq-vaZ2}
\begin{split}
\Delta G^1(x') := \frac{1}{|\Xi|} \intY_{\Xi} \Delta g^1(x',y')\dd y'\;,\quad x'\in \Gamma_0\;,
\end{split}
\end{equation}
where $\Delta g^{1}=g^{1+}-g^{1-}$ has been introduced above.

Assume for a while that $\Delta P$ and $\Delta G^1$ are given on $\Gamma_0$.
Then, since $B_\alpha=0$, the model constituted by \eq{eq-f12},\eq{eq-p9} and
\eq{eq-cc5} would yield two separate problems which are described in the next
two sections. Also we recall the relationship between the plate thickness $h$
and the perforation size $\veps_0$, $h = \veps_0 \hh$, whereby $\delta_0 =
\veps_0 \vkappa$ determines the actual transmission layer thickness.

\subsubsection{Tangential acoustic wave coupled with in-plane plate vibration}

The first subproblem arising from \eq{eq-p9} and \eq{eq-f12} is independent of
the jump $\Delta P$. As explained below, it couples in-plane plate vibrations
described by $\ol{\ub}^0$ with surface acoustic waves propagating in the fluid
in the tangential direction \wrt the plate, thus, being described by the
acoustic potential $p^0$.

To obtain a symmetric system, we multiply \eq{eq-p9} by factor $\hh/\rho_0$ and
consider $v_3 = 0$ and $\vtheta_\alpha = 0$. The fluid equation \eq{eq-f12} is
multiplied by $1/c^2$. The following separate problem can be distinguished: For
a given $\Delta G^1 \in L^2(\Gamma_0)$, find $(p^0,\ol{\ub}^0) \in
H^1(\Gamma_0)\times (H_0^1(\Gamma_0))^2$ such that (by $\ol{\Ib} =
(\delta_{ij})$ we denote the 2D identity)
\begin{equation}\label{eq-mac-inpl1}
\begin{split}
\ipGm{\Ab \gradpl p^0}{\gradpl q} -\om^2\ipGm{(\frac{\zeta^*}{c^2}+\rho_0 K)p^0}{q} & \\
+ \imu\om \left(\ipGm{\Db\ol{\ub}^0}{\gradpl q}
+ \hh\ipGm{\Hb:\gradplS \ol{\ub}^0}{q}\right) & = -\imu\om \ipGm{\Delta G^1}{q}\;, \\
 -\imu\om \left(\ipGm{\gradpl p^0}{\Db\ol{\vb}} + \hh\ipGm{p^0}{\Hb:\gradplS \ol{\vb}} \right) & \\
+ \frac{\hh}{\rho_0}\ipGm{\Eop^\hom \gradplS \ol{\ub}^0}{\gradplS \ol{\vb}}
-\om^2\ipGm{(\frac{\hh\ol{\rho}_S}{\rho_0}\ol{\Ib}+\ol{\Tb})\ol{\ub}^0}{\ol{\vb}}
 & = 0\;,
\end{split}
\end{equation}
for all $(q,\ol{\vb}) \in H^1(\Gamma_0)\times (H_0^1(\Gamma_0))^2$ .

\subsubsection{Plate deflection coupled with transversal acoustic momentum}

The second subproblem governs the transversal plate vibrations described by the
couple $(u_3^0,\thetabf^0)$ with the transversal acoustic momentum $g^0$ of the
fluid in response to a given jump $\Delta P$, whereas $\Delta G^1$ is not
involved. We consider $v_\alpha = 0$ and $\alpha = 1,2$ in \eq{eq-p9} which
again is multiplied by factor $\hh/\rho_0$ . The resulting equation is coupled
with \eq{eq-cc5} multiplied by $\imu\om$, so that the following separate
problem can be distinguished: For a given $\Delta P \in L^2(\Gamma_0)$ find
$(g^0,u_3^0,\thetabf^0) \in L^2(\Gamma_0)\times (H_0^1(\Gamma_0))^3$ such that
\begin{equation}\label{eq-mac-trans1}
\begin{split}
-\om^2\ipGm{Fg^0}{\psi} + \om^2 \ipGm{C_3 u_3^0}{\psi} 
& =  \imu\om\ipGm{\Delta P}{\psi}\;, \\
\om^2\ipGm{g^0}{C_3 v_3}
-\om^2 \ipGm{(\frac{\hh\ol{\rho}_S}{\rho_0}+T_{33})u_3^0}{v_3} -\om^2 \frac{\hh h^2}{12 \rho_0}\ipGm{\frac{\ol{\rho}_S}{\rho_0} \thetabf^0}{\vthetabf}\\
+ \frac{\hh}{\rho_0}\ipGm{\Sb^\hom (\gradpl u_3^0 - \thetabf^0)}{\gradpl v_3 - \vthetabf} +  \frac{\hh h^2}{12 \rho_0}\ipGm{\Eop^\hom \gradplS \thetabf^0}{ \gradplS \vthetabf}  & = 0\;,
\end{split}
\end{equation}
for all $(\psi,v_3,\vthetabf)\in L^2(\Gamma_0)\times (H_0^1(\Gamma_0))^3$.

\subsubsection{Rigid plate with general shape of perforations}

Assuming arbitrary shaped pores perforating the plate allowing for a more
general geometry of the fluid cell $Y^*$ and leading to a nonsymmetry of the
response $\xi$, see \cite{rohan-lukes-waves07}, \eq{eq-f12} is coupled with
\eq{eq-cc5}. For a given $\Delta P \in L^2(\Gamma_0)$ and $\Delta G^1 \in
L^2(\Gamma_0)$, find a couple $(p^0,g^0) H^1(\Gamma_0)\times L^2(\Gamma_0)$,
such that

\begin{equation}\label{eq-cr}
\begin{split}
\int_{\Gamma_0} ( \Ab \gradpl_x p^0 )\cdot \gradpl_x q
- \zeta^*\frac{\om^2 }{c^2} \int_{\Gamma_0} p^0 q + \imu \om \int_{\Gamma_0} g^0 \Bb\cdot\gradpl_x q 
= & -\imu \om \int_{\Gamma_0} q \Delta G^1\;,\\
\int_{\Gamma_0} \psi \left( \Bb\cdot\gradpl_x p^0 - \imu \om F g^0 
\right) & =  \int_{\Gamma_0} \psi \Delta P\;,
\end{split}
\end{equation}
for all $(q,\psi) \in H^1(\Gamma_0)\times L^2(\Gamma_0)$. Thus, both $\Delta P$
and $\Delta G^1$ are involved.

\subsection{Dirichlet-to-Neumann (DtN) operator and the transmission conditions}

For a given $\veps_0>0$, by virtue of the approaximation introduced above, let
us consider a problem arizing from \eq{eq-G2-01} for $\dlt\rightarrow 0$. The
interface condition is now replaced by
\begin{equation}\label{eq-G3-ifc}
\begin{split}
\pdiff{\hat P|_{\hat\Om^\pm}}{n^\pm} \approx \pdiff{\hat P^{\dlt_0\pm}}{n^\pm}= \imu\om \intY_\Xi \Tcal_{\veps_0}(\hat g^{\veps_0,\pm}) \quad \mbox{ on } \Gamma_0\;,
\end{split}
\end{equation}
where $\pdiff{p}{n} = \nb \cdot \nabla p $ denotes the normal derivative of $p$
on $\Gamma_0$; note that $\nb^+ = -\nb^-$ is outward to $\hat\Om^+$. Therefore,
we introduce the averaged momentum fluxes $\hat G_0^\pm$ defined by averaging
the unfolded expressions \eq{eq-aes2} over the period $\Xi$, so that
\begin{equation}\label{eq-G3-00}
\begin{split}
\hat G_0^\pm(x') & = \pm\frac{1}{|\Xi|}\int_\Xi \Tcal_{\veps_0}(\hat g^{\veps_0,\pm}) = \frac{1}{|\Xi|}\int_\Xi \left(g^0(x') + \veps_0 g^{1,\pm}(x',y')\right)\dd y' \\
& = g^0(x') + \veps_0 G^{1\pm}(x')\;,\quad\mbox{ where }  G^{1\pm} = \frac{1}{|\Xi|}\int_\Xi g^{1\pm}(x',y')\dd y'\;.
\end{split}
\end{equation}
Hence
\begin{equation}\label{eq-G3-04}
\begin{split}
g^0 & \approx \frac{1}{2}\left(\hat G_0^+ + \hat G_0^-\right) = g^0 + \frac{\veps_0}{2}(G^{1+} + G^{1-})\;,\\
\Delta G^1 & = G^{1+} - G^{1-} = \frac{1}{\veps_0}\left(\hat G_0^+ - \hat G_0^-\right)\;,
\end{split}
\end{equation}
where $\Delta G^1$ was defined in \eq{eq-vaZ2} with $\Delta g^1 = g^{1,+} -
g^{1,-}$. Moreover, if we assume that $G^{1+} = - G^{1-}$, then equality holds
in \eq{eq-G3-04}$_1$ for any $\veps_0$

Now for a while, we may consider the following problem: For given $\bar p$,
$\hat G_0^+$ and $\hat G_0^+$, find $\hat P$ in $\Om^G$ satisfying
\begin{equation}\label{eq-G3-01}
\begin{split}
c^2 \nabla^2 \hat P + \om^2 \hat P & = 0 \quad \mbox{ in } \hat\Om^+\cup\hat\Om^-\;,\\
\mbox{ \textbf{ interface condition:} } & \\
\pdiff{\hat P|_{\hat\Om^+}}{n^+} = \imu\om \hat G_0^+ \quad \mbox{ on } \Gamma_0\;,\\
\pdiff{\hat P|_{\hat\Om^-}}{n^-} = -\imu\om \hat G_0^- \quad \mbox{ on } \Gamma_0\;,\\
 \mbox{ outer boundary conditions }& \\
r \imu\om c \hat P + c^2 \pdiff{\hat P}{n} & = s 2\imu\om c\bar p
\quad  \mbox{ on } \pd_\ext \Om^G\;.
\end{split}
\end{equation}
see problem~\eq{eq-G2-01} concerning constants $r$ and $s$ involved in the
boundary conditions on $\pd_\ext \Om^G$.

Problem \eq{eq-G3-01} arizes from \eq{eq-G2-01} for $\dlt\rightarrow 0$,
however, the Dirichlet type interface conditions are now replaced by the
Neumann ones represented by \eq{eq-G3-ifc} with \eq{eq-G3-00}.

Problem \eq{eq-G3-01}, as well as problems \eq{eq-mac-inpl1} and
\eq{eq-mac-trans1} are artificial; in fact, neither $\hat G_0^\pm$, nor $\Delta
P$ or $\Delta G^1$ are known a~priori; they all are coupled to the solution
$\hat P$ due to the interface vibroacoustic interaction conditions
\eq{eq-mac-inpl1} and \eq{eq-mac-trans1} which involve $\Delta P,g^0$ and
$\Delta G^1$. While $g^0$ and $\Delta G^1$ are related to $\hat G_0^\pm$
directly by \eq{eq-G3-04},
$\Delta P$ and $p^0$ are related to traces $\hat P^\pm$ by \eq{eq-G3-03}.

The problems \eq{eq-mac-inpl1} and \eq{eq-mac-trans1}, and the conditions
\eq{eq-G3-03} and \eq{eq-G3-04} present an implicit form of the
\emph{Dirichlet-to-Neumann operator} (DtN) associated with the ``outer''
acoustic problem defined for $\hat P$ in $\hat\Om^\pm$.

\subsection{Global acoustic problem with homogenized perforated plate}

In order to write the weak formulation of problem \eq{eq-G3-01} with the DtN
mapping introduced above, we shall employ the following bilinear
forms involving the homogenized coefficients introduced in the preceding sections,
\begin{equation}\label{eq-sa-SP-0}
\begin{split}
\Acal(p,q) & = \ipGm{\Ab \gradpl p}{\gradpl q}\;,\\
\Scal((w,\thetabf),(v,\vthetabf)) & = \frac{\hh}{\rho_0}\ipGm{\Sb^\hom (\gradpl w - \thetabf)}{\gradpl v - \vthetabf}\;,\\
\Ecal(\thetabf,\vthetabf) & = \frac{\hh}{\rho_0}\ipGm{\Eop^H\gradplS \thetabf}{\gradplS \vthetabf}\;,\\
\Fcal(g,\psi) & = \ipGm{Fg}{\psi}\;,\\
\Hcal(\ol{\vb},p) & =\hh\ipGm{\Hb:\gradplS\ol{\vb}}{p} \;,\\
\Ccal(u,\psi) & = \ipGm{C_3 u}{\psi}\;,\\
\Kcal(p,q) & = \ipGm{(\frac{\zeta^*}{c^2}+\rho_0 K)p}{q}\;,\\
\Lcal(\thetabf,\vthetabf) & = \frac{\hh}{\rho_0}\ipGm{\frac{\ol{\rho}_S}{\rho_0} \thetabf}{\vthetabf}\;,\\
\Mcal(\ol{\ub},\ol{\vb}) & = \ipGm{(\frac{\hh\ol{\rho}_S}{\rho_0}\ol{\Ib}+\ol{\Tb})\ol{\ub}}{\ol{\vb}}\;,\\
\Ncal(u,z) & = \ipGm{\left(\hh\frac{\ol{\rho}_S}{\rho_0} + T_{33}\right)u}{z}\;,\\
\Dcal(\ol{\vb},p) & = \ipGm{\Db\ol{\vb}}{\gradpl p}\;,\\
\Bcal(p,\psi) & = \ipGm{\Bb\cdot\gradpl p}{\psi}\;.
\end{split}
\end{equation}
Above the bilinear forms $\Fcal,\Ccal,\Ncal$ and $\Mcal$ are related to the
inertia and fluid-structure interaction effects, while $\Scal$ and $\Ecal$ are
related to the plate stiffness.

The global acoustic field $\hat P \in H^1(\hat \Om^+\cup\hat \Om^+)$ satisfies
\begin{equation}\label{eq-DtN-0}
\begin{split}
c^2\int_{\hat \Om^+\cup\hat \Om^-}\nabla \hat P \cdot \nabla Q - \om^2  \int_{\hat \Om^+\cup\hat \Om^-}\hat P Q 
+ \int_{\pd_\ext \Om^G} r \imu\om c \hat P Q & \\
- \imu\om c^2 \left(\ipGm{\hat G_0^+}{Q^+} - \ipGm{\hat G_0^-}{Q^-}\right) 
= \int_{\pd_\ext \Om^G} s\imu\om c \bar{p} Q &
\end{split}
\end{equation}
for all $Q \in H^1(\hat\Om^+\cup\hat\Om^-)$, whereby $Q^\pm$ denotes the trace
of $Q$ on $\pd \hat\Om^\pm$.

The DtN operator involving functions $(\hat P^\pm,p^0,\ol{\ub}) \in
[L^2(\Gamma_0)]^2 \times H^1(\Gamma_0)\times [H_0^1(\Gamma_0)]^2$ and $(\hat
G_0^\pm,u_3,\thetabf) \in [L^2(\Gamma_0)]^2\times H_0^1(\Gamma_0)\times
[H_0^1(\Gamma_0)]^2$ is represented by the following equalities arizing from
\eq{eq-mac-inpl1} and \eq{eq-mac-trans1} where we employ \eq{eq-G3-03} and
\eq{eq-G3-04}, so that we have
 \begin{equation}\label{eq-DtN-1}
\begin{split}
\Acal(p^0,q) - \om^2\Kcal(p^0,q) + \imu\om\left(\Dcal(\ol{\ub},q) + \Hcal(\ol{\ub},q)\right)
+  \frac{\imu\om}{\veps_0}\ipGm{[\hat G_0^+ - \hat G_0^-]}{q}  & = 0\;,\\
-\imu\om\left(\Dcal(\ol{\vb},p^0) + \Hcal(\ol{\vb},p^0)\right)
+ \Ecal(\ol{\ub},\ol{\vb}) - \om^2\Mcal(\ol{\ub},\ol{\vb}) & = 0\;,
\end{split}
\end{equation}
for all $(q,\ol{\vb}) \in H^1(\Gamma_0)\times [H_0^1(\Gamma_0)]^2$, and
\begin{equation}\label{eq-DtN-2}
\begin{split}
\frac{\om^2}{2}\Fcal([\hat G_0^+ + \hat G_0^-],\psi) - \om^2\Ccal(u_3,\psi) + \frac{\imu\om}{\veps_0}\ipGm{[\hat P^+ - \hat P^-]}{\psi} & = 0\;,\\
\frac{\om^2}{2}\Ccal([\hat G_0^+ + \hat G_0^-],v_3) - \om^2\left(\Ncal(u_3,v_3)
 + \frac{h^2}{12}\Lcal(\thetabf,\vthetabf) \right) & \\
+\Scal((u_3,\thetabf),(v_3,\vthetabf)) + \frac{h^2}{12}\Ecal(\thetabf,\vthetabf)  & = 0\;,
\end{split}
\end{equation}
for all $(\psi,v,\vthetabf) \in L^2(\Gamma_0)\times H_0^1(\Gamma_0)\times
[H_0^1(\Gamma_0)]^2$. There is the coupling equation:
\begin{equation}\label{eq-DtN-3}
\begin{split}
\ipGm{2 p^0 - [\hat P^+ + \hat P^-]}{q} & = 0\quad \forall q \in L^2(\Gamma_0)\;.
\end{split}
\end{equation}

Now we can state the main result of the paper.

\paragraph{Global acoustic problem with the homogenized perforated plate}
Given the incident acoustic wave represented by $\bar p$ on $\Gamma_{\rm in}$, find the acoustic potential $\hat P$ defined in $\hat\Om^G = \hat\Om^+\cup\hat\Om^-$ and other functions $(p^0,\hat G_0^\pm,\ub,\thetabf)$
defined on $\Gamma_0$ such that the variational equalities
\eq{eq-DtN-0}-\eq{eq-DtN-3} hold.

\begin{myremark}{rem-rig}
It is left as an easy excersie for raders that, if a rigid plate is considered,
the model reduces. From \eq{eq-cr}, insetad of \eq{eq-DtN-1} and \eq{eq-DtN-2},
the DtN operator is established in terms of $\hat P^\pm$ and $(p^0,\hat G_0^\pm)$
which satisfy the following variational equalities,
\begin{equation}\label{eq-DtN-12r}
\begin{split}
\Acal(p^0,q) - \om^2\Kcal_R(p^0,q) +  \frac{\imu\om}{2} \Bcal({q},{[\hat G_0^+ + \hat G_0^-]})
+  \frac{\imu\om}{\veps_0}\ipGm{[\hat G_0^+ - \hat G_0^-]}{q}  & = 0\;,\\
\frac{\om^2}{2}\Fcal([\hat G_0^+ + \hat G_0^-],\psi) + \imu\om\Bcal({p^0},{\psi}) - \frac{\imu\om}{\veps_0}\ipGm{[\hat P^+ - \hat P^-]}{\psi} & = 0\;,
\end{split}
\end{equation}
for all $(q,\psi) \in H^1(\Gamma_0)\times L^2(\Gamma_0)$, whereby \eq{eq-G3-03}
holds. Note that $\Kcal_R$ is defined according to \eq{eq-sa-SP-0}, but with $K = 0$, whereby $\Bcal$ vanishes for simple perforations.
\end{myremark}

\section{Validation of the homogenized model}\label{sec-valid}
The homogenized model derived in this paper provides an approximation of the vibroacoustic interaction in a vicinity of the perforated plate. This approximation is introduced as the limit behaviour of the wave propagation in the heterogeneous structure when the transmission layer thickness and the characteristic size of the perforations diminish with $\veps\rightarrow 0$. However, by virtue of coupling the limit layer model \eq{eq-f12} and \eq{eq-p9} with the ``outer'' acoustic problem, the Global problem described in Section~\ref{sec-global} is featured by the specific scale parameter $\veps_0$ associated with a given plate thickness and the perforation size. In this section, we examine how numerical responses of the proposed homogenized vibroacoustic model corresponds with solutions of the ``original'' problem \eq{eq-G1} associated with the  3D  heterogeneous solid structure representing the plate.

\begin{figure}
    \centering
    \includegraphics[width=0.95\linewidth]{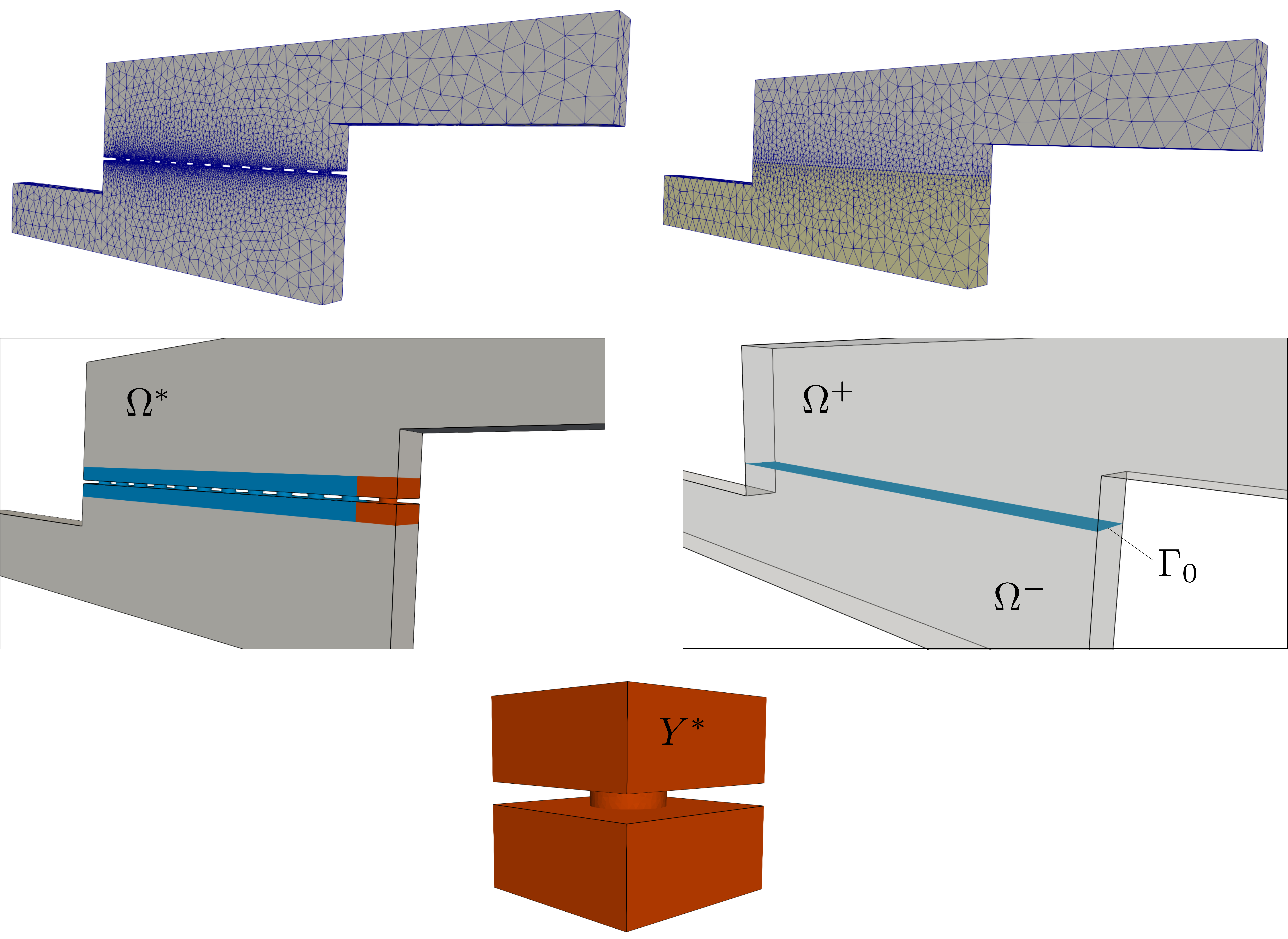}
    \caption{Geometries and FE meshes of the fluid domains related to the reference (left)
             and homogenized (right) models. Geometry of the periodic
             cell (bottom) employed in the multiscale simulation
             and in construction of the reference geometry.}
    \label{fig-valid-ac-geom}
\end{figure}

To this aim, the reference model is established as the finite element (FE) approximation of problem introduced in Section~\ref{sec-problem}. For this model, the heterogeneous structure  of the transmission layer is built up as the periodic lattice by copies of the reference periodic
cell according to \eq{eq-hh}-\eq{100}. The geometries associated with the homogenized and the reference models are illustrated in Fig.~\ref{fig-valid-ac-geom}, where the unit cell $Y^\ast$ represents the fluid domain.
By virtue of the asymptotic homogenization, the layer presenting a ``fictitious'' interface, highlighted by red and blue colors in
Fig~\ref{fig-valid-ac-geom}(left), is  replaced by  homogenized transmission
conditions imposed on the interface $\Gamma_0$ in the multiscale simulation, see
Fig~\ref{fig-valid-ac-geom}(right).

The validation of the homogenized model \eq{eq-DtN-0}-\eq{eq-DtN-3} is performed in two steps. First, we compare the acoustic fields computed by the
reference and the homogenized models, whereby the perforated plate is assumed to be rigid, see Remark~\ref{rem-rig}. Secondly
we compare the responses of the homogenized vibrating plate with the deflections
obtained by direct numerical simulations (DNS) of the heterogeneous 3D elastic structure. In this case, the plate surface is loaded by a given acoustic pressure distribution, thus, the vibroacoustic problem is decoupled. 
The reason for such a simplification arises as the consequence of the FE mesh complexity increasing with the number of the perforating holes, thus, inducing a discretized problem with large number of the degrees of freedom (DOFs).

\begin{figure}
    \centering
    \includegraphics[width=0.65\linewidth]{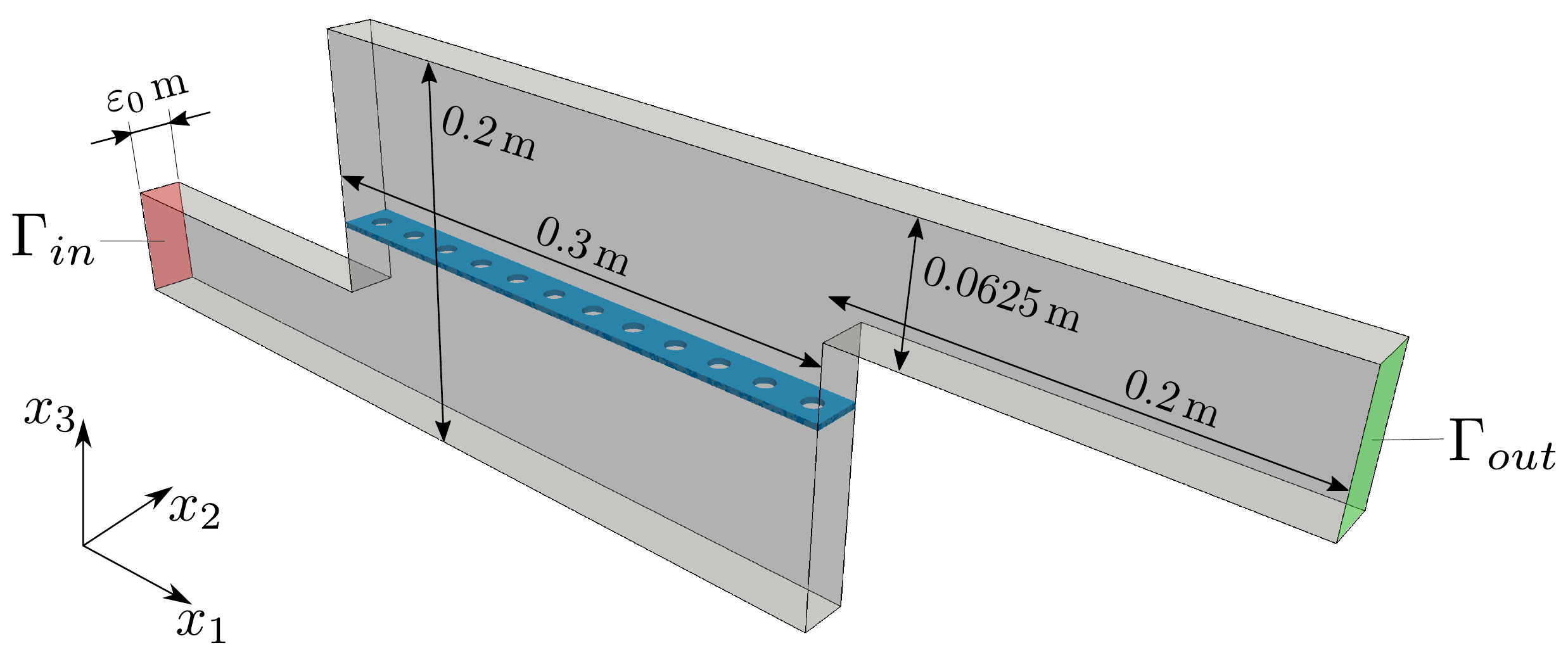}
    \caption{The acoustic domain with the embedded perforated plate.}
    \label{fig-valid-domain}
\end{figure}

For the purpose of this validation test, we consider the waveguide $\Om$ represented by the ``S''-shaped slice of thickness $\veps_0$\,m, as measured in the $x_2$-axis direction, see  Fig.~\ref{fig-valid-domain}, where the slice dimensions are indicated.
The waveguide is symmetric \wrt the center of the perforated plate structure which splits the acoustic domain into two mutually symmetric parts. The thickness of
the perforated plate is  $0.12 \veps_0$\,m, where $\veps_0 = 0.3 / N$ varies with $N$,
the number of the perforation periods (holes) drawn in the $x_1$-axis direction.

The homogenized models and the reference model presented in this paper have
been implemented in {\it SfePy} -- Simple Finite Elements in Python
\cite{sfepy-multiscale-2019}, a software developed for an efficient solving
of multiscale problems by means of the finite element method. In the validation tests and the coupled problem simulation, by the ``multiscale simulations'' we mean solutions of the homogenized (macroscopic) problem supplemented by the reconstruction procedure which allows us to respect the local fluctuations superimposed to the solutions of the macroscopic problem when the scale parameter $\veps_0$ is given.

\subsection{Validation test -- acoustic field in fluid}\label{sec-vaild-f}
In this test, the perforated plate is rigid, so that the distribution of the reference acoustic pressure field is governed by equations \eq{eq-G1} modified for the rigid plate, i.e. $\ub = 0$ in $\Sigma$. Accordingly, the homogenized layer presents the coupling conditions \eq{eq-DtN-12r} for the acoustic field in the wave guide which is governed by \eq{eq-DtN-0}.
Recall that the homogenized coefficients $\Ab$, $\Bb$, $F$ are given by expressions
\eq{eq-f11}, \eq{eq-cc4} involving solutions of the local problems \eq{eq-va20} defined in
$Y^\ast$. 

For both the homogenized and the reference models an incident wave with amplitude $\bar p = 300$\,Pa is imposed on
$\Gamma_{in}$, whereby  the anechoic condition on $\Gamma_{out}$ is considered.
The periodic conditions are prescribed on the two faces orthogonal to the $x_2$-axis direction
(front and back faces of the waveguide) for the geometry  depicted in Fig.~\ref{fig-valid-domain}.

Responses of the reference and the homogenized models are compared using the global acoustic
properties expressed by the transmission loss (\TL{}), and by the local
distributions of the acoustic pressure.  
These responses were computed for the fluid characterized by the acoustic speed $c = 343$\, m/s, the density $\rho_0 = 1.2$\,kg\,m$^3$. 


\begin{figure}
    \centering
    \includegraphics[width=0.49\linewidth]{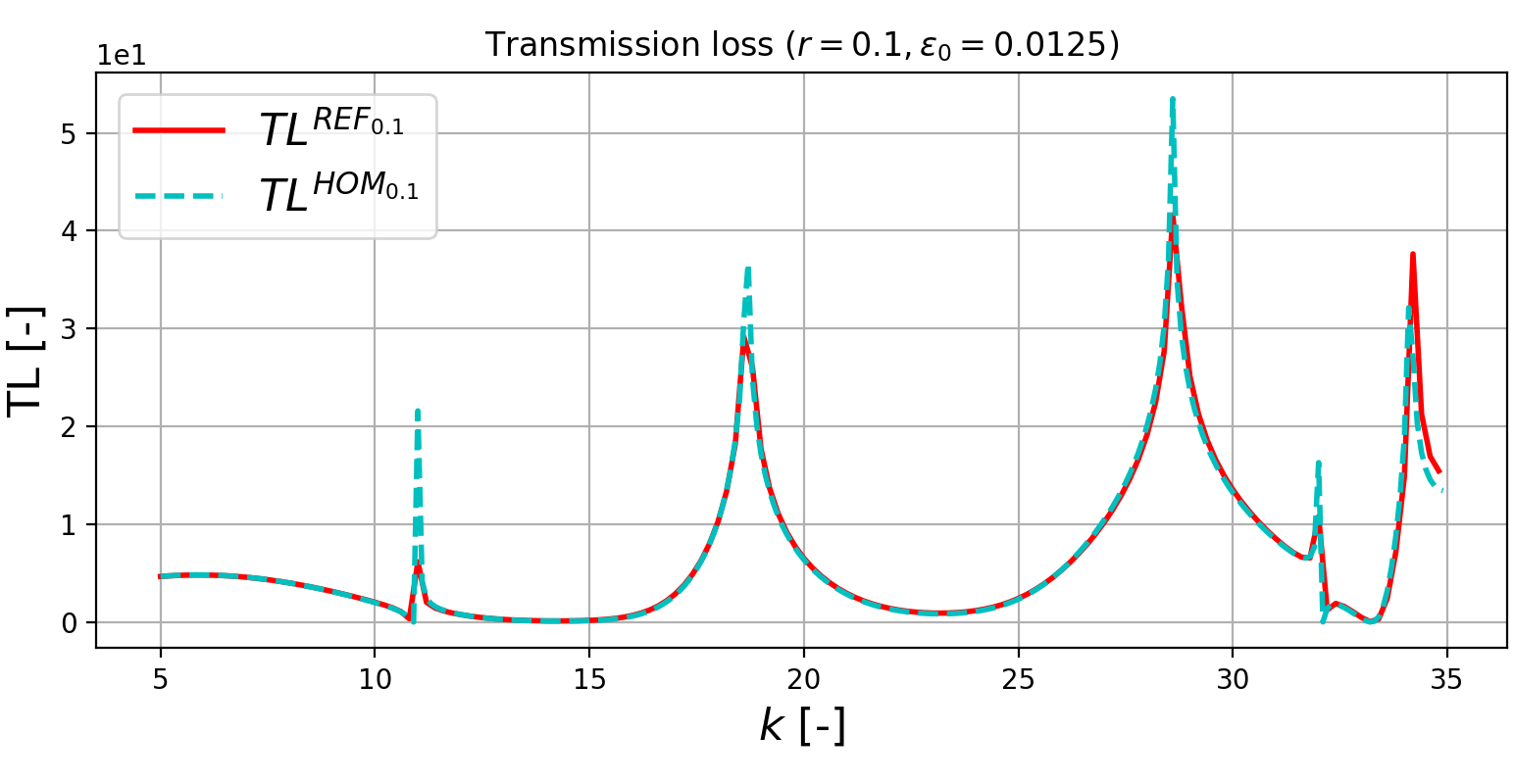}\hfil
    \includegraphics[width=0.49\linewidth]{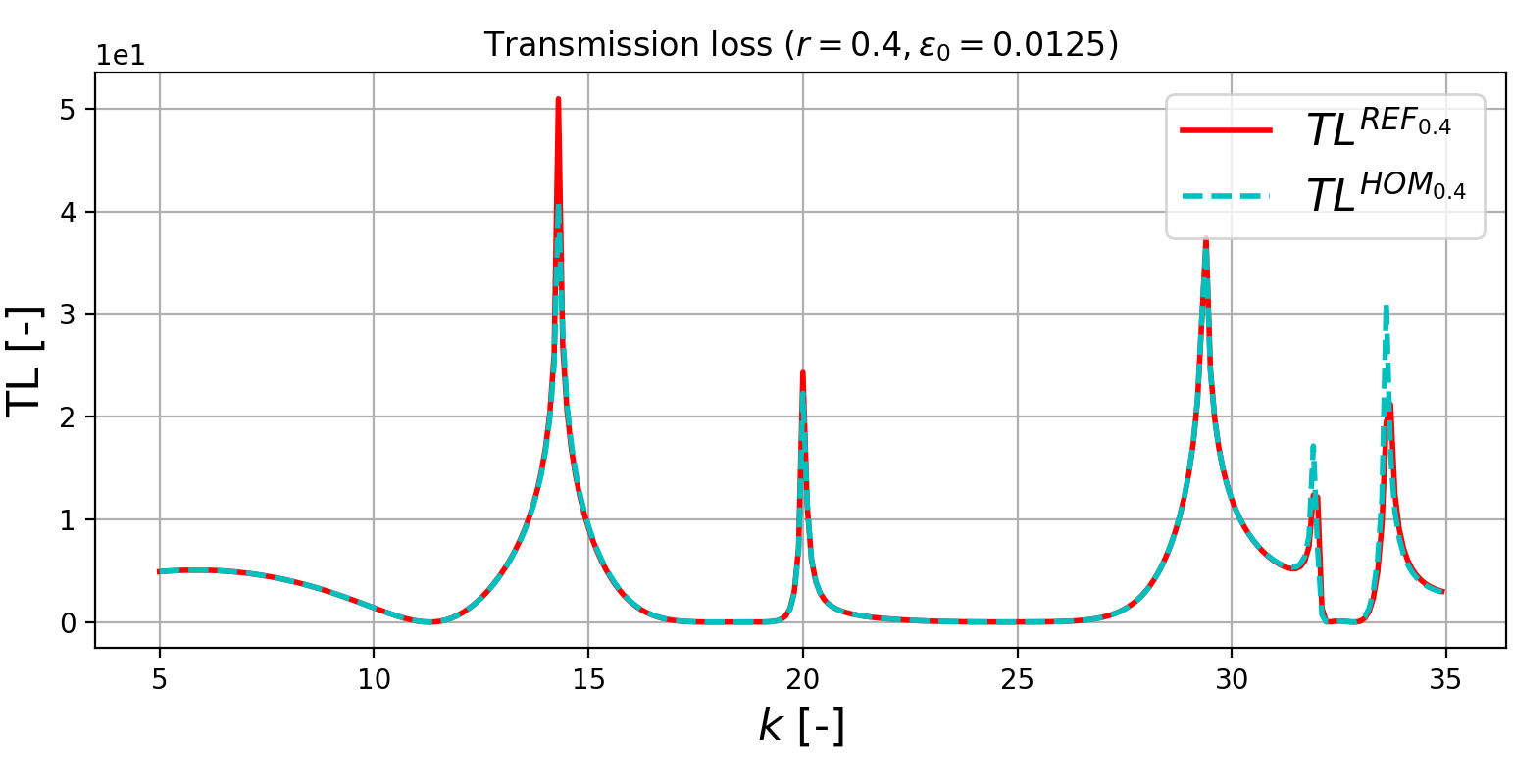}\\
    \includegraphics[width=0.49\linewidth]{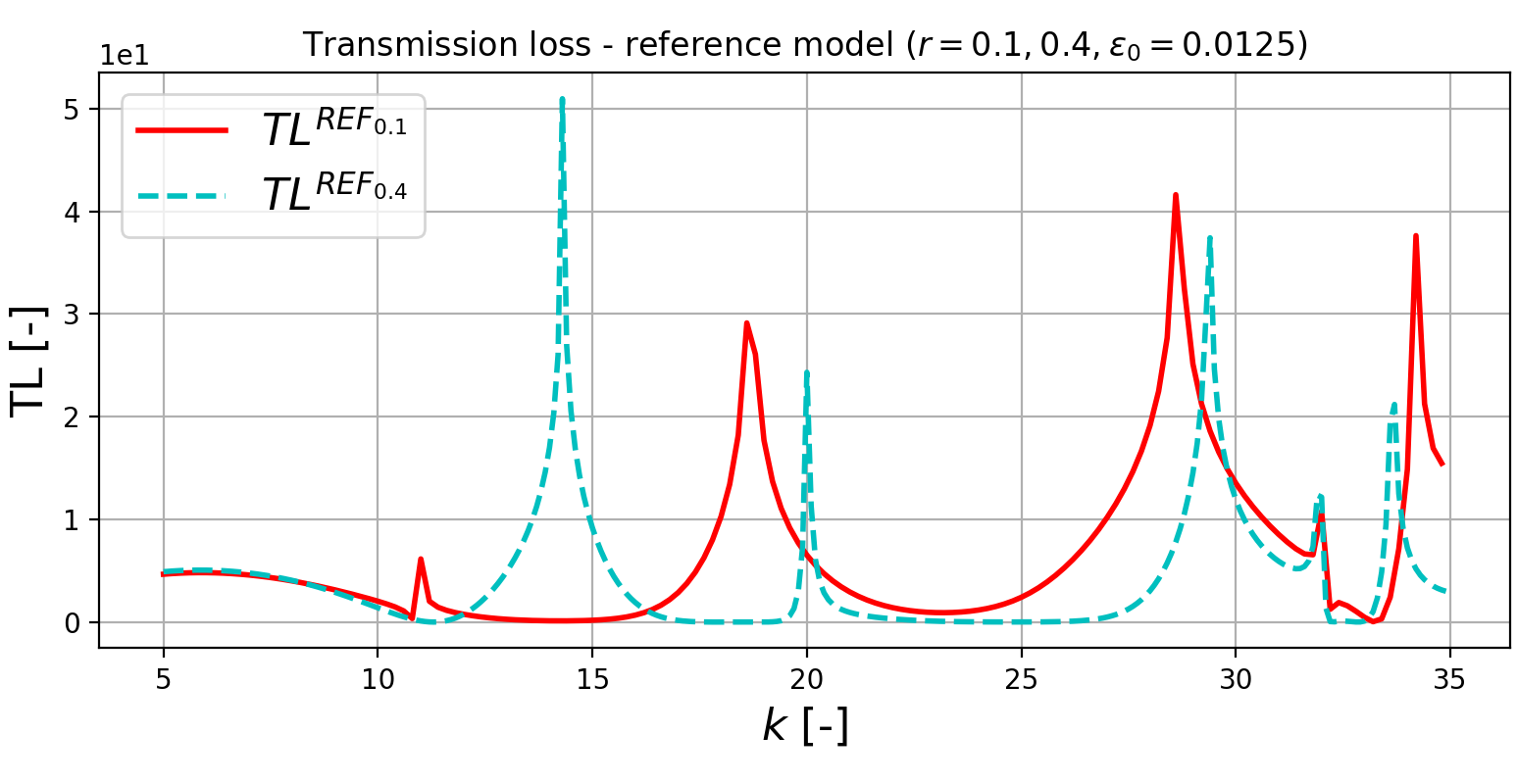}
    \caption{Transmission loss curves for the perforations with the hole
             radius of $r=0.1 \veps_0$\,m and $r=0.4 \veps_0$\,m
             obtained by the reference ({\it REF})
             and homogenized ({\it HOM}) models, N = $24$.}
    \label{fig-valid-ac-TL}
\end{figure}

The \TL{} curves obtained for both the models are compared
in Fig.~\ref{fig-valid-ac-TL}. Perforations with cylindrical holes were
examined for two radii $r$. Results for $r=0.1 \veps_0$\,m are depicted in
Fig.~\ref{fig-valid-ac-TL}(a) and for $r=0.4 \veps_0$\,m in
Fig.~\ref{fig-valid-ac-TL}(b). From these graphs it is apparent that the result
differ only in the vicinity of wave numbers yielding the \TL{} peaks; in those
regions associated with higher wave numbers also the shift of the peak
positions can be observed. However, this effect can be caused by different FE
discretizations of both the models. The difference of the two \TL{} curves is
displayed in Fig.~\ref{fig-valid-ac-TL}(c) for the two dimensions of the holes.
The calculations are performed for an interval of the wave number $k \in
[5,35]$ and for $\veps_0 = 0.0125$ which corresponds to $N = 24$.

\begin{figure}
    \centering
    \includegraphics[width=0.35\linewidth]{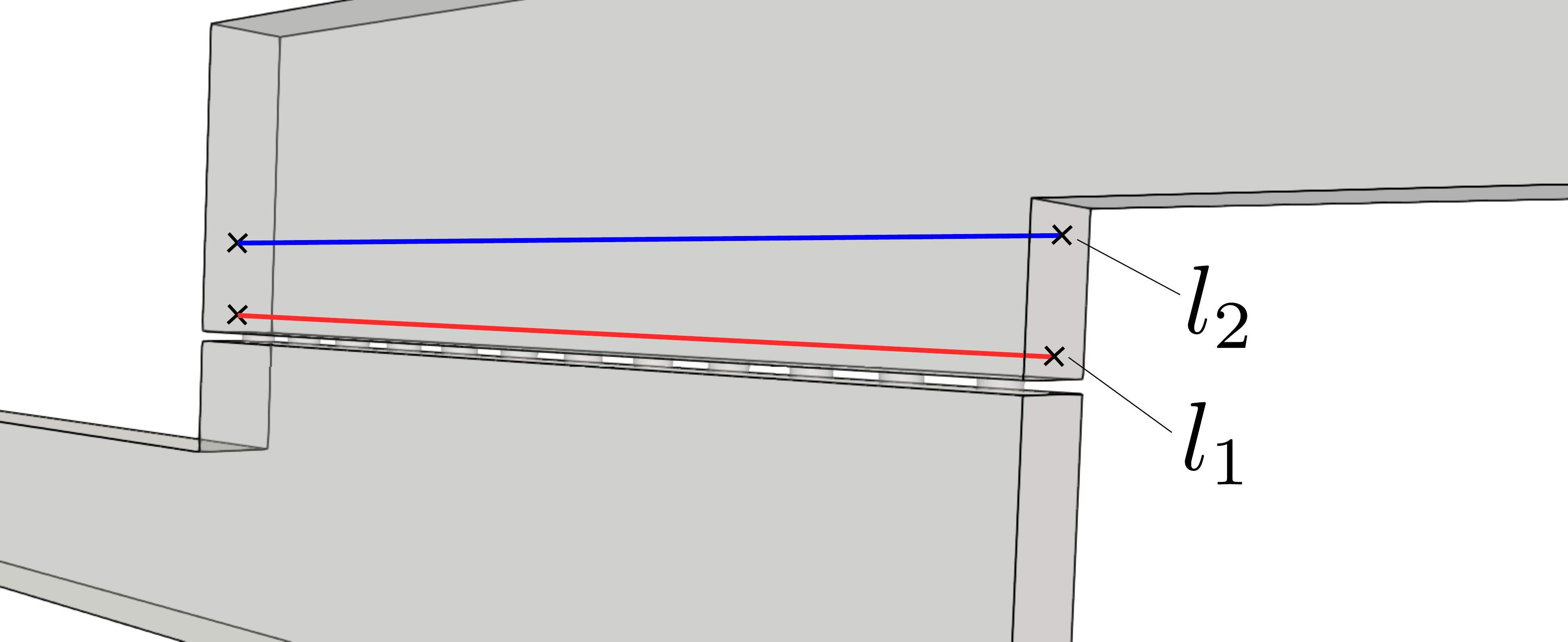}\\
    \begin{tabular}{cc}
    \includegraphics[width=0.49\linewidth]{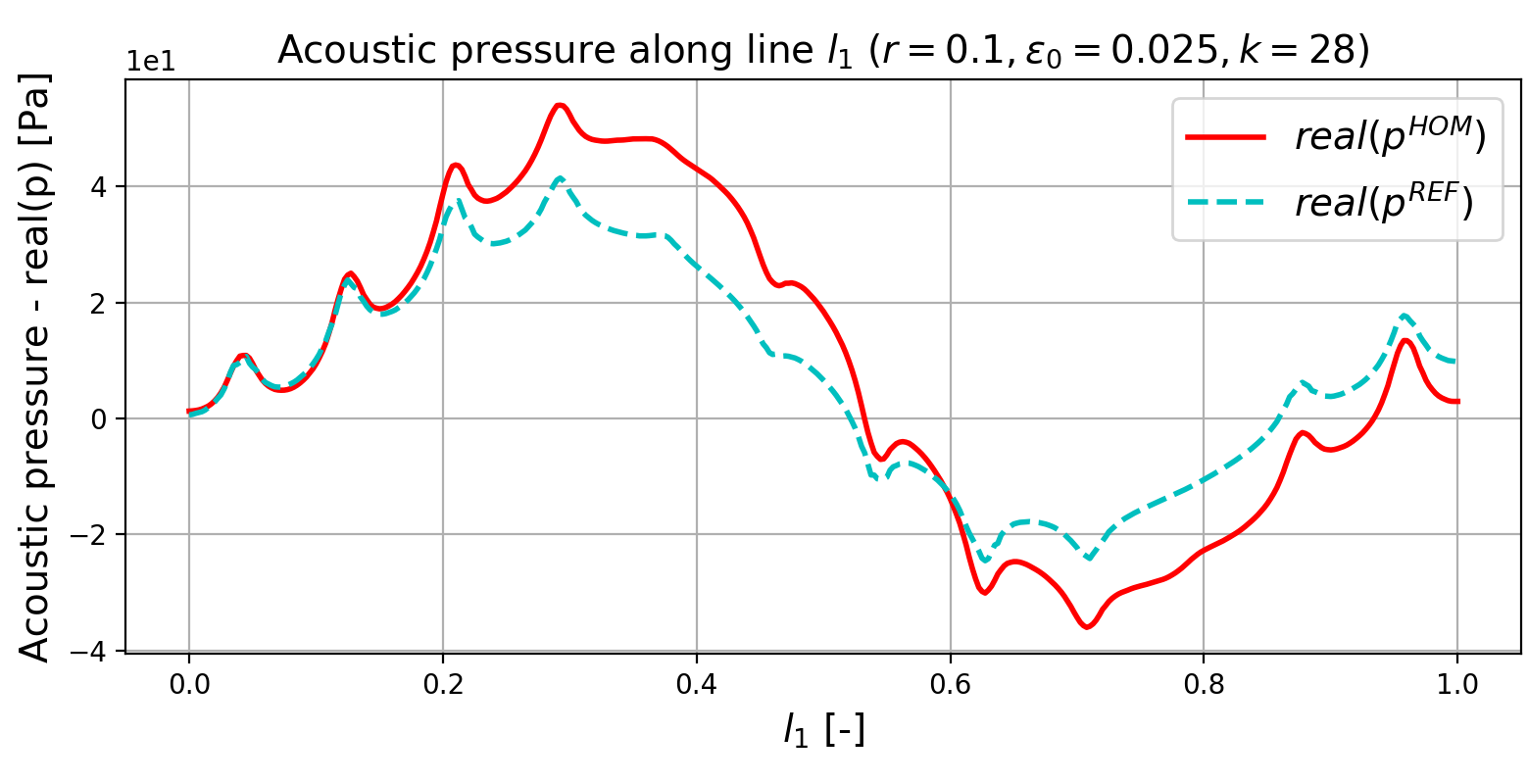} &
    \includegraphics[width=0.49\linewidth]{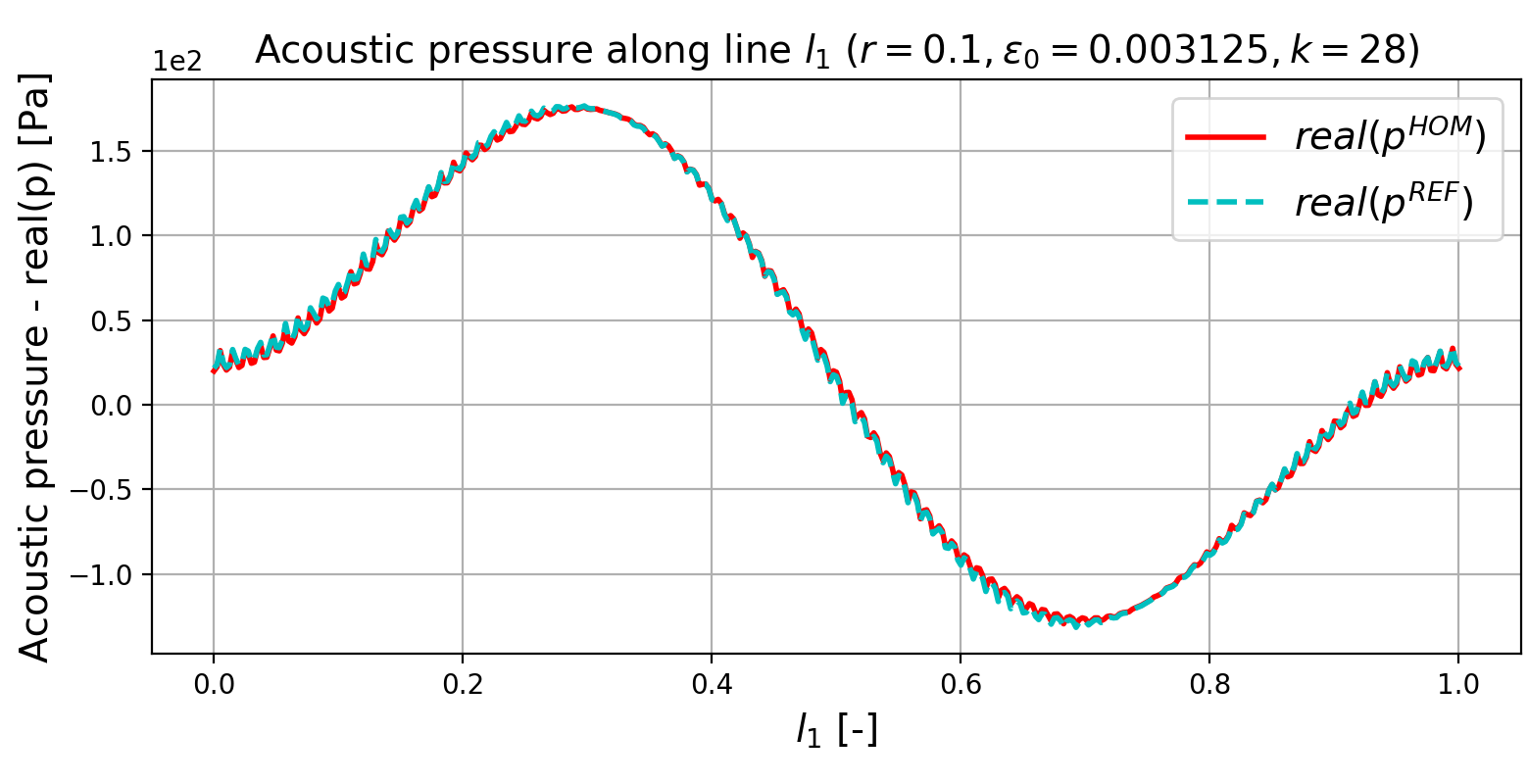}\\
    (a) \chV{$l_1$,}{} $N=12$ & (b) \chV{$l_1$,}{} $N=96$\\
    \includegraphics[width=0.49\linewidth]{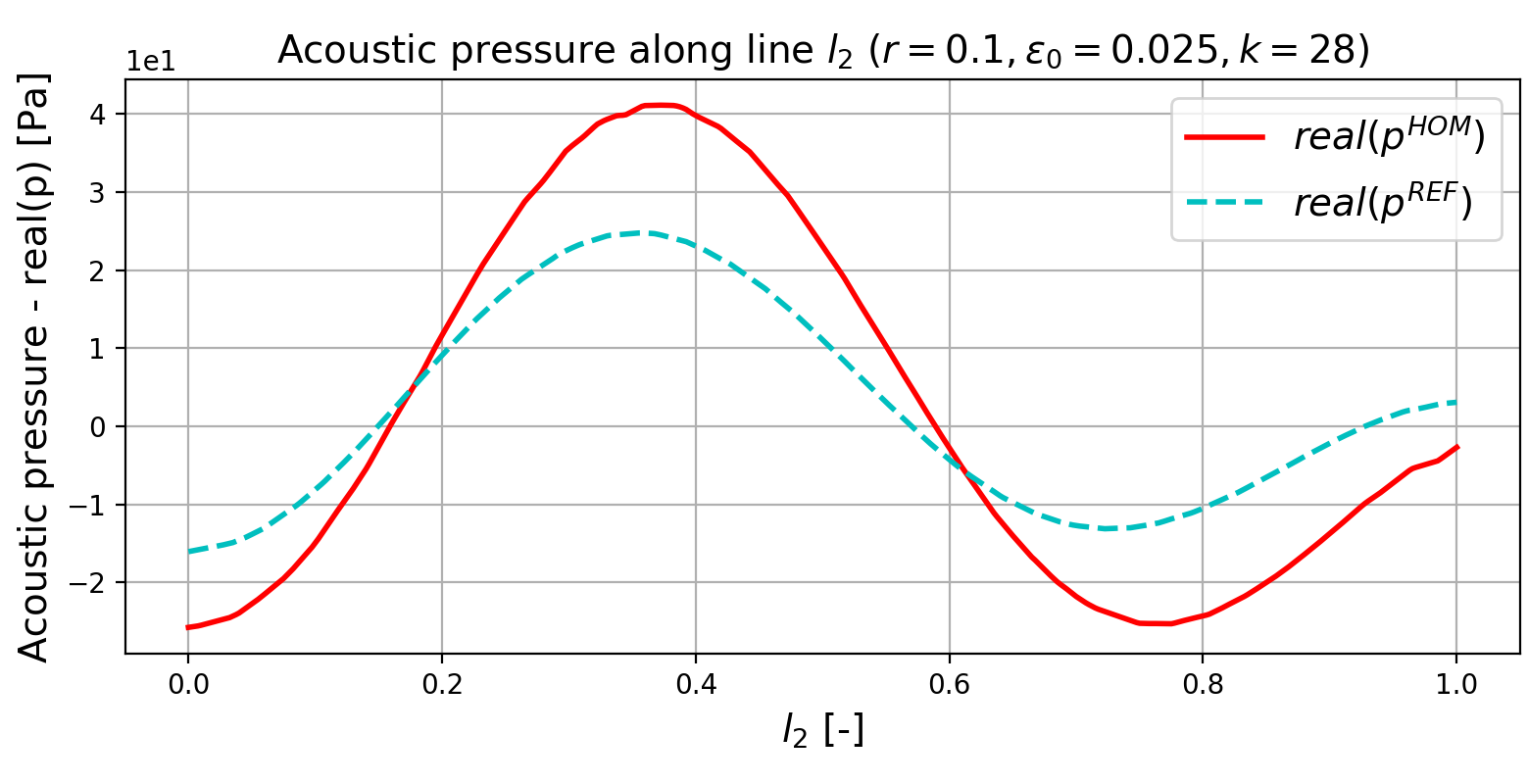} &
    \includegraphics[width=0.49\linewidth]{ac_p_compare_00312_l1_real}\\
    (c) \chV{$l_2$,}{} $N=12$ & (d) \chV{$l_2$,}{} $N=96$\\
    \end{tabular}
    \caption{The real parts of the acoustic pressure along lines
    $l_1$ (a,b), $l_2$ (c,d) obtained by the reference ($p^{REF}$)
     and homogenized ($p^{HOM}$) models for $N=12$ (a,c) and $N=96$ (b,d), holes with radius
     $r=0.1\veps_0$\,m.}
    \label{fig-valid-ac-p}
\end{figure}

\begin{figure}
    \centering
    \includegraphics[width=0.49\linewidth]{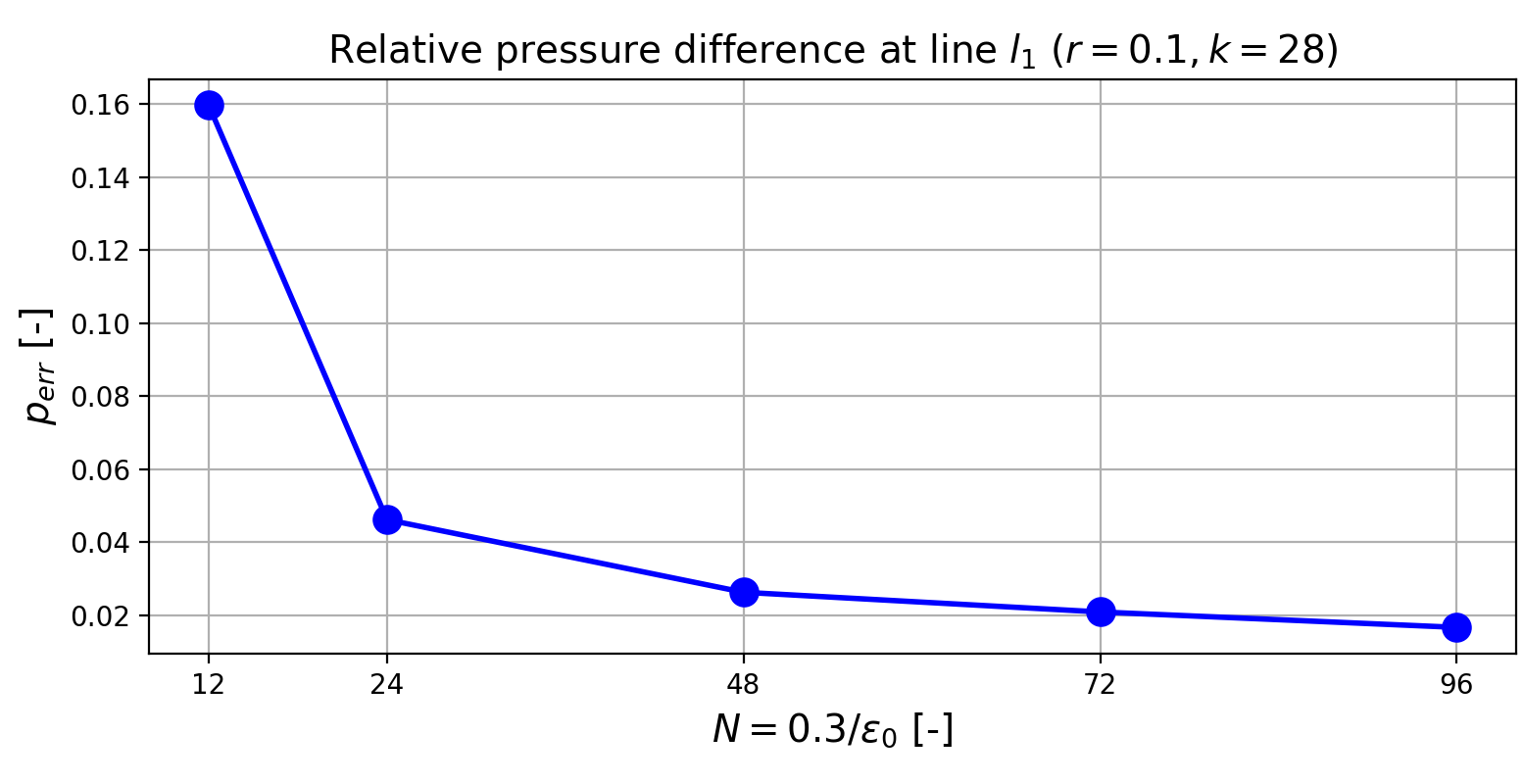}\hfil
    \includegraphics[width=0.49\linewidth]{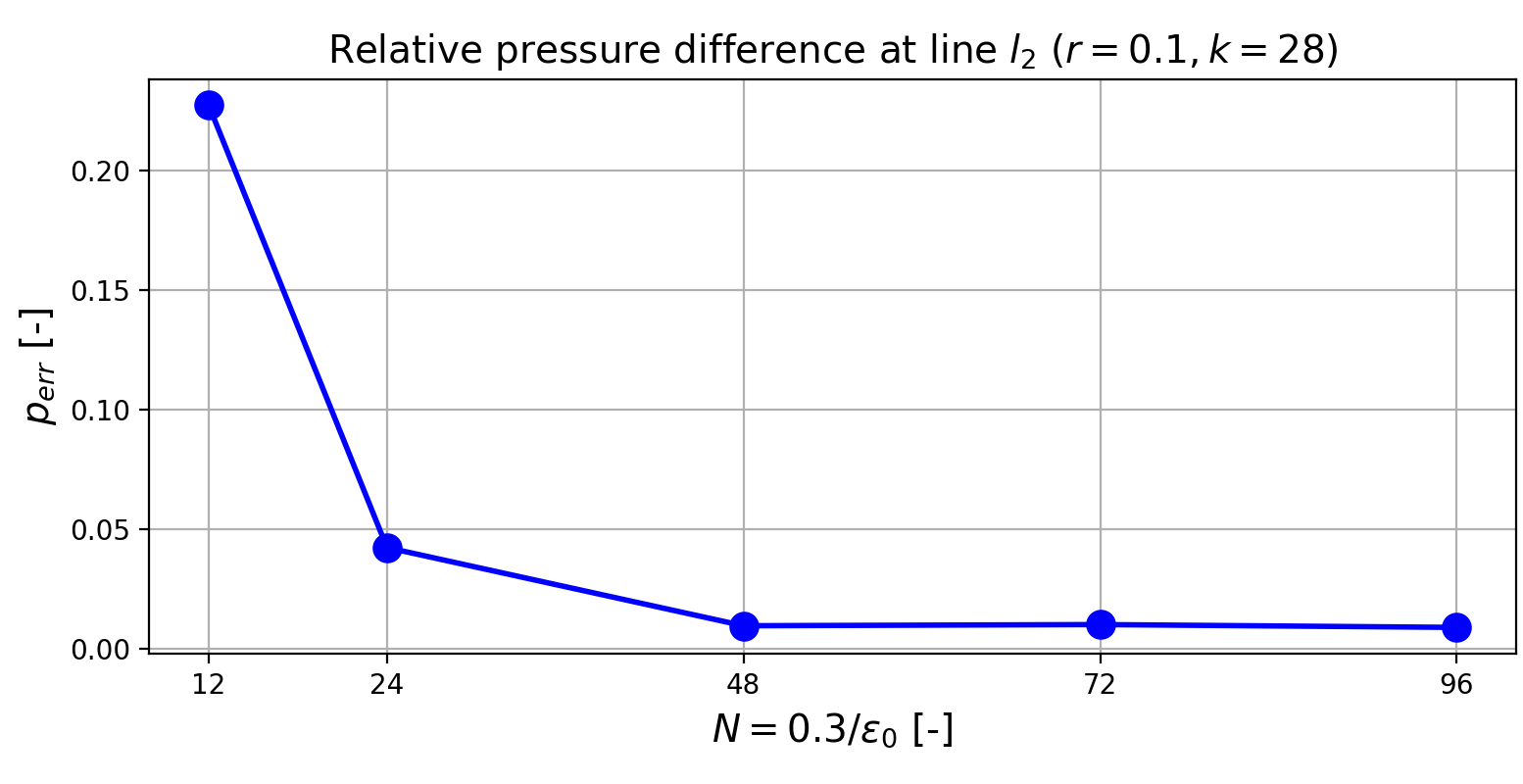}\\
    \caption{Change of the relative pressure error $p_{err}$
     at lines $l_1$ (left) and $l_2$ (right)
      with the increasing number of perforations.}
    \label{fig-valid-ac-perr}
\end{figure}

\begin{figure}
  \centering
  \includegraphics[width=0.49\linewidth]{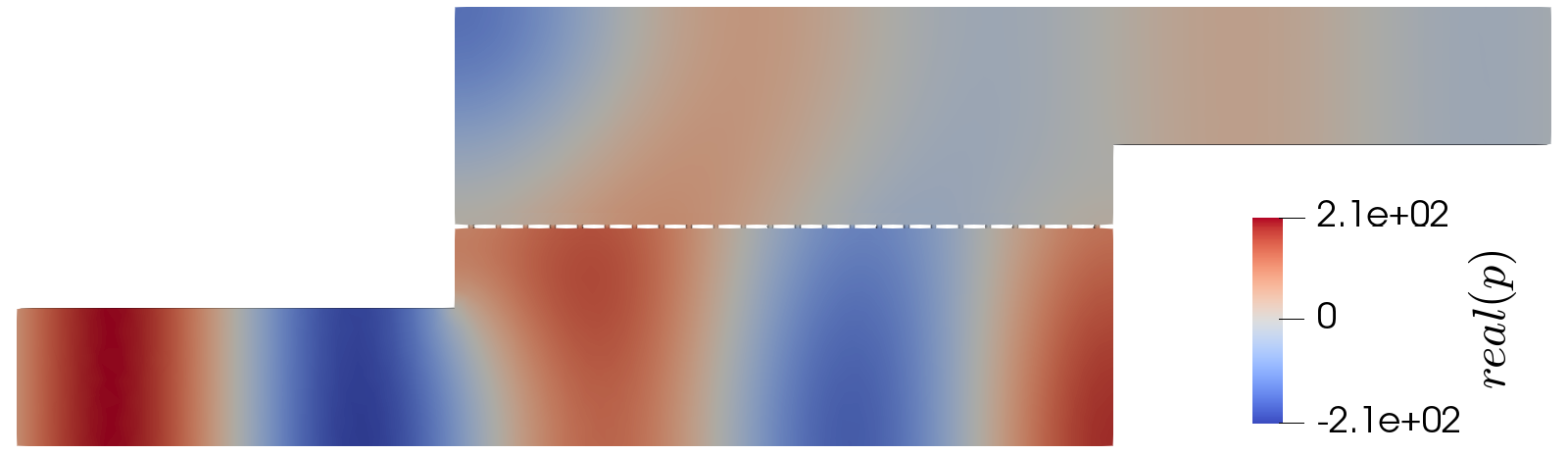}\hfil
  \includegraphics[width=0.49\linewidth]{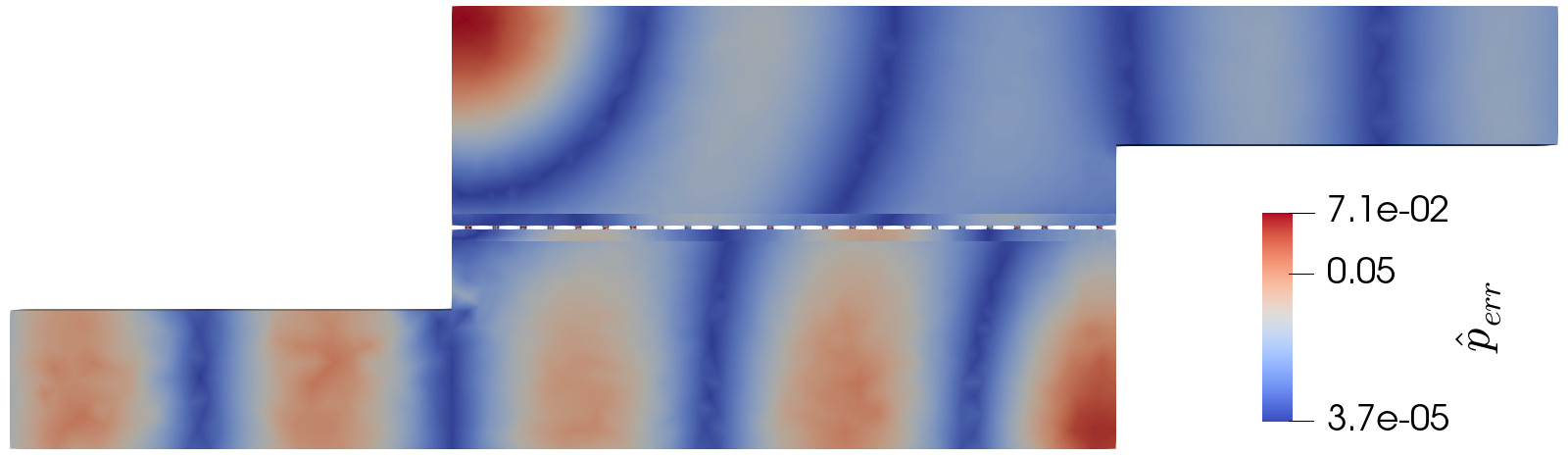}\\
  \caption{Distribution of the acoustic pressure (left)
   and the relative pressure error $\hat p_{err}$ (right).}
  \label{fig-valid-ac-perr-omega}
\end{figure}

We also examined responses of the two models in terms of the acoustic fields
fluctuations in the wave guide near to and far from the perforated plate. For
this, the acoustic pressure distributions were traced along lines $l_1$ and
$l_2$, see Fig.~\ref{fig-valid-ac-p}(top). Line $l_2$ has a fixed place outside
the fictitious layer while $l_1$ lies in the layer, so that its position
depends on $\veps_0$. We perform a series of calculations for perforations with
holes of radii $r=0.1\veps_0$ and wave number $k=28$, whereby $\veps_0 = 0.3/
N$ varies for $N = 12, 24, 48, 72, 96$. The real parts of the acoustic pressure
fields are compared in Fig.~\ref{fig-valid-ac-p}, distributions of the
imaginary parts are similar \chV{}{way} as the real parts. The pressure field
$p^{HOM}$ is reconstructed using the results of the multiscale simulation and
the expressions introduced in \Appx{sec-appendixB}. We define the relative
pressure error associated with a given line $l$ as $p_{err} := \left\vert\,
\norm{p^{REF}}_{l} - \norm{p^{HOM}}_{l}\,\right\vert / \norm{p^{REF}}_{l}$,
where $\norm{~}_l$ is the $L^2(l)$-norm which is well defined due to the
conforming FE approximation of $p$. This error is illustrated in
Fig.~\ref{fig-valid-ac-perr} for the increasing number of perforations $N$. It
can be seen that $p_{err}$ on both lines $l_1$ and $l_2$ is quite high, above
16\%, for $N=12$, however, with growing $N$ the error decreases down to
$\approx 1$\% for $N=96$. The relative pressure error distribution in the whole
domain $\Omega$ is shown for $N=24$ in Fig.~\ref{fig-valid-ac-perr-omega}
(right), the left figure shows the distribution of the pressure $p^{REF}$.

\subsection{Validation test -- compliant perforated plate} 
\begin{figure}
    \centering
    \includegraphics[width=0.95\linewidth]{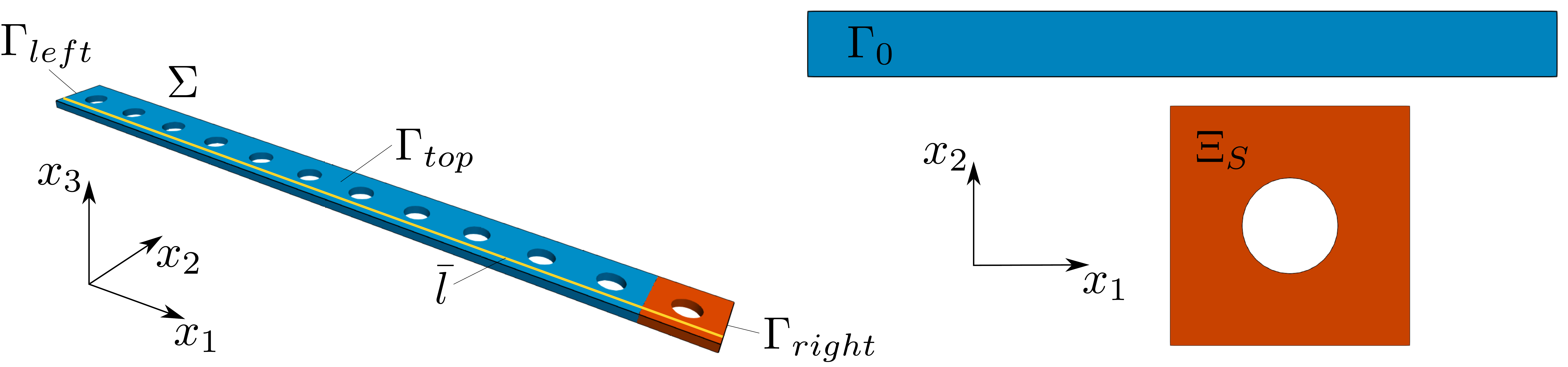}
    \caption{
    Left: The 3D elastic structure $\Sigma$ is used in the direct numerical
    simulation (DNS) as the plate representation; \chV{responses depicted}{responses of the reference
    and homogenized models are compared} in Fig.~\ref{fig-valid-va-w} are
    compared using traces of solutions on the line $\bar l \in \Gamma_{top}$.
    Right: Geometric 2D representation of the Reissner-Mindlin perforated
    plate.}
      \label{fig-valid-va-geom}
\end{figure}

\begin{figure}
    \centering
    \includegraphics[width=0.49\linewidth]{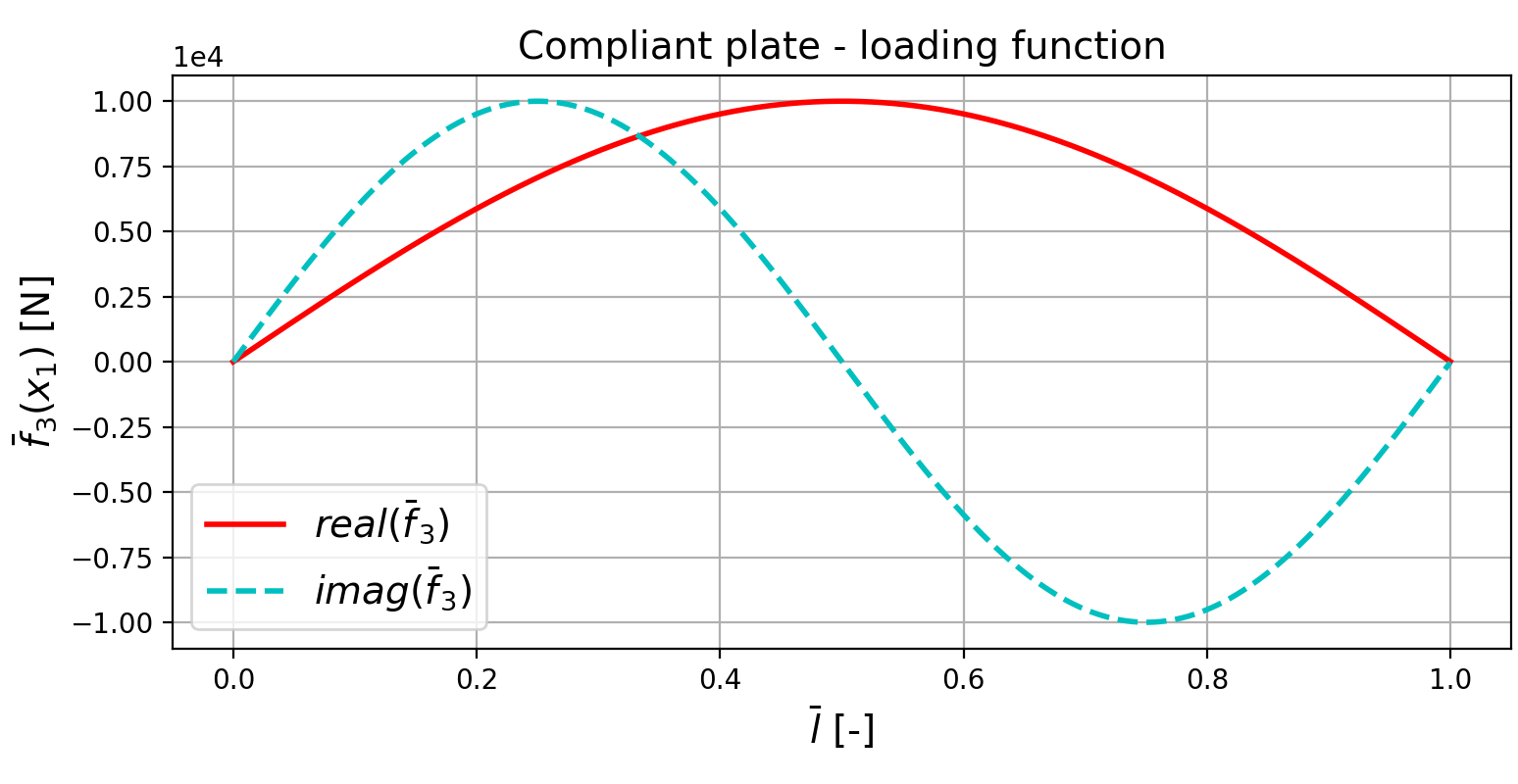}
    \caption{Loading function $\bar f_3$ applied to the compliant plate
        in the $x_3$-axis direction.}
    \label{fig-valid-va-loading}
\end{figure}

The second part of
the validation test concerns the homogenized model of a perforated plate of the
Reissner-Mindlin type. The aim is to compare responses of the homogenized plate
model with the ones of the associated 3D elastic structure with the geometry
depicted in Fig.~\ref{fig-valid-va-geom}. This structure representing the plate
is loaded on its top surface $\Gamma_{top}$ by a prescribed complex loading
traction stress, see Fig.~\ref{fig-valid-va-loading}, which mimics the action
of the acoustic pressure, so that the loading traction is applied in the
out-of-plane direction, axis $x_3$. The perforated plate is fixed at its both ends:
$\ub = \bm{0}$ on $\Gamma_{left}$ and $\Gamma_{right}$, and the periodic
boundary conditions are applied in the $x_2$-axis direction, as it is considered in
the acoustic problem above. The equivalent boundary conditions and loading
function are used in the homogenized model, where the 3D structure is
represented by the plate model described as a 2D structure, see
Fig.~\ref{fig-valid-va-geom}. The \chV{}{elastic material of the 3D structure} material
properties of the plate are given by the Young modulus $E = 70$\,GPa, the
Poisson ration $\nu = 0.35$, and by the density $\rho = 2700$\,kg\,m$^{3}$.


\begin{figure}
    \centering
    \includegraphics[width=0.49\linewidth]{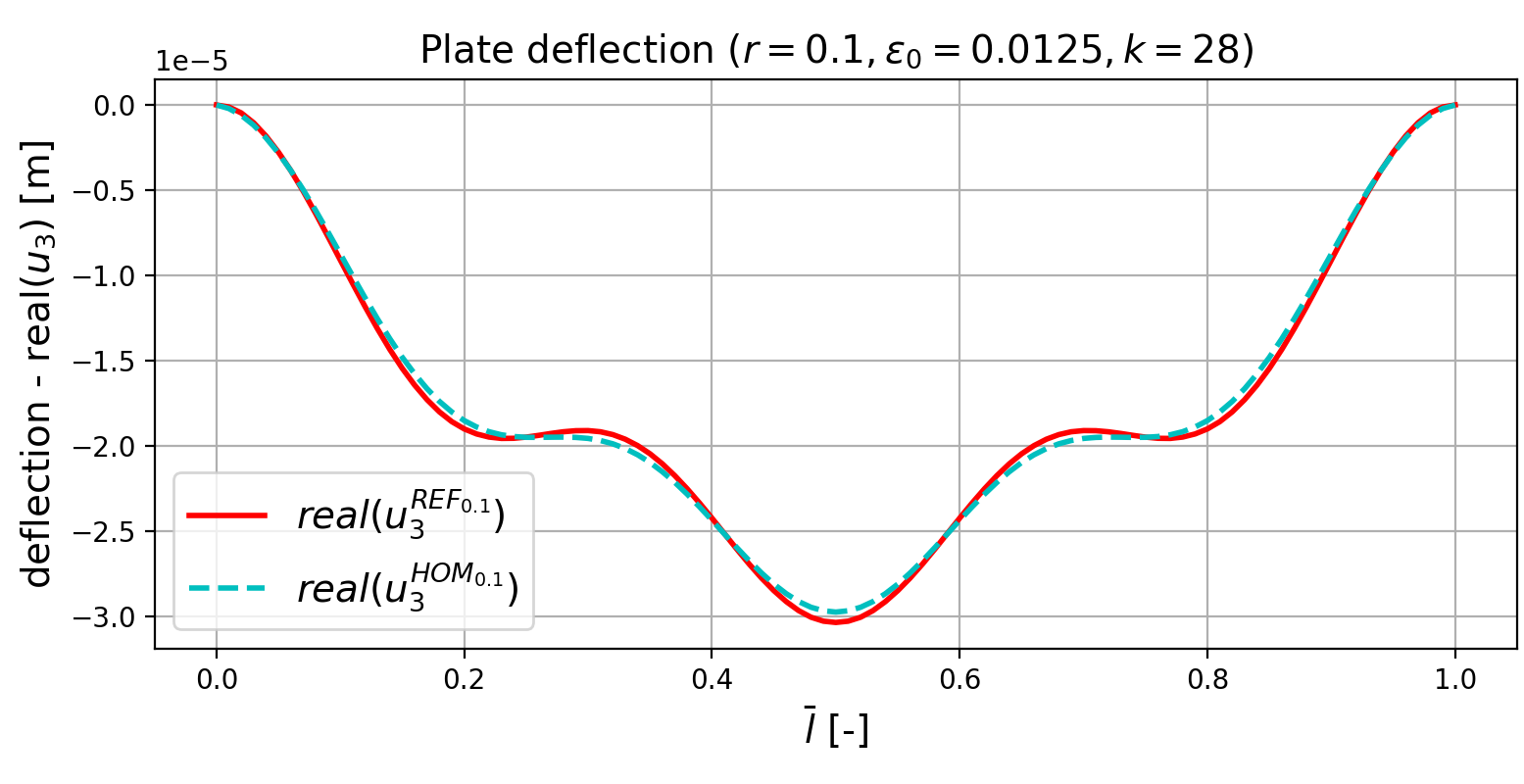}\hfil
    \includegraphics[width=0.49\linewidth]{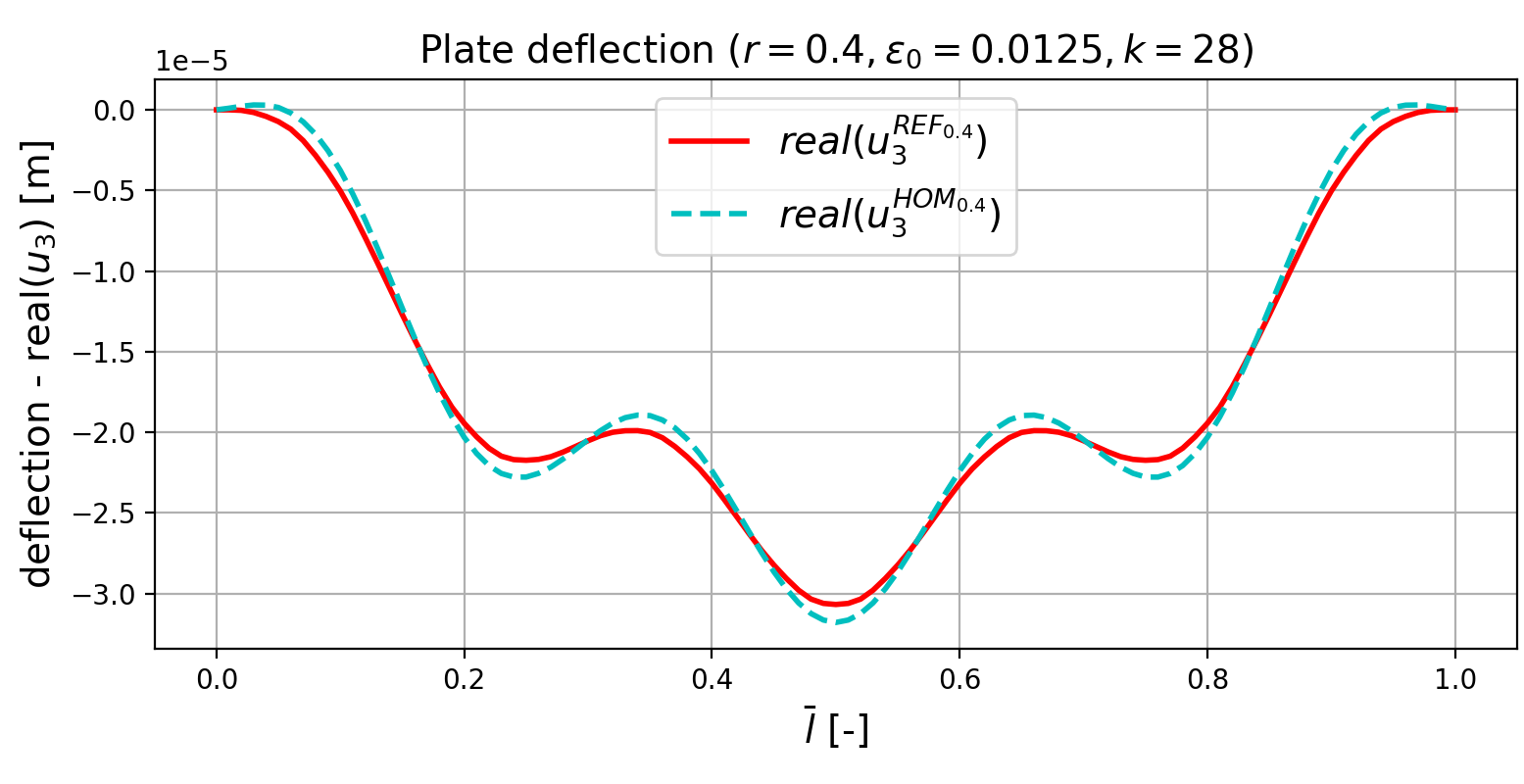}\\
    
    \caption{The real parts of the plate deflection along the central line
    $\bar l$ of domains $\Sigma$ ($u_z^{REF}$) and 
    $\Gamma_0$ ($u_z^{HOM}$) for
    $r=0.1\veps_0$\,m (left) and $r=0.4\veps_0$\,m (right), $N=24$.}
    \label{fig-valid-va-w}
\end{figure}

The plate deflections are computed for the two models, \ie using the DNS of the 3D structure and using the multiscale simulations of the plate. The responses are compared for  a fixed  wave number $k=28$, where $\veps_0 = 0.3/24$, and for
$r=0.1\veps_0$\,m and $r=0.4\veps_0$\,m. As seen in
Fig.~\ref{fig-valid-va-w}, the difference of the results is less than 
5\%, even for the relatively small number of perforations $N = 24$. The values
are plotted along line $\bar l$ which is parallel to axis $x_1$, as shown in
Fig.~\ref{fig-valid-va-geom} left.

\begin{figure}[t]
  \centering
  \includegraphics[width=0.49\linewidth]{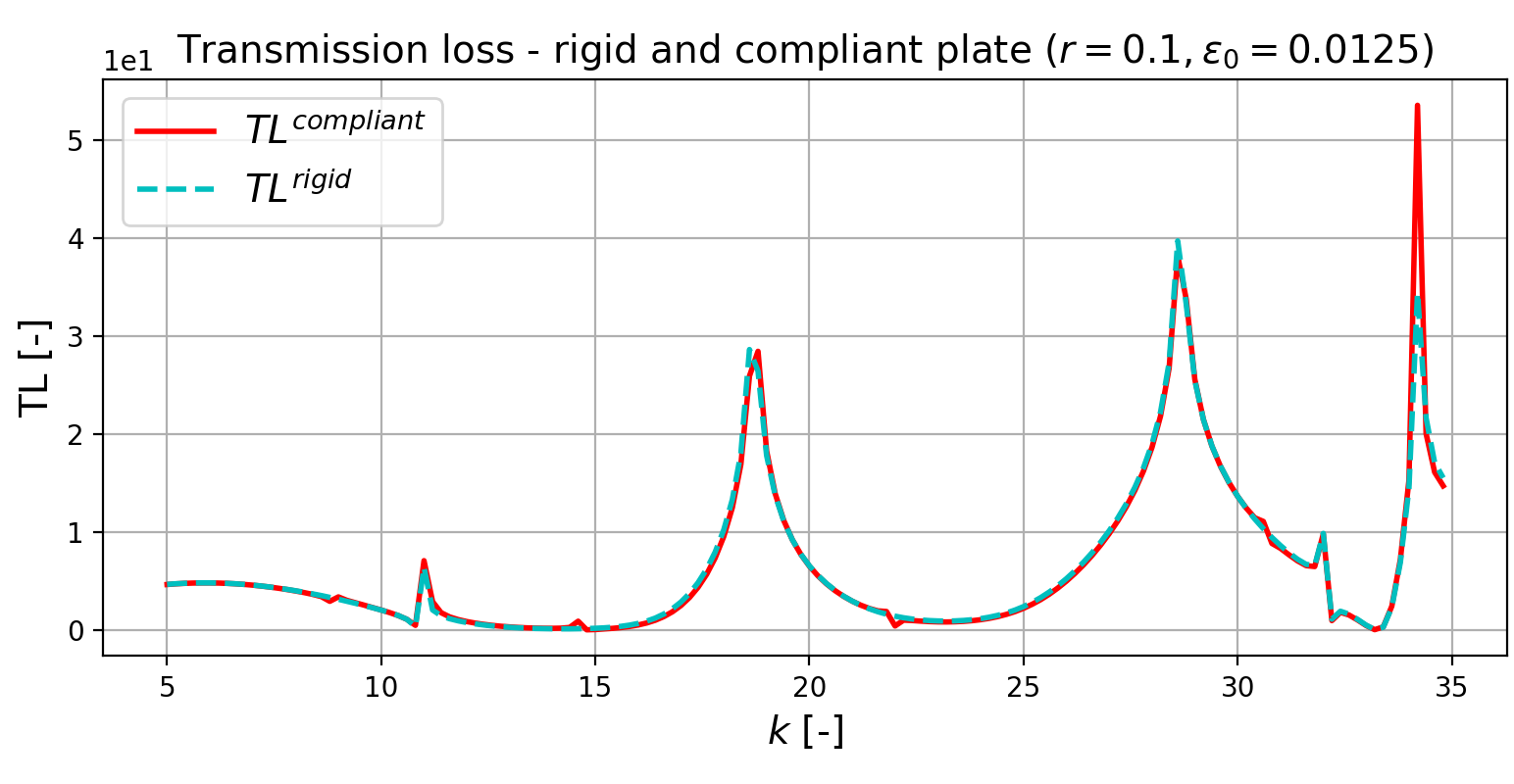}\hfill
  \includegraphics[width=0.49\linewidth]{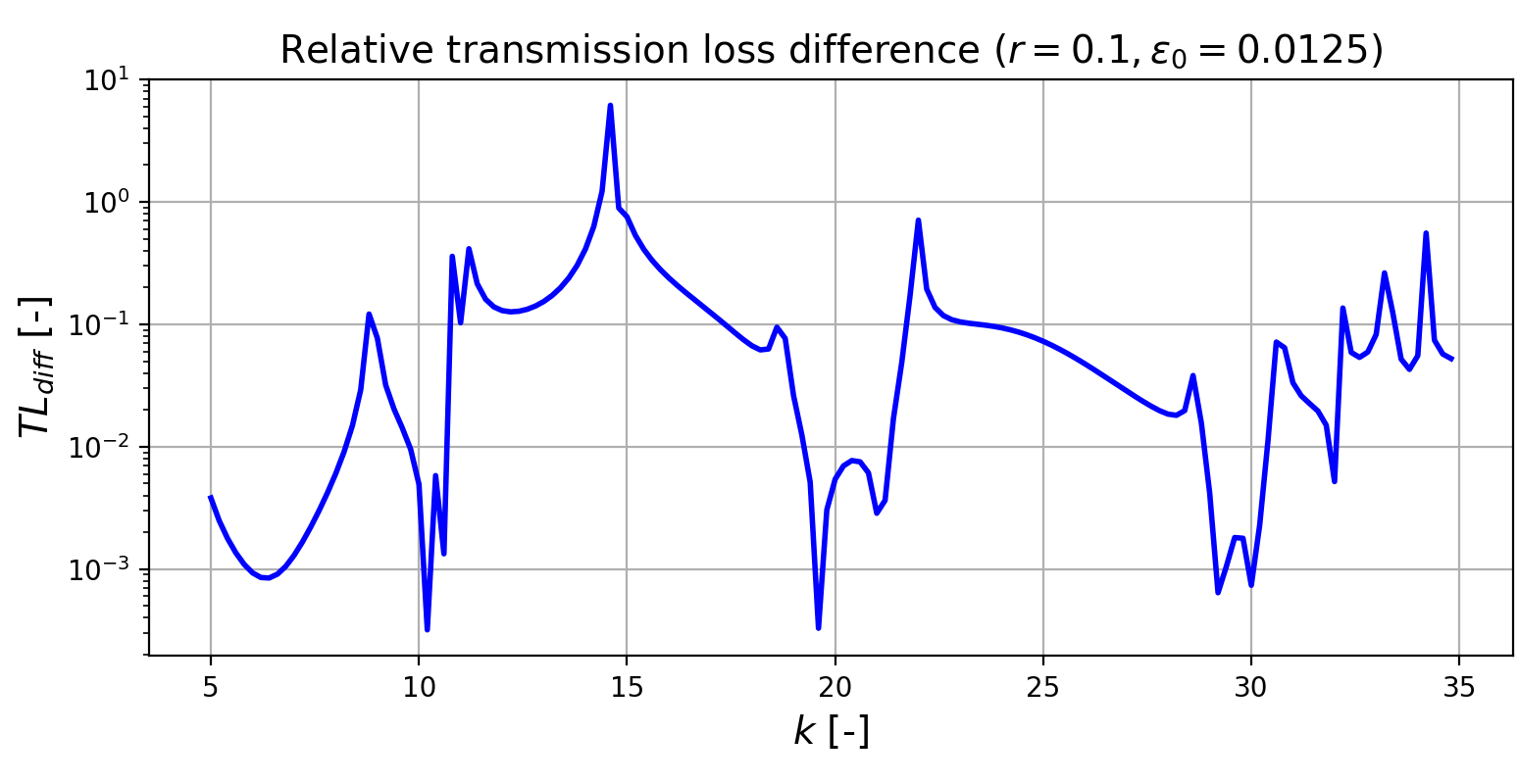}
  
  \caption{Relative difference of the transmission loss values for the rigid
     and compliant plate, $r=0.1\veps_0$\,m, $N=24$.}
  \label{fig-vac-tl}
\end{figure}

The effect of the plate compliance is illustrated in Fig.~\ref{fig-vac-tl},
where we compare the  values computed for the rigid and
compliant perforated interfaces. The relative difference of the values is
defined as $\TL_{diff} = \vert \TL^{rigid} - \TL^{compl} \vert / \TL^{rigid}$.
The influence of the plate compliance on the transmission loss in the waveguide is sensitive on frequency intervals, nevertheless this phenomenon will deserve a further study.

\begin{figure}[h]
  \centering
  \includegraphics[width=0.99\linewidth]{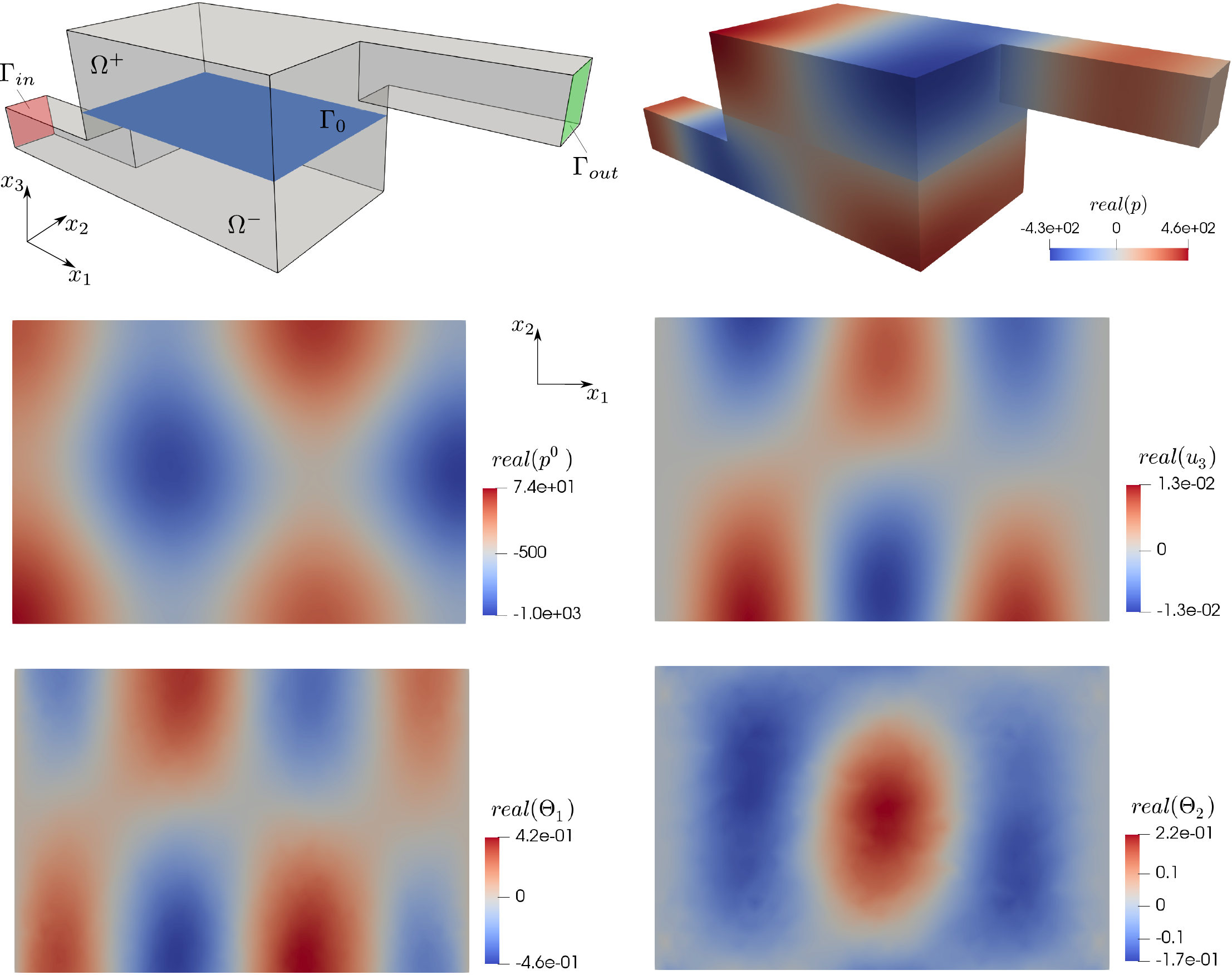}
  
  \caption{
  Acoustic pressure fields and plate deformations induced by the incident wave
  prescribed at the inlet of the waveguide. The plate periodic perforation \chV{is}{and
  its thickness are} characterized by the hole radius, $r=0.1\veps_0$\,m, where
  $\veps_0 = 0.0025$.}
  \label{fig-vac-all}
\end{figure}

\section{Coupled numerical simulation}
\label{sec-simul}

The purpose of this part is to illustrate, how the homogenized vibroacoustic
transmission model derived in this paper can be used for numerical simulations.
acoustic waves using the two-scale 
in this paper. To this aim we consider an analogous problem as the one
specified in Section~\ref{sec-vaild-f}, whereby the mathematical model given by
the coupled equations \eq{eq-DtN-0}-\eq{eq-DtN-3}. The geometry of the
waveguide $\Om^G$ is depicted in Fig.~\ref{fig-vac-all}(top). The boundary
conditions at the inlet and outlet parts of the domain boundary, $\Gamma_{in}$
and $\Gamma_{out}$, are defined as in the validation test reported in
Section~\ref{sec-vaild-f}. On the rest of the boundary $\pd (\hat\Omega^+ \cup
\hat\Omega^-)$ represents the rigid wall, thus, for the acoustic pressure field
the zero Neumann condition applies. The plate represented by interface
$\Gamma_0$ is anchored in the waveguide walls, so that quantities $\ub = 0$ and
$\thetabf = 0$ on $\pd\Gamma_0$. Also the same acoustic fluid and the same
elastic solid are considered, as in the validation tests. The heterogeneous
structure -- the plate perforations -- are specified by the circular holes with
$r=0.1\veps_0$\,m, whereby $\veps_0 = 0.0025$, which corresponds to the number
$N=120$ of holes (counted along the longer side of the plate).

The computed macroscopic responses are shown in
Fig.~\ref{fig-vac-all} which depicts the acoustic pressure field $\hat P$
in $\hat\Omega^+ \cup \hat\Omega^-$ and the distributions of the following
quantities defined in the interface $\Gamma_0$: the ``in-layer'' pressure $p^0$, the
plate deflection $u_3$ and the plate rotations $\thetabf = (\theta_1,\theta_2)$. 

\section{Conclusion}
\label{sec-concl}
In this paper, we derive transmission conditions which serve for coupling acoustic fluid pressure fields on an interface which represents a compliant perforated elastic plate. For this, we consider a fictitious layer which embeds the elastic plate with periodic perforation, such that the perforation period is proportional to the layer and plate thicknesses. To derive the transmission conditions, the layer is decoupled form the ``outer'' acoustic field which is respected by introducing Neumann fluxes (the acoustic momentum). The layer is then treated by the asymptotic analysis based on the periodic unfolding homogenization method. As the result, the layer reduces to the 2D planar manifold $\Gamma_0$ where the homogenized model presents a coupled system of PDEs governing the ``in-layer'' variables: the mean pressure field and the plate deflection and rotations. Further averaging procedure based on a weighted integration in the transversal direction \wrt the layer mid-plane yields additional relationships which enable us to couple the ``outer''  acoustic field with the ``in-layer'' variables. In this way, the Dirichlet-to-Neumann operator is constructed which couples traces of the ``outer''  acoustic pressure with its normal-projected derivatives on both sides of the interface.

The numerical examples reported here illustrate the validation tests which have been performed to explore the modelling errors associated with the homogenization and the ``3D-to-2D'' dimension reduction of the layer which is replaced by the interface coupling conditions.
We used the circular shape of holes, however, arbitrary shaped cylindrical holes can be considered.
The validation tests were based on the comparison of responses computed using the homogenized models with the corresponding responses of reference model, here presented by direct numerical simulations (DNS) of the non-homogenized vibroacoustic problem. In these test, the convergence $\veps\rightarrow 0$ was examined by increasing the number of the periods with proportionally smaller holes. It has been demonstrated that for a sufficiently small $\veps_0$, justifying the scale separation, numerical results obtained using the homogenized
vibroacoustic model are quite close to the corresponding results of the DNS. 
This observation underlines the main advantage of the homogenized model: it provides very good approximation of the reference solution, but at a considerably lower computational cost than the DNS solution. To illustrate the computational effort reduction, in the presented
examples, the piecewise linear FE approximation of the homogenized problem has only about $2\times 10^4$ degrees of freedom (DOFs) at the
macroscopic level, whereas the microscopic subproblems are solved each with about
$1.5\times 10^4$ DOFs to get the homogenized coefficients. To compute
relevant results even  for much simpler geometry employed
in the validation test, the DNS requires more than $2\times 10^5$ DOFs for the
approximation of the acoustic field in the layer  and about the similar number of DOFs for
the compliant elastic structure.

 Among the topics of the future research, the homogenization-based modelling of the compliant plate with arbitrarily shaped periodic perforations presents one of the most interesting issues since such structures provide significantly bigger potential to modify the vibroacoustic transmission. First steps towards optimal design of perforated plates in the acoustic transmission problems were reported in \cite{isma2012,rohan-lukes-icovp2013}.

\paragraph{Acknowledgment}
This research was supported by project  GACR 17-01618S of the Scientific Foundation of the Czech Republic and due to
 the European Regional Development Fund-Project ``Application of Modern Technologies in Medicine and Industry'' {(No.~CZ.02.1.01/0.0/ 0.0/17~048/0007280)}, and in part by project
 LO 1506 of the Czech Ministry of Education, Youth and Sports.

\begin{appendices} 

\section{Appendix}

\subsection{A priori estimates}\label{sec-appendixA}

To derive the \emph{a~priori} estimates on the solution of the problem
\eq{eq-wf6}-\eq{eq-wp7}, we substitute there $q^\veps=p^\veps$, $\vb^\veps =
\ub^\veps$, $v_3 = u_3$ and $\thetabf^\veps = \vthetabf^\veps$. Cosequently,
upon obvious division of the two equations by $\hh$ and $\rho_0$, respectively,
and summation the resulting identities we get
\begin{equation}\label{eq-aes01}
\begin{split}
-\om^2  \int_{\Gamma^\veps} \frac{\rho_S}{\rho_0}
\left[|\ub^\veps|^2 
+  \frac{h^2}{12} |\thetabf^\veps|^2 \right]
-\frac{\om^2}{c^2\hh}\int_{\hat\Om^{*\veps}} |p^\veps|^2 \\
+ \frac{h^2}{12\rho_0} \int_{\Gamma^\veps}
\Eop^\veps\gradplS\thetabf^\veps :\gradplS\thetabf^\veps 
+  \frac{1}{\rho_0} \int_{\Gamma^\veps}
\Eop^\veps\gradplS\ol{\ub}^\veps:\gradplS\ol{\ub}^\veps\\
+ \frac{1}{\rho_0} \int_{\Gamma^\veps}
\shp^\veps|\gradpl u_3^\veps - \thetabf^\veps|^2 
+ \frac{1}{\hh}\int_{\hat\Om^{*\veps}} |\nablad p^\veps|^2 = 
- \frac{\imu \om }{\veps\hh}\int_{\Gamma_\delta^\pm}g^{\veps\pm}p^\veps\;.
\end{split}
\end{equation}
Note that, due to the above substitution of the test functions in
\eq{eq-wf6}-\eq{eq-wp7}, the \rhs interaction terms expressed in terms of the
integrals on $\pd_\circ \Gamma^\veps$ and $\Gamma^\veps$ vanish upon the
summation of the two equalities. Using the asumption on the elastic constants
$\Eop^\veps$, $\Sb^\veps = \shp^\veps\Ib$, there exist constants $c_\Eop$ and
$c_\shp$ (independent of $\veps$), such that we get
 \begin{equation}\label{eq-aes02}
\begin{split}
\left[\frac{h^2}{12\rho_0}  c_\Eop
\nrm{\gradplS\thetabf^{\veps}}{L^2(\Gamma^\veps)}^2 + 
 \frac{1}{\rho_0} \big(
c_\Eop\nrm{\gradplS\ol{\ub}^{\veps}}{[L^2(\Gamma^\veps)]^2}^2 +
c_\shp \nrm{\gradpl u_3^{\veps}-\thetabf^{\veps}}{[L^2(\Gamma^\veps)]^2}^2\big)
\right]\\ + \frac{1}{\hh}\int_{\hat\Om^{*\veps}} |\nablad p^\veps|^2
\leq \om^2  \frac{\bar\rho_S}{\rho_0}\left(
\nrm{u_3^{\veps}}{L^2(\Gamma^\veps)}+\nrm{\ol{\ub}^{\veps}}{[L^2(\Gamma^\veps)]^2}
+\frac{h^2}{12}\nrm{\theta^{\veps}}{[L^2(\Gamma^\veps)]^2}^2
\right)\\
+ \frac{\om^2}{c^2\hh}\nrm{p^\veps}{L^2(\hat\Om^{*\veps})}^2
+\frac{\om}{\veps\hh} \big|\int_{\Gamma^\pm}\hat{g}^{\veps\pm}p^\veps\big|\;.
\end{split}
\end{equation}
To estimate the last \rhs integral, we use the unfolding operation, see \eg
\cite{Cioranescu-etal-2008}, and a smooth extension $\pop{p}$ from
$\hat\Om^{*\veps}$ to whole $\hat\Om$, \ie, for unfolded functions, there is an
extension from $Y^*$ to whole $Y$. For any $p \in H^1(\Om^\veps)$ there exists
$\pop{p} \in H^1(\Om)$ such that $\pop{p}-p = 0$ in $\Om^\veps$ and the
following estimates hold:
\begin{equation}\label{eq-aes03}
\begin{split}
\nrm{\pop{p} - \int_\Om\pop{p} }{L^2(\hat\Om)} & \leq C \nrm{\nabla \pop{p}}{[L^2(\hat\Om)]^3}\;, \\
\nrm{\nabla \pop{p}}{[L^2(\hat\Om)]^3} & \leq C \nrm{\nabla {p}}{[L^2(\hat\Om^{*\veps})]^3}\;.
\end{split}
\end{equation}
The trace theorem yields the following modification of the
Poincar\'e--Wirtinger inequality: for any $f \in H^1(Y)$
\begin{equation}\label{eq-aes04}
\begin{split}
\nrm{f - \MeanY{f}}{L^2(\pd Y)} \leq C \nrm{\nabla_y f}{L^2(Y)} \;.
\end{split}
\end{equation}
Let us define
 \begin{equation}\label{eq-aes05}
\begin{split}
G^\veps(x') = \intY_{I_y^\pm} \Tuf{g^{\veps\pm}}(x',y')\,\dd y'\;,\quad x' \in \Gamma_0\;.
\end{split}
\end{equation}
Now, we can now estimate the last \rhs integral in \eq{eq-aes02},
\begin{equation}\label{eq-aes06}
\begin{split}
&\frac{1}{\veps}\big| \int_{\Gamma^\pm}\hat{g}^{\veps\pm}p^\veps\big|
= \frac{1}{\veps} \int_{\Gamma_0} \intY_{I_y^\pm} |\Tuf{\hat{g}^{\veps\pm}p^\veps}|\\
& \leq \frac{1}{\veps} \int_{\Gamma_0} 
\big|\MeanYe{\Tuf{p^\veps}} \intY_{I_y^\pm} \Tuf{\hat{g}^{\veps\pm}}\big|\\
& \quad + \frac{1}{\veps} \left( \int_{\Gamma_0} \intY_{I_y^\pm}|\Tuf{p^\veps} - \MeanYe{\Tuf{p^\veps}}|^2\right)^{1/2}
 \left( \int_{\Gamma_0}\intY_{I_y^\pm}\big|\Tuf{\hat{g}^{\veps\pm}}\big|\right)^{1/2}\\
& \leq C'
\nrm{p^\veps}{L^2(\hat\Om^{*\veps})}\nrm{\frac{1}{\veps}G^\veps}{L^2(\Gamma)} + 
\frac{C_Y}{\veps}\left(
\int_{\Gamma_0} \intY_{Y} |\nabla_y \Tuf{\pop{p^\veps}}|^2\right)^{1/2}
\nrm{g^{\veps\pm}}{L^2(\Gamma_\delta^\pm)} \\
& \leq C'\nrm{p^\veps}{L^2(\hat\Om^{*\veps})}\nrm{\frac{1}{\veps}G^\veps}{L^2(\Gamma)} + C\left(
 \int_\Om |\hat \nabla_x \pop{p^\veps}|^2\right)^{1/2}
\nrm{\hat{g}^{\veps\pm}}{L^2(\Gamma_\delta^\pm)}\\
& \leq C'\nrm{p^\veps}{L^2(\hat\Om^{*\veps})}\nrm{\frac{1}{\veps}G^\veps}{L^2(\Gamma)} + C\nrm{\hat \nabla_x {p^\veps}}{[L^2(\hat\Om)]^3}
\nrm{\hat{g}^{\veps\pm}}{L^2(\Gamma_\delta^\pm)}\;,
\end{split}
\end{equation}
where all the positive constants $C,C'$ $C_Y$ are independent of $\veps$. Due
to the assumed form of the fluxes $\hat{g}^{\veps\pm}$ given by \eq{eq-aes2}
and assuming the bounded solution $(p^\veps,\ub^\veps,\thetabf^\veps)$ in the
$L^2$ norms, \eq{eq-aes02} with \eq{eq-aes06} yields the desired estimates on
the gradients of the solution, see the Theorem~\ref{thm-aes}.

\subsection{Reconstruction of responses at microlevel}\label{sec-appendixB}

The two-scale field reconstruction of the homogenized model response is based on the coordinate split related to
the periodic lattice. For $\veps_0 > 0$, using the rescaled cell $Z^{\veps_0}$
we introduce its local copies $Z^{K,\veps_0}$ labeled by index $K$ whereby
$\{\bar{x}^K\}_K$ is the set of centers $\bar{x}^K \in \Gamma_0$ of each
$Z^{K,\veps_0}$. For the sake of simplicity, we consider only such domains
$\Gamma_0$ for which the transmission layers $\Om^{\dlt_0}$ are generated as a
union of non overlapping $Z^{K,\veps_0}$, thus (recall that $\ol{Z}$ is the
closure of $Z$)
\begin{equation}\label{eq:Omega}
\ol{\Om^{\dlt_0}} = \ol{\bigcup}_{K \in \Jcal_\Om^{\veps_0}}  Z^{K,\veps_0}\;,\quad
Z^{K,\veps_0} = Z^{\veps_0} + \xibf^K\;,
\end{equation}
where $\Jcal_\Om^{\dlt_0}$ is the set of indices $K$ associated to the lattice
vector $\kb = (k_i) \in \ZZ^2$ such that $\xibf^K = \veps_0 k_i a_i$, recalling
the definition $Y = \prod_i ]-a_i/2, a_i/2[$ and $|Y| = 1$.

For any global position $x \in Z^{K,\veps_0}$, the local ``mesoscopic''
coordinate
\begin{equation}\label{eq:xy}
y = (x - \bar{x}^K)/\veps_0\;,
\end{equation}
can be introduced, such that $y \in Y$. The folding procedure can be
summarized, as follows: for each ``real sized'' cell $Z^{K,\veps_0}$ with its
center $\bar{x}^K \in \Gamma_0$ evaluate the local responses given below as
two-scale functions $f(x',y)$, where $(x',y/\veps_0) \in Z^{K,\veps_0}$, thus,
$y \in Y$ is given, as described above. In what follows, we drop the index $K$
labelling the copy of the local cell.

Due to the homogenization result, by virtue of the characteristic responses, it
is possible to reconstruct the acoustic pressure in the fictitious transmission
layer. For this we use the truncated expansion \eq{eq-va13a}, where $p^1(x',y)$
is depends on $\gradplx p^0$, $g^0$, and $\ub^0$ through \eq{eq-f7}. Thus, for
a given plate thickness $h = \veps_0 \bar h$ yielding $\veps_0$, the acoustic
pressure field in $Z^{*,\veps_0}(\veps_0[x'/\veps_0]_\Xi) = \veps_0 Y^* + \kb
\veps_0$, $\kb = (k_1,k_2) \in \ZZ^2$ which is the local copy of $\veps_0 Y^*$
placed at $x'\in \Gamma_0$, is given as follows,
\begin{equation}\label{eq-f7-R}
\begin{split}
p^{\veps_0} & \approx p^0(x') + \veps_0 p^1(x',y)\;,\quad \mbox{ where }\\
p^1(x',y) & = \pi^\beta(y)\pd_\beta^x p^0(x') + \imu\om \xi(y) g^0(x') +  \imu\om \eta^k(y) u_k^0(x')\\
& = \pi^\beta(y)\frac{1}{2}\pd_\beta^x(\hat P^+(x') + \hat P^-(x')) 
+ \imu\om \xi(y) \frac{1}{2}(\hat G_0^+(x')+\hat G_0^-(x')) + \imu\om \eta^k(y) u_k^0(x')
\;,
\end{split}
\end{equation}
where $y = (y',z) \in Y^*$. We recall that $p^0$ and $g^0$ are expressed in
terms of the global fields $\hat P^\pm$ and $\hat G_0^\pm$ according to
\eq{eq-G3-03}$_2$ and \eq{eq-G3-04}$_1$. It is worth noting, that the position
$x' \in \Gamma_0 \cap Z^{*,\veps_0}(\veps_0[x'/\veps_0]_\Xi)$ varies and $y$ is
determined by \eq{eq:xy}.

\end{appendices}


\bibliographystyle{plain}
\bibliography{bib-acoustics}

\begin{thebibliography}{10}

\bibitem{bb2005}
A.S. Bonnet-Bendhia, D.~Drissi, and N.~Gmati.
\newblock Mathematical analysis of the acoustic diffraction by a muffler
  containing perforated ducts.
\newblock {\em Math. Models and Methods in Appl. Sci.}, 15(7):1059--1090, 2005.

\bibitem{sfepy-multiscale-2019}
R.~Cimrman, V.~Luke\v{s}, and E.~Rohan.
\newblock Multiscale finite element calculations in python using sfepy.
\newblock {\em Advances in Computational Mathematics}, 2019.
\newblock Accepted for publication.

\bibitem{Cioranescu2008-Neumann-sieve}
A.~Cioranescu, D.~Damlamian, G.~Griso, and D.~Onofrei.
\newblock The periodic unfolding method for perforated domains and neumann
  sieve models.
\newblock {\em Journal de Math\'ematiques Pures et Appliqu\'ees},
  89(3):248--277, 2008.

\bibitem{Cioranescu-etal-2008}
D.~Cioranescu, A.~Damlamian, and G.~Griso.
\newblock The periodic unfolding method in homogenization.
\newblock {\em SIAM Journal on Mathematical Analysis}, 40(4):1585--1620, 2008.

\bibitem{Clayes-Delourm-AA2013}
Xavier Claeys and B\'erang{\`e}re Delourme.
\newblock High order asymptotics for wave propagation across thin periodic
  interfaces.
\newblock {\em Asymptotic Analysis}, pages 35--82, 2013.

\bibitem{DELOURME201228}
B\'erang{\`e}re Delourme, Houssem Haddar, and Patrick Joly.
\newblock Approximate models for wave propagation across thin periodic
  interfaces.
\newblock {\em Journal de Math\'ematiques Pures et Appliqu\'ees}, 98(1):28 --
  71, 2012.

\bibitem{Dorlemann2017}
Christina D{\"o}rlemann, Martin Heida, and Ben Schweizer.
\newblock Transmission conditions for the helmholtz-equation in perforated
  domains.
\newblock {\em Vietnam Journal of Mathematics}, 45(1):241--253, 2017.

\bibitem{Jung-etal-2007-JKPS}
S.~S. Jung and et.al.
\newblock Sound absorption of micro-perforated panel.
\newblock {\em Journal of the Korean Physical Society}, 50:1044--1051, 2007.

\bibitem{LIU2016149}
Yu~Liu and Chuanbo He.
\newblock Analytical modelling of acoustic transmission across double-wall
  sandwich shells: Effect of an air gap flow.
\newblock {\em Composite Structures}, 136:149 -- 161, 2016.

\bibitem{Marigo-Maurel-JASA2016}
Jean-Jacques Marigo and Agn{\`e}s Maurel.
\newblock Homogenization models for thin rigid structured surfaces and films.
\newblock {\em J. Acoust. Soc. Am.}, 140(1):260--273, 2016.

\bibitem{Marigo-Maurel-PRSA2016}
Jean-Jacques Marigo and Agn{\`e}s Maurel.
\newblock Two-scale homogenization to determine effective parameters of thin
  metallic-structured films.
\newblock {\em Proceedings of the Royal Society of London A: Mathematical,
  Physical and Engineering Sciences}, 472(2192), 2016.

\bibitem{Maxit-JASA2012}
L.~{Maxit}, C.~{Yang}, L.~{Cheng}, and J.-L. {Guyader}.
\newblock {Modeling of micro-perforated panels in a complex vibro-acoustic
  environment using patch transfer function approach}.
\newblock {\em Acoustical Society of America Journal}, 131:2118, 2012.

\bibitem{Rohan2015bg-plates}
E.~Rohan, R.~Cimrman, and B.~Miara.
\newblock {Modelling response of phononic Reissner-Mindlin plates using a
  spectral decomposition}.
\newblock {\em Appl. Math. Comput.}, 258:617--630, may 2015.

\bibitem{rohan-lukes-waves07}
E.~Rohan and V.~Luke\v{s}.
\newblock Homogenization of the acoustic transmission through perforated layer.
\newblock {\em J. of Comput. and Appl. Math.}, 234:1876--1885, 2010.

\bibitem{isma2012}
E.~Rohan and V.~Luke\v{s}.
\newblock Sensitivity analysis for optimal design of perforated plates in
  vibro-acoustics: homogenization approach.
\newblock In {\em Proceedings of {ISMA} 2012 -- {USD} 2012}, pages 4201--4214.
  KU Leuven, 2012.

\bibitem{rohan-lukes-icovp2013}
E.~Rohan and V.~Luke\v{s}.
\newblock Homogenized perforated interface in acoustic wave propagation --
  modeling and optimization.
\newblock In Z.~Dimitrovov\'{a} et.al., editor, {\em Proc. of the 11th
  International Conference on Vibration Problems, ICOVP 2013}, pages 1--10,
  Lisbon, Portugal, 2013.

\bibitem{rohan-miara-CRAS2011}
E.~Rohan and B.~Miara.
\newblock Band gaps and vibration of strongly heterogeneous
  {R}eissner-{M}indlin elastic plates.
\newblock {\em Comptes Rendus Mathematique}, 349:777--781, 2011.

\bibitem{rohan-miara-ZAMM2015}
Eduard Rohan and Bernadette Miara.
\newblock {Elastodynamics of strongly heterogeneous periodic plates using
  Reissner-Mindlin and Kirchhoff-Love models}.
\newblock {\em ZAMM - J. Appl. Math. Mech. / Zeitschrift f{\"{u}}r Angew. Math.
  und Mech.}, 96(3):304--326, mar 2016.

\bibitem{Sakagami2010}
K.~Sakagami, K.~Matsutani, and M.~Morimoto.
\newblock Sound absorption of a double-leaf micro-perforated panel with an
  air-back cavity and a rigid-back wall: Detailed analysis with a
  helmholtz-kirchhoff integral formulation.
\newblock {\em Applied Acoustics}, 71:411--417, 2010.

\bibitem{Stremtan2012}
F.-A. Stremtan and I.~Lupea.
\newblock Assessing the sound absorption of micro-perforated panels by using
  the transfer function and the impedance tube.
\newblock {\em RJAV}, IX(2):94--99, 2012.
\newblock ISSN 1584-7284.

\bibitem{Takahashi-Tanaka2002}
D.~Takahashi and M.~Tanaka.
\newblock Flexural vibration of perforated plates and porous elastic materials
  under acoustic loading.
\newblock {\em The Journal of the Acoustical Society of America},
  112(4):1456--1464, 2002.

\bibitem{Toyoda-JSV2005}
M.~Toyoda and D~Takahashi.
\newblock Reduction of acoustic radiation by impedance control with a
  perforated absorber system.
\newblock {\em Journal of Sound and Vibration}, 286:601--614, 2005.

\bibitem{Zhou2013}
J.~Zhou, A.~Bhaskar, and X.~Zhang.
\newblock Sound transmission through a double-panel construction lined with
  poroelastic material in the presence of mean flow.
\newblock {\em Journal of Sound and Vibration}, 332:3724--3734, 2013.

\end{thebibliography}

\end{document}